\documentclass[11pt,a4paper,aps,nofootinbib]{article}
\usepackage{jcappub}
\usepackage[labelfont=bf]{caption}
\usepackage[utf8]{inputenc}
\usepackage{color}
\usepackage{amsmath, bm, physics}
\usepackage{amssymb, mathtools}
\usepackage{subcaption}
\usepackage{hyperref}
\usepackage{comment}
\usepackage{multirow}
\usepackage{float}
\usepackage{booktabs}
\usepackage[normalem]{ulem}
\newcommand{\mpl}{M_{\rm Pl}}
\newcommand{\calP}{{\cal P}}

\date{\today}

\title{Mean-Field Backreaction in Single-Field Inflation} 

\author[a]{Matteo Braglia,}
\author[b,c]{Gabriele Franciolini,}
\author[b,c]{Nadir Ijaz,}
\author[b,c]{and Marco Peloso}

\affiliation[a]{Theoretical Physics Department, CERN, 1211 Geneva 23, Switzerland}
\affiliation[b]{Dipartimento di Fisica e Astronomia ``G. Galilei", Universit\`a degli Studi di Padova, via Marzolo 8, I-35131 Padova, Italy}
\affiliation[c]{INFN, Sezione di Padova, via Marzolo 8, I-35131 Padova, Italy}

\emailAdd{matteo.braglia@cern.ch}
\emailAdd{gabriele.franciolini@unipd.it}
\emailAdd{nadir.ijaz@phd.unipd.it}
\emailAdd{marco.peloso@pd.infn.it}

\abstract{The validity of linear perturbation theory and the role of backreaction in single-field inflationary models that enhance power at small scales have recently been under debate. We investigate scalar backreaction semi-analytically in a mean-field approach within the decoupling limit, considering popular models with a transient ultra-slow-roll (USR) stage or an oscillatory feature (OF) in the potential. We find that ultraviolet (UV) modes in the Bunch--Davies vacuum can substantially affect backreaction estimates unless removed by adiabatic subtraction. Our mean-field treatment can reproduce recent lattice results, but reveals a spurious sensitivity to UV contributions in some cases. This sensitivity is reduced in the OF model when copious particle production separates the amplified physical modes from the UV vacuum tail, making lattice results more reliable in this regime. In USR, instead, adiabatically subtracted spectra yield milder backreaction on the background evolution than previously inferred.}

\begin{document}

\begin{flushright}
CERN-TH-2026-239
\end{flushright}
\maketitle

\section{Introduction}
\label{sec:intro}
Inflation is arguably the leading candidate framework to solve the shortcomings of the hot Big Bang scenario and produce the observed spectrum of initial density perturbations that seed the Cosmic Microwave Background (CMB) anisotropies and the Large Scale Structure (LSS) observed in our Universe. Cosmological observations are in striking agreement with the so-called slow-roll vanilla models, where a scalar field slowly rolls down an almost flat potential, sourcing nearly scale-invariant spectra for the density perturbations~\cite{Planck:2018jri}. On the other hand, the constraints become much weaker at smaller scales, which cross the horizon at later stages during inflation. This raises the possibility that the scalar power spectrum could differ significantly from the one at large scales and, in fact, be amplified by several orders of magnitude.

Such a tantalizing possibility would have profound phenomenological implications, including the generation of Primordial Black Holes (PBHs)---which could be dark matter candidates~\cite{Byrnes:2025tji}--and a Stochastic Gravitational Wave Background (SGWB), sourced at second order in perturbation theory by the large scalar fluctuations~\cite{Tomita:1967wkp,Matarrese:1992rp,Domenech:2021ztg}. This has stimulated model builders to devise scenarios capable of generating such large fluctuations, whose phenomenological consequences are routinely investigated. Given such observable consequences, it is desirable to have theoretical predictions under control.

The standard procedure for computing these spectra is to define a cosmological background and compute the evolution of the perturbations using cosmological linear perturbation theory. While this is certainly justified for slow-roll inflation, it has recently been questioned in the context of models amplifying perturbations at small scales, see e.g.~\cite{Cheng:2021lif,Inomata:2022yte,Kristiano:2022maq,Riotto:2023hoz,Firouzjahi:2023aum,Motohashi:2023syh,Franciolini:2023lgy,Tasinato:2023ukp,Cheng:2023ikq,Fumagalli:2023hpa,Maity:2023qzw,Davies:2023hhn,Tada:2023rgp,Iacconi:2023ggt,Inomata:2024lud,Firouzjahi:2024psd,Ballesteros:2024zdp,Kristiano:2024vst,Kawaguchi:2024rsv,Fumagalli:2024jzz,Ruiz:2024weh,Firouzjahi:2024sce,Sheikhahmadi:2024peu,Inomata:2025bqw,Fang:2025vhi,Firouzjahi:2025gja,Firouzjahi:2025ihn,Inomata:2025pqa,Iacconi:2026uzo,Li:2026vrn,Fang:2025kgf,Fang:2026off} for an incomplete list of works. While most of these works addressed the question of perturbativity in the context of in-in perturbation theory at 1-loop, a few works also explored non-perturbative methods on the lattice~\cite{Caravano:2024tlp,Caravano:2024moy,Caravano:2025diq,Caravano:2026hca,Mizuguchi:2024kbl} and even attempted full numerical general relativity (GR) simulations~\cite{Launay:2025kef}. A related question is the extent to which perturbations can backreact on the background inflationary dynamics.

Lattice simulations are particularly promising, as they can evolve the system of equations with full non-linearity without relying on a perturbative expansion. The question then becomes how classical lattice techniques can account for quantum effects when these are relevant, and whether the contributions of quantum UV modes have been consistently subtracted. These issues motivate recent discussions of the limitations of inflationary lattice simulations~\cite{Barker:2026yyg,Ballesteros:2026evr,del-Corral:2026lsf}. A full renormalization of inflationary observables also presents several technical and conceptual challenges~\cite{Ballesteros:2024cef,Braglia:2025cee,Braglia:2025qrb,Ballesteros:2025nhz,Braglia:2026fle,Ballesteros:2026fwo}. It would therefore be desirable to complement numerical simulations with an analytical framework that helps clarify the interplay between background dynamics, perturbations, and their quantum contributions. In this paper, we adopt a mean-field approximation, also referred to as homogeneous backreaction, where we study the evolution of the perturbations in a background consistently evolved while taking into account their backreaction. We take inspiration from the related context of axion inflation, where tachyonically amplified helical gauge fields backreact on the background dynamics~\cite{Peloso:2022ovc,vonEckardstein:2023gwk,Caravano:2024xsb,Barbon:2025wjl,Franciolini:2026cps}.  Here we apply this approach to single-field models of inflation amplifying small-scale perturbations, derive the quantities backreacting on the background dynamics, and identify the perturbative regime in which our method can be applied.
Our approach differs from stochastic inflation~\cite{Starobinsky:1986fx,Starobinsky:1994bd}, where the coarse-grained field excludes subhorizon modes by construction and their effect on long-wavelength dynamics is encoded in stochastic noise. By retaining the explicit evolution of subhorizon modes, our treatment can be more informative about their contribution to backreaction and the need for UV regularization\footnote{A similar approach has already been adopted in Refs.~\cite{Syu:2019uwx,Cheng:2021lif,Cheng:2023ikq}, where, however, the authors did not take subhorizon modes into account, as they did not adopt a regularization prescription.}.

Before introducing the applications, we regularize the quantities sourcing backreaction by adiabatic subtraction. These arise at 1-loop in perturbation theory and contain UV divergences, which we remove by adiabatic subtraction based on the Wentzel--Kramers--Brillouin (WKB) expansion~\cite{Zeldovich:1971mw,Parker:1974qw,Parker:2009uva}, routinely used in cosmological applications~\cite{Finelli:2004bm,Finelli:2007fr,Finelli:2008zg,Durrer:2009ii,Marozzi:2011da,delRio:2014cha,Ferreiro:2018qzr,Ferreiro:2019cyp,Kobayashi:2014zza,Chu:2016kwv,Adshead:2018oaa,Adshead:2019aac,Ballardini:2019rqh,Animali:2022lig,Pla:2024xsv,Negro:2024bbf,Negro:2024iwy,Hoory:2025qgm,Hoory:2026clf}\footnote{See also~\cite{Kristiano:2025ajj,Ballesteros:2025nhz,Braglia:2026fle,Ballesteros:2026fwo} for applications of the WKB expansion not in the context of the adiabatic subtraction.}. We then apply our method to two models that enhance scalar perturbations through intrinsically different physical mechanisms: a transient USR stage and the OF model. We show examples of regions of parameter space where backreaction remains perturbatively under control. For the USR benchmarks considered, backreaction remains small after adiabatic subtraction, whereas the OF model can exhibit sizeable backreaction associated with efficient particle production. Our analysis helps distinguish this physical contribution from a spurious dependence on the UV cutoff in lattice simulations.

The paper is organized as follows. In Section~\ref{sec:system} we introduce the mean-field framework and derive the backreacted background equations in the decoupling limit. In Section~\ref{sec:adiabatic} we discuss the adiabatic regularization of the fluctuation spectra and the validity of the WKB approximation. In Section~\ref{sec:example} we apply the formalism to representative USR and OF benchmarks, and highlight how the different physics at play influences the backreaction. In Section~\ref{sec:results} we compare bare-mode calculations with lattice simulations and examine their UV sensitivity, before discussing the range of validity of the mean-field approximation. We summarize our conclusions in Section~\ref{sec:conslusions}. The Appendices collect technical derivations, details of the benchmark models, and the all-scale expressions for the backreaction including metric perturbations, thus relaxing the decoupling-limit approximation.

\section{Mean-field backreaction equations in single field inflation}
\label{sec:system}

We consider the action
\begin{equation}
S_{\rm EH}+S_\phi =  \int d^4 x \sqrt{-g} \left[\frac{\mpl^2}{2}R+ \frac{1}{2} g^{\mu \nu} \partial_\mu \phi \partial_\nu \phi - V \left( \phi \right) \right] \;, 
\label{S-phi}
\end{equation}
describing General Relativity plus the minimally coupled inflaton field $\phi$. We work in the unperturbed flat Friedmann--Lema\^itre--Robertson--Walker (FLRW) metric, with line element $d s^2 = d t^2 - a^2 \left( t \right) d x^i d x^i = a^2 \left( \tau \right) \left[ d \tau^2 - d x^i d x^i \right]$ where $t$ and $\tau$ are the cosmic and the conformal time, respectively, $a$ is the scale factor, and $H\equiv\dot a/a$ is the Hubble rate. The Latin indices $i=1,2,3$ run over the spatial coordinates and the Einstein convention for implicit summation over repeated indices is employed. From this action one obtains the equation of motion for the scalar field 
\begin{equation}
\frac{1}{\sqrt{-g}} \partial_\mu \left( \sqrt{-g} g^{\mu \nu} \partial_\nu \phi \right) + \frac{\partial V}{\partial \phi} = 0 \;, 
\label{eom}
\end{equation} 
as well as the energy-momentum tensor 

\begin{equation} 
T_{\mu \nu } \equiv \frac{2}{\sqrt{-g}} \, \frac{\delta S_\phi}{\delta g^{\mu \nu}} = \partial_\mu \phi \, \partial_\nu \phi - g_{\mu \nu} \left( \frac{1}{2} g^{\alpha \beta} \, \partial_\alpha \, \phi \, \partial_\beta \phi - V \right) \;. 
\end{equation} 
which is employed in the Einstein equations $G_{\mu \nu} = {T_{\mu \nu}}/{M_{\rm Pl}^2}$, where $M_{\rm Pl}$ indicates the reduced Planck mass. 

We decompose the scalar field according to 
\begin{equation}
\label{eq:definition_phi_background}
\phi \left( t ,\, \vec{x} \right) \equiv \varphi \left( t \right) + \delta \phi \left( t ,\, \vec{x} \right) \;\;\;,\;\;\; \varphi \equiv \left\langle \phi \right\rangle 
\end{equation}
where  $\left\langle \dots \right\rangle$ denotes a spatial average (we note that, by construction, $\delta \phi $ has vanishing spatial average).

Our goal is to compute the backreaction of $\delta \phi$ on the background dynamics to leading order in perturbation theory in models for which the inflaton fluctuation is highly amplified. 
The perturbations of the scalar field are influenced by gravity in two ways. On one hand, they are affected by the background expansion. On the other, the non-linear nature of gravity directly couples them to the metric perturbations, which significantly complicates the analysis.  Luckily, a decoupling limit exists~\cite{Cheung:2007st}, where one can ignore metric perturbations, and the dominant gravitational effects on the evolution of $\delta\phi$, as well as on the backreaction, are induced by the background expansion. This regime corresponds to a slowly varying background, and it is quantified by
\begin{equation}
    \epsilon_H\equiv-\frac{\dot{H}}{H^2}\ll1.
    \label{def-veh}
\end{equation}
It is also useful to introduce the second Hubble slow-roll parameter $ \eta_H\equiv-{\ddot{H}}/({2H\dot{H}})$. Their relation to the inflaton dynamics and the suppression of metric contributions are discussed in Appendix~\ref{app:linear-dynamics}. In the main text, we will consider scenarios in which it is justified to work within this decoupling limit, where calculations are significantly simplified and the backreaction quantities take a compact form. Nevertheless, we extend our formulae by including metric perturbations in Appendix~\ref{sec:Bphi_allscale}, where we will also confirm the results derived here in the decoupling limit.

By retaining the leading-order quadratic contributions of $\delta \phi$ to the spatial average of Eq.~\eqref{eom} and of the $00$ and diagonal spatial Einstein equations, which correspond to the backreaction at 1-loop in perturbation theory, we get: 
\begin{eqnarray}
&& \ddot{\varphi}  + 3 H \dot{\varphi}   + \frac{\partial V}{\partial \varphi}  + \frac{1}{2}  \frac{\partial^3 V}{\partial \varphi^3} \left\langle \delta \phi^2 \right\rangle \simeq 0 \;, \label{eq:KG_backreaction}\\ 
&& H^2 = \frac{\left\langle T_{00} \right\rangle}{3 M_{\rm Pl}^2} \simeq \frac{1}{3M_{\rm Pl}^2} \left[ \frac{1}{2} \dot{\varphi}^2 + V + \left\langle \frac{1}{2} \delta \dot{\phi}^2 + \frac{1}{2 a^2} \partial_i \delta \phi \partial_i \delta \phi + \frac{1}{2} \frac{d^2 V}{d \varphi^2} \delta \phi^2 \right\rangle \right] \;, \label{eq:H2_backreaction}\\ 
&&  \dot{H} = -\frac{\left\langle T_{00} + \frac{1}{3 a^2} T_{ii} \right\rangle}{2 M_{\rm Pl}^2} \simeq - \frac{1}{2 M_{\rm Pl}^2} \left[ \dot{\varphi}^2 + \left\langle \delta \dot{\phi}^2 + \frac{1}{3 a^2} \partial_i \delta \phi \partial_i \delta \phi \right\rangle \right] \;,  \label{eq:Hdot_backreaction}
\end{eqnarray} 
 where the first of these equations matches the results derived in~\cite{Inomata:2024lud}, where backreaction on the Hubble expansion was neglected. The backreaction on $H^2$ was considered in Refs.~\cite{Cheng:2021lif,Cheng:2023ikq}.

 To evaluate the backreaction terms, we also need the evolution equation for $\delta \phi$, obtained from the difference between Eq.~\eqref{eom} and its spatial average 
\begin{equation}
\delta \ddot{\phi}  + 3 H \delta \dot{\phi} - \frac{1}{a^2} \Delta \delta \phi  + \frac{\partial^2 V}{\partial \varphi^2} \, \delta \phi + \frac{1}{2}  \frac{\partial^3 V}{\partial \varphi^3} \left(  \delta \phi^2 - \left\langle \delta \phi^2 \right\rangle \right) = 0 \;. 
\label{eq-dphi}
\end{equation} 
One can verify that Eq.~\eqref{eq:Hdot_backreaction} can be obtained from the time derivative of Eq.~\eqref{eq:H2_backreaction}, combined with Eq.~\eqref{eq:KG_backreaction} and with~\eqref{eq-dphi}. As is well known, the fact that one equation is not independent is a consequence of a nontrivial Bianchi identity. Equation~\eqref{eq-dphi} is derived by retaining terms quadratic in $\delta\phi$, at the same order in the fluctuation expansion as the background equations~\eqref{eq:KG_backreaction}--\eqref{eq:Hdot_backreaction}. At 1-loop, however, the expectation values sourcing the background depend only on the leading-order perturbations. In our mean-field treatment, we evaluate these by evolving the linear mode functions self-consistently on the backreacted background. We therefore solve the linear part of Eq.~\eqref{eq-dphi}
\begin{equation}
\delta \ddot{\phi}  + 3 H \delta \dot{\phi} - \frac{1}{a^2} \Delta \delta \phi  + \frac{\partial^2 V}{\partial \varphi^2} \, \delta \phi \simeq 0 \;,
\label{eq-dphi}
\end{equation} 
which means that the perturbations respond to the modified background, but we do not evolve their explicit non-linear interactions. Computing the non-linear corrections to the inflaton perturbations is left for future work.\footnote{A first attempt to do this within a Hartree approximation is provided in
Refs.~\cite{Cheng:2021lif,Cheng:2023ikq}, which, as mentioned in the introduction, restricted the integrals to infrared modes, 
with a physical UV cutoff of order the Hubble scale that excludes 
subhorizon fluctuations.  Within this approximation, quartic interactions generate a self-consistent 
correction to the effective mass entering the fluctuation equation. We caution, however, that the Hartree 
approximation does not correspond to a complete one-loop calculation of the 
two-point function: in particular, it does not include contributions involving 
two cubic interaction vertices, which arise at the same loop order and can 
participate in cancellations among the different one-loop contributions to the 
power spectrum~\cite{Pimentel:2012tw,Fumagalli:2024jzz,Kawaguchi:2024lsw,
Braglia:2024zsl,Inomata:2025pqa,Inomata:2025bqw,Braglia:2025cee,Braglia:2025qrb}.}

We Fourier transform and quantize the inflaton fluctuations in the standard manner 
\begin{equation}
\delta \phi \left(t, \vec{x} \right) = \int \frac{d^3 k}{\left( 2 \pi \right)^{3/2}} {\rm e}^{ i \vec{k} \cdot \vec{x}} \left[ \delta \phi_k \left( t \right) \, {\hat a}_{\vec{k}} + \delta \phi_k^* \left( t \right) \, {\hat a}_{-\vec{k}}^\dagger \right] \;, 
\label{fourier}
\end{equation}  
where ${\hat a}_{\vec{k}}$ annihilates the vacuum, and $\left[ {\hat a}_{\vec{k}} ,\, {\hat a}_{\vec{p}}^\dagger \right] = \delta^{(3)} \left( \vec{k} - \vec{p} \right)$. Quantum expectation values of two-point functions of~\eqref{fourier} are used to evaluate the spatial averages appearing in~\eqref{eqs-bck}, under the ergodic assumption and the expected quantum-to-classical transition~\cite{Polarski:1995jg} of the perturbations: 
\begin{eqnarray} 
&& \ddot{\varphi} + 3 H \dot{\varphi} + \frac{d V}{d \varphi} + \frac{1}{2} \frac{d^3 V}{d \varphi^3} \int d \ln k \; \calP_{\delta \phi} \simeq 0  \;, \nonumber\\ 
&& H^2 \simeq \frac{1}{3 M_{\rm Pl}^2} \left\{ \frac{1}{2} \dot{\varphi}^2 + V +\rho_{\delta\phi}\right\} \;, \nonumber\\ 
&& \dot{H} \simeq - \frac{1}{2 M_{\rm Pl}^2} \left\{ \dot{\varphi}^2 + \rho_{\delta\phi}+p_{\delta\phi} \right\} \;, 
\label{eqs-bck}
\end{eqnarray} 
where we  have defined the backreaction contribution to the energy and pressure densities as
\begin{eqnarray} 
&& \rho_{\delta\phi} \simeq  \frac{1}{2} \int d \ln k \left[ \calP_{\delta \dot\phi} + \left( \frac{k^2}{a^2} + \frac{d^2 V}{d \varphi^2} \right) \calP_{\delta \phi} \right] \;, \nonumber\\ 
&& p_{\delta\phi} \simeq  \frac{1}{2} \int d \ln k \left[ \calP_{\delta \dot\phi} - \left( \frac13\frac{k^2}{a^2} + \frac{d^2 V}{d \varphi^2} \right) \calP_{\delta \phi} \right]  \;, 
\label{eq:rho-P-backreaction}
\end{eqnarray} 
in terms of the dimensionless power spectra 
\begin{equation}
\calP_{\delta \phi} \equiv \frac{k^3}{2 \pi^2} \left\vert \delta \phi_k \right\vert^2 \;\;\;,\;\;\; \calP_{\delta \dot\phi} \equiv \frac{k^3}{2 \pi^2} \left\vert \delta \dot{\phi}_k \right\vert^2 \;,
\label{eq:definition_power_spectra}
\end{equation}  
and we note that the backreacted energy and pressure densities satisfy
\begin{equation}
    \label{eq:consk}
    \frac{d}{dt}\rho_{\delta\phi}+3 H (\rho_{\delta\phi}+p_{\delta\phi})=\frac{\dot\varphi}{2}   \frac{d^3 V}{d \varphi^3} \int d \ln k \; \calP_{\delta \phi} ,
\end{equation}
which implies that the total energy density still satisfies the conservation law:
\begin{equation}
    \label{eq:cons}
    \frac{d}{dt}(\rho_\varphi+\rho_{\delta\phi})+3 H (\rho_{\varphi}+p_{\varphi}+\rho_{\delta\phi}+p_{\delta\phi})=0.
\end{equation}

The power spectra need to be regularized in the UV. To see this, we introduce the canonical variable $v_k\equiv a\,\delta\phi_k$. The linearized evolution equation, obtained from Eq.~\eqref{eq-dphi} by dropping the quadratic term, takes the form
\begin{equation}
v_k''+\omega_k^2v_k=0\;,\qquad \omega_k^2\equiv k^2+\mu^2\;,
\label{eq:MS}
\end{equation}
where a prime denotes differentiation with respect to conformal time. At the order required for the linear mode functions, the effective mass can be written in terms of the potential and the background field as
\begin{equation}
\mu^2(\varphi,\dot\varphi,H)
=-a^2H^2\left\{
2-\frac{\dot\varphi^2}{2M_{\rm Pl}^2H^2}
-\frac{1}{H^2}\frac{d^2V}{d\varphi^2}
-\alpha\left[
\frac{3\dot\varphi^2}{M_{\rm Pl}^2H^2}
-\frac{\dot\varphi^4}{2M_{\rm Pl}^4H^4}
+\frac{2\dot\varphi}{M_{\rm Pl}^2H^3}\frac{dV}{d\varphi}
\right]\right\} \;.
\label{eq:mu}
\end{equation}
Here $\alpha=0$ corresponds to neglecting metric perturbations, while $\alpha=1$ includes their linear contribution in spatially flat gauge, following the standard bookkeeping convention (e.g.,~\cite{Barker:2026yyg}). As shown in Appendix~\ref{sec:eq:MS}, the terms multiplying $\alpha$ are proportional to $\epsilon_H$ or $\epsilon_H\eta_H$ and are suppressed in the decoupling regime considered here. We have verified that the results for the models we consider in this paper are unaffected by this choice, therefore justifying the decoupling limit. We therefore set $\alpha=0$ in the mean-field evolution, which will make our comparison with the results on the lattice in Section~\ref{sec:results} more transparent. We evaluate Eq.~\eqref{eq:mu} using the potential and the dynamically evolved $\dot\varphi$ and $H$, retaining its full field and potential dependence. The equivalent expression in code variables is given in Appendix~\ref{sec:code}.

As emphasized in Ref.~\cite{Barker:2026yyg}, neglecting metric perturbations can significantly distort the curvature spectrum when $\epsilon_H$ is large. In their USR example, $\epsilon_H$ becomes ${\rm O}(1)$ before the USR phase, enhancing this effect. The benchmarks studied here remain well within the decoupling limit, so this source of large corrections is suppressed.

To illustrate the UV behavior, consider the de Sitter limit. Then $H$ is nearly constant, $a \simeq -1/(H\tau)$, and Eq.~\eqref{eq:mu} reduces to 
\begin{equation}
\mu^2=-\frac{\nu^2-\frac14}{\tau^2}\;,
\qquad
\nu^2 \simeq\frac94\;.
\end{equation}
The Bunch--Davies solution of Eq.~\eqref{eq:MS} is
\begin{equation}
v_k=\frac{\sqrt{\pi}}{2}\,e^{i\pi(2\nu+1)/4}\sqrt{-\tau}\,H_\nu^{(1)}(-k\tau)\;,
\label{eq:de-sitter-mode}
\end{equation}
where $H_\nu^{(1)}$ is the Hankel function of the first kind.
Substituting this solution into Eq.~\eqref{eq:definition_power_spectra} gives the subhorizon limits
\begin{equation}
\lim_{k \gg a H} \calP_{\delta \phi} = \frac{H^2}{4 \pi^2} \left( \frac{k}{a H} \right)^2 \;\;\;,\;\;\; \lim_{k \gg a H} \calP_{\delta \dot\phi} = \frac{H^4}{4 \pi^2} \left( \frac{k}{a H} \right)^4 \;. 
\end{equation} 
This de Sitter example is used only to display the UV asymptotics; the numerical evolution uses Eq.~\eqref{eq:mu} on the evolving background. However, the scaling of the spectrum in the UV remains untouched. 
These UV-divergent contributions show that the backreaction quantities require UV regularization. In the next section we regularize these quantities via adiabatic subtraction in the UV, so as to obtain finite contributions in Eqs.~\eqref{eqs-bck}.

\section{Adiabatic regularization}
\label{sec:adiabatic}

Adiabatic regularization~\cite{Zeldovich:1971mw,Parker:1974qw} removes the ultraviolet vacuum contribution from the power spectra~\eqref{eq:definition_power_spectra} by subtracting a mode-by-mode adiabatic approximation. We write the corresponding canonical mode as
\begin{equation}
v_{{\rm ad},\,k}(\tau) \equiv \frac{1}{\sqrt{2 W_k(\tau)}} \; {\rm e}^{-i \int^\tau d \tau' W_k(\tau')}  \;,
\label{eq:MS_WKB}
\end{equation}
where the real function $W_k(\tau)$ satisfies the following non-linear, second-order differential equation
\begin{equation}
W^2 - \frac{3 (W')^2}{4 W^2} + \frac{W''}{2 W} = \omega^2 \;,
\label{eq-W}
\end{equation}
where we have suppressed the explicit dependence on $k$ and $\tau$ for simplicity. The key point in the WKB approximation is that this equation is not solved exactly, but rather in an iterative fashion. 
For this purpose, recall that $\mu^2=\omega^2-k^2$, and define the conformal Hubble rate $\mathcal H\equiv a'/a$. The quantity $\mu^2$ is of second adiabatic order, while $\mathcal H$ is of first order. We make this counting explicit with a bookkeeping parameter $\epsilon$, unrelated to the slow-roll parameter $\epsilon_H$: $\mu^2\to\epsilon^2\mu^2$, $\partial_\tau\mu^2\to\epsilon^3\partial_\tau\mu^2$, $\partial_\tau^2\mu^2\to\epsilon^4\partial_\tau^2\mu^2$, and $\mathcal H\to\epsilon\mathcal H$, where $\partial_\tau$ denotes the derivative with respect to conformal time. We set $\epsilon=1$ at the end. Expanding
$W=k+\epsilon W_1+\epsilon^2W_2+\epsilon^3W_3+\epsilon^4W_4+{\rm O}(\epsilon^5)$ in Eq.~\eqref{eq-W} and matching equal powers of $\epsilon$ gives $W_1=W_3=0$ and
\begin{equation}
W_2 = \frac{\mu^2}{2 k} \;\;\;,\;\;\; W_4 = - \frac{\partial_\tau^2\mu^2 + \mu^4}{8 k^3} \;.
\label{W24}
\end{equation}
Let $W_{(n)}$ denote this series truncated at order $n$. For even $n$, the residual in Eq.~\eqref{eq-W} starts at order $n+2$. Hence $W_{(n)}$ reproduces the large-$k$ expansion of the exact mode through order $n$. This is the sense in which increasing the adiabatic order improves the ultraviolet subtraction.

In analogy with $\delta \phi_k=v_k/a$, we define $\delta \phi_{{\rm ad},k}\equiv v_{{\rm ad},\,k}/a$. The associated power spectra are
\begin{equation}
\calP_{\delta \phi,{\rm ad}} \equiv \frac{k^3}{2 \pi^2} \left\vert \delta \phi_{{\rm ad},k}  \right\vert^2 \;\;\;,\;\;\; \calP_{\delta \dot\phi,{\rm ad}} \equiv \frac{k^3}{2 \pi^2} \left\vert \delta \dot{\phi}_{{\rm ad},k} \right\vert^2 \;.
\label{eq:definition_power_spectra-ad}
\end{equation}
The regularized spectra are then defined by
\begin{equation}
\calP_{\delta \phi,\,{\rm reg}} \equiv \calP_{\delta \phi} - \calP_{\delta \phi,{\rm ad}} \;\;\;,\;\;\;
\calP_{\delta \dot\phi,\,{\rm reg}} \equiv \calP_{\delta \dot\phi} - \calP_{\delta \dot\phi,{\rm ad}} \;.
\label{P01-sub}
\end{equation}
These replace the bare spectra wherever they appear in the background equations~\eqref{eqs-bck}. The adiabatic expansion is organized by the number of time derivatives of the background quantities: each derivative raises the adiabatic order by one.

Inserting the ansatz~\eqref{eq:MS_WKB} into Eq.~\eqref{eq:definition_power_spectra-ad}, and using $\delta \phi_{{\rm ad},k}=v_{{\rm ad},\,k}/a$, the regularized spectra~\eqref{P01-sub} read
\begin{equation}
\calP_{\delta \phi,\,{\rm reg}} = \frac{k^3}{2 \pi^2 a^2} \left( \left\vert v_k \right\vert^2 - \left\vert v_{{\rm ad},\,k} \right\vert^2 \right) \;,
\label{eq:P0_reg}
\end{equation}
while the derivative spectrum is
\begin{equation}
\calP_{\delta \dot\phi,\,{\rm reg}} = \frac{k^3}{2 \pi^2 a^4} \left( \left\vert v_k' - \mathcal H v_k \right\vert^2 - \left\vert v_{{\rm ad},\,k}' - \mathcal H v_{{\rm ad},\,k} \right\vert^2 \right) \;,
\label{eq:P1_reg}
\end{equation}
where, explicitly,
\begin{equation}
\left\vert v_{{\rm ad},\,k} \right\vert^2 = \frac{1}{2W} \;\;\;,\;\;\;
\left\vert v_{{\rm ad},\,k}' - \mathcal H v_{{\rm ad},\,k} \right\vert^2 = \frac{W}{2}+\frac{1}{2W}\left(\frac{W'}{2W}+\mathcal H\right)^2 \;.
\label{eq:F}
\end{equation}
The exact mode $v_k$ is evolved numerically, while the adiabatic spectra are evaluated with the truncated frequency $W_{(n)}$. In practice, at order $n$, $X\vert_{\leq n}$ is obtained by inserting $W_{(n)}$ in $X$, re-expanding it in $\epsilon$, retaining terms through $\epsilon^n$, and only then setting $\epsilon=1$. With this prescription, Eqs.~\eqref{eq:P0_reg} and~\eqref{eq:P1_reg} take, through the retained order, the practical form
\begin{equation}
\calP_{\delta \phi,\,{\rm reg}} \Big\vert_n = \calP_{\delta \phi} - \calP_{\delta \phi,{\rm ad}} \Big\vert_{\leq n} \;\;\;,\;\;\;
\calP_{\delta \dot\phi,\,{\rm reg}} \Big\vert_n = \calP_{\delta \dot\phi} - \calP_{\delta \dot\phi,{\rm ad}} \Big\vert_{\leq n} \;,
\label{Phat-generic}
\end{equation}
independently of the background chosen. For $n=2$ and $n=4$, the counterterms are
\begin{equation}
\begin{aligned}
\calP_{\delta \phi,{\rm ad}} \Big\vert_{\leq 2} &= \frac{k^3}{2 \pi^2 a^2} \left( \frac{1}{2 k} - \frac{\mu^2}{4 k^3} \right) \;, \\[1mm]
\calP_{\delta \dot\phi,{\rm ad}} \Big\vert_{\leq 2} &= \frac{k^3}{2 \pi^2 a^4} \left( \frac{k}{2} + \frac{\mathcal H^2}{2 k} + \frac{\mu^2}{4 k} \right) \;, \\[1mm]
\calP_{\delta \phi,{\rm ad}} \Big\vert_{\leq 4} &= \calP_{\delta \phi,{\rm ad}} \Big\vert_{\leq 2} + \frac{k^3}{2 \pi^2 a^2} \, \frac{3 \mu^4 + \partial_\tau^2\mu^2}{16 k^5} \;, \\[1mm]
\calP_{\delta \dot\phi,{\rm ad}} \Big\vert_{\leq 4} &= \calP_{\delta \dot\phi,{\rm ad}} \Big\vert_{\leq 2} - \frac{k^3}{2 \pi^2 a^4} \, \frac{\partial_\tau^2\mu^2 + \mu^4 - 4 \mathcal H \, \partial_\tau\mu^2 + 4 \mu^2 \mathcal H^2}{16 k^3} \;.
\end{aligned}
\label{explicit-FQ}
\end{equation}

For the positive-frequency adiabatic initial state used here and a smooth background, $\calP_{\delta \phi,{\rm ad}}\vert_{\leq n}$ and $\calP_{\delta \dot\phi,{\rm ad}}\vert_{\leq n}$ match the large-$k$ expansion of the exact mode through order $n$. The leading remainder therefore starts generically at order $n+2$. Including the prefactors in Eq.~\eqref{Phat-generic}, the regularized spectra scale as
\begin{equation}
\lim_{k \gg a H} \calP_{\delta \phi,\,{\rm reg}} \Big\vert_n \propto H^2 \left( \frac{a H}{k} \right)^n \;\;\;,\;\;\;
\lim_{k \gg a H}\calP_{\delta \dot\phi,\,{\rm reg}} \Big\vert_n \propto H^4 \left( \frac{a H}{k} \right)^{n-2}\;.
\label{UV-generic}
\end{equation}
Equivalently, both are suppressed by $(aH/k)^{n+2}$ relative to their bare ultraviolet spectra~\eqref{eq:definition_power_spectra}. The powers of $k$ in Eq.~\eqref{UV-generic} are universal, while their coefficients depend on the background.

To display these coefficients explicitly, take the near de Sitter limit, $\mathcal H \simeq -1/\tau$, and additionally treat $\nu$ as constant, so that $\mu^2=-\left( \nu^2 - \frac{1}{4} \right)/\tau^2$. Equation~\eqref{explicit-FQ} then becomes
\begin{align}
\calP_{\delta \phi,{\rm ad}} \Big\vert_{\leq 2}
&= \frac{H^2}{4 \pi^2} \left( \frac{k}{a H} \right)^2
\left[ 1 + \frac{\nu^2 - \frac{1}{4}}{2}
\left( \frac{a H}{k} \right)^2 \right] \;,
\nonumber \\[1mm]
\calP_{\delta \phi,{\rm ad}} \Big\vert_{\leq 4}
&= \calP_{\delta \phi,{\rm ad}} \Big\vert_{\leq 2}
+ \frac{3 \left( \nu^2 - \frac{1}{4} \right)
\left( \nu^2 - \frac{9}{4} \right)}{8}
\, \frac{H^2}{4 \pi^2} \left( \frac{a H}{k} \right)^2 \;,
\label{F-dS} \\[2mm]
\calP_{\delta \dot\phi,{\rm ad}} \Big\vert_{\leq 2}
&= \frac{H^4}{4 \pi^2} \left( \frac{k}{a H} \right)^4
\left[ 1 - \frac{\nu^2 - \frac{9}{4}}{2}
\left( \frac{a H}{k} \right)^2 \right] \;,
\nonumber \\[1mm]
\calP_{\delta \dot\phi,{\rm ad}} \Big\vert_{\leq 4}
&= \calP_{\delta \dot\phi,{\rm ad}} \Big\vert_{\leq 2}
- \frac{\left( \nu^2 - \frac{1}{4} \right)
\left( \nu^2 - \frac{9}{4} \right)}{8}
\, \frac{H^4}{4 \pi^2} \;.
\label{FQ-dS}
\end{align}
At second order, the leading ultraviolet remainder is given by the fourth-order terms in Eqs.~\eqref{F-dS} and~\eqref{FQ-dS}. In the limit $k/(aH)\to\infty$, Eq.~\eqref{Phat-generic} gives
\begin{equation}
\begin{aligned}
\lim_{k \gg a H}
\calP_{\delta \phi,\,{\rm reg}} \Big\vert_2
&\sim \frac{3 \left( \nu^2 - \frac{1}{4} \right) \left( \nu^2 - \frac{9}{4} \right) H^2}{32 \pi^2} \left( \frac{a H}{k} \right)^2 \;,\\
\lim_{k \gg a H} \calP_{\delta \dot\phi,\,{\rm reg}} \Big\vert_2
&\sim - \frac{\left( \nu^2 - \frac{1}{4} \right) \left( \nu^2 - \frac{9}{4} \right) H^4}{32 \pi^2} \;.
\end{aligned}
\label{P0P1-n2}
\end{equation}
in agreement with Eq.~\eqref{UV-generic}. The first spectrum falls as $k^{-2}$, but the second tends to a constant. Moreover, the gradient contribution $(k^2/a^2)\calP_{\delta\phi,\,{\rm reg}}$ in Eq.~\eqref{eqs-bck} also tends to a constant. The corresponding $d\ln k$ integrals appearing in Eq.~\eqref{eqs-bck} therefore remain logarithmically divergent at second order.

At fourth order, the leading remainder comes from the sixth-order terms in the adiabatic spectra, which we do not display. In the same limit, they give
\begin{equation}
\begin{aligned}
\lim_{k \gg a H} \calP_{\delta \phi,\,{\rm reg}} \Big\vert_4
&\sim \frac{5 \left( \nu^2 - \frac{1}{4} \right) \left( \nu^2 - \frac{9}{4} \right) \left( \nu^2 - \frac{25}{4} \right) H^2}{64 \pi^2} \left( \frac{a H}{k} \right)^4 \;,\\
\lim_{k \gg a H} \calP_{\delta \dot\phi,\,{\rm reg}} \Big\vert_4
&\sim - \frac{\left( \nu^2 - \frac{1}{4} \right) \left( \nu^2 - \frac{9}{4} \right) \left( \nu^2 - \frac{49}{4} \right) H^4}{64 \pi^2} \left( \frac{a H}{k} \right)^2 \;.
\end{aligned}
\label{P0P1-n4}
\end{equation}
Now $\calP_{\delta\phi,\,{\rm reg}}\propto k^{-4}$, while both $\calP_{\delta\dot\phi,\,{\rm reg}}$ and $(k^2/a^2)\calP_{\delta\phi,\,{\rm reg}}$ fall as $k^{-2}$.  All ultraviolet integrals in Eq.~\eqref{eqs-bck} therefore converge. The field variance entering the inflaton equation is already finite after second-order subtraction,
\begin{equation}
\left\langle\delta\phi^2\right\rangle_{\rm reg}
=
\int d\ln k\,
\left.\calP_{\delta\phi,\,{\rm reg}}\right|_2
\;.
\label{eq:reg-field-variance}
\end{equation}
The equations for $H^2$ and $\dot H$ instead depend on the energy density and pressure of the fluctuations, which we regularize by adiabatic subtraction as follows
\begin{align}
\rho_{\delta\phi}^{\rm reg}
&=
\frac{1}{2}\int d\ln k\,
\left[
\calP_{\delta\dot\phi,\,{\rm reg}} +\frac{k^2}{a^2} \calP_{\delta\phi,\,{\rm reg}} +V_{,\phi\phi} \calP_{\delta\phi,\,{\rm reg}} \right]\biggr\vert_4\;,
\label{eq:reg-energy-density}
\\
p_{\delta\phi}^{\rm reg} &= \frac{1}{2}\int d\ln k\,\left[\calP_{\delta\dot\phi,\,{\rm reg}}-\frac{k^2}{3a^2}\calP_{\delta\phi,\,{\rm reg}}-V_{,\phi\phi}\calP_{\delta\phi,\,{\rm reg}}\right]\biggr\vert_4\;.
\label{eq:reg-pressure}
\end{align}
Let us note that, while the kinetic and gradient contributions require fourth-order subtraction, the terms proportional to $V_{,\phi\phi}$ require in practice the field spectrum only through second order, since $V_{,\phi\phi}$ itself carries second adiabatic order. We also note that the use of different adiabatic orders in Eq.~\eqref{eq:reg-field-variance} and Eqs.~\eqref{eq:reg-energy-density} and~\eqref{eq:reg-pressure} is therefore consistent with the minimal adiabatic prescription: for each complete observable, the subtraction is carried out only to the lowest order required to remove its ultraviolet divergences. We stress that our procedure is not intended to constitute a complete renormalization program. We remove the ultraviolet-divergent vacuum contributions and adopt a definite prescription for extending the subtraction to finite momenta, but we do not impose physical
renormalization conditions fixing the finite parts of the corresponding
local counterterms by matching renormalized parameters to observables.
The finite backreaction therefore retains a residual prescription and
scheme dependence. Our numerical results should accordingly be
interpreted as estimates within the prescription adopted here: reasonable
changes of the finite subtraction may modify their magnitude by factors
of order unity. In the absence of parametrically large or finely tuned
finite counterterms, however, we do not expect these ambiguities to alter
the qualitative conclusions or produce changes by orders of magnitude.

The adiabatic subtraction terms should nevertheless be used only where the WKB expansion from which they are derived remains reliable. We shall discuss in which regime this happens in the following. 

\subsection{Validity of the adiabatic approximation}
\label{subsec:wkb-filter}

The ultraviolet analysis above fixes the adiabatic order required to remove the UV divergences from the backreaction integrals. It does not, however, determine how far toward the infrared the corresponding counterterms can be trusted. In fact, as can easily be seen from the expressions for $W_{(n)}$ above, the inclusion of higher adiabatic corrections introduces new IR divergences. The WKB series is an asymptotic expansion in derivatives of the background, and its finite-order truncations need not provide a controlled approximation for modes whose evolution is nonadiabatic. Applying the counterterms in this regime can therefore introduce unphysical finite contributions. We address this issue by supplementing the adiabatic subtraction with a smooth, order-dependent restriction based directly on the accuracy of the WKB expansion. We note that a discussion of the validity of the WKB approximation can be found in~\cite{Braglia:2026fle}. Our treatment here is similar to that of~\cite{Corba:2022ugu}, although significantly simplified.

As discussed above, insertion of the frequency $W_{(n)}$, truncated at adiabatic order $n$, into Eq.~\eqref{eq-W} leaves a residual defined as
\begin{equation}
    R[W_{(n)}] \equiv 
W_{(n)}^2
-\frac{3(W_{(n)}')^2}{4W_{(n)}^2}
+\frac{W_{(n)}''}{2W_{(n)}}
-\omega_k^2
=
\varepsilon^{n+2}R_{n+2}
+\mathcal{O}(\varepsilon^{n+4}) \, ,
\end{equation}
where $R_{n+2}$ denotes the first non-vanishing coefficient
in the adiabatic expansion of the residual, which is of adiabatic order $n+2$.
We then define the dimensionless WKB-validity estimator
\begin{equation}
\delta_n(k,\tau)\equiv
\frac{\left|R_{n+2}(k,\tau)\right|}{k^2}\;,
\qquad n=0,2,4\;.
\label{eq:wkb-validity-estimator}
\end{equation}
The normalization by the leading term $k^2$ in the WKB equation makes $\delta_n$ a dimensionless measure of the first omitted contribution to that equation. For the first three truncations one obtains the exact expressions
\begin{align}
\delta_0
&=\frac{\left|\mu^2\right|}{k^2}\;,
\nonumber\\
\delta_2
&=\frac{\left|\mu^4+\partial_\tau^2\mu^2\right|}{4k^4}\;,
\nonumber\\
\delta_4
&=\frac{
\left|
\partial_\tau^4\mu^2+6\mu^2\,\partial_\tau^2\mu^2
+5\left(\partial_\tau\mu^2\right)^2+2\mu^6
\right|
}{16k^6}\;.
\label{eq:wkb-validity-explicit}
\end{align}
Consequently, $\delta_0\ll1$ identifies the regime in which the leading WKB solution is reliable, while $\delta_2<\delta_0$ and $\delta_4<\delta_2$ indicate that the second- and fourth-order terms, respectively, improve the approximation. At fixed background quantities, these estimators have the UV scaling
\begin{equation}
\delta_n\propto k^{-(n+2)}\;,
\qquad n=0,2,4\;.
\label{eq:wkb-estimator-uv}
\end{equation}

For the background considered in the next section, we show the numerical estimators in Fig.~\ref{fig:wkb-validity} at fixed comoving momentum $k=k_{\rm in}$. Their asymptotic behavior at both $k\ll a H$ and $k\gg a H$ is well approximated by
\begin{equation}
\begin{aligned}
\delta_n^{\rm fit}(N,k)
&=c_n\left(\frac{a(N)H(N)}{k}\right)^{n+2}\;,
&& n=0,2,4.
\end{aligned}
\label{eq:wkb-estimator-fit}
\end{equation}
The overall momentum dependence follows from Eq.~\eqref{eq:wkb-estimator-uv}, while the coefficients are determined numerically. The fit captures the overall evolution of the estimators while smoothing their localized oscillations occurring around horizon crossing. These oscillations can produce isolated apparent violations of the adiabatic hierarchy: a contribution at a given order may cross zero, while the next-order term, which generally oscillates with a different frequency and has different zero crossings, remains finite. Ratios of successive contributions then develop poles and can suggest that the higher-order term dominates, even when its overall amplitude remains suppressed. Such localized features need not signal a breakdown of the expansion and can be neglected when assessing the hierarchy of its envelopes. We therefore use the fitted estimators to construct smooth filters for the adiabatic subtraction.

The leading filter determines whether the WKB approximation is sufficiently accurate for the subtraction to be applied, while the remaining filters control the inclusion of successive adiabatic orders. We define
\begin{align}
S_0
&\equiv\left[
1+
c_0^{2/\sigma_0}\left(\frac{k}{a H}
\right)^{-4/\sigma_0}
\right]^{-1}\;,
\nonumber\\
S_2
&\equiv\left[
1+\left(c_2/c_0\right)^{2/\sigma_2}\left(\frac{k}{a H}
\right)^{-4/\sigma_2}\right]^{-1}\;,
\nonumber\\
S_4
&\equiv\left[
1+
\left(c_4/c_2\right)^{2/\sigma_4}\left(\frac{k}{a H}
\right)^{-4/\sigma_4}\right]^{-1}\;,
\label{eq:wkb-smooth-filters}
\end{align}
where the positive parameters $\sigma_n$ determine the width of the transitions. Deep inside the horizon, the WKB hierarchy is well satisfied and all three filters approach unity. As a mode approaches the regime where the expansion is no longer controlled, the filters progressively suppress the corresponding adiabatic counterterms.

For either of the two adiabatic spectra defined in Eq.~\eqref{eq:definition_power_spectra-ad}, generically denoted by $X$, the standard adiabatic expansion is $X_{\rm WKB}(k,\tau)=X_0(k,\tau)+\varepsilon^2X_2(k,\tau)+\varepsilon^4X_4(k,\tau)+\cdots$, where $X_n$ denotes the contribution of adiabatic order $n$. We modify this expansion by introducing the filters defined above:
\begin{equation}
\begin{aligned}
X_{\rm WKB}(k,\tau)
=S_0(k,\tau)\big[
&X_0(k,\tau)
+S_2(k,\tau)\varepsilon^2X_2(k,\tau)\\
&+S_4(k,\tau)\varepsilon^4X_4(k,\tau)
+\cdots\big]\;,
\qquad X=\calP_{\delta \phi,{\rm ad}},\,\calP_{\delta \dot\phi,{\rm ad}}\;,
\end{aligned}
\label{eq:wkb-filtered-counterterm}
\end{equation}
where we have multiplied the overall expression by the $0^{\rm th}$ filter to make sure that we do not subtract any quantity outside the regime of validity of the WKB expansion, thereby preventing the introduction of spurious IR divergences. 
The filtered adiabatic expressions are then subtracted from the corresponding bare quantities in Eq.~\eqref{Phat-generic}, with $\varepsilon$ set to unity at the end of the expansion. In the UV, all three filters approach unity and the standard adiabatic subtraction is recovered. Outside the regime of validity of the WKB expansion, the counterterms are smoothly suppressed. 

The parameters $\sigma_n$ that characterize the filters must satisfy certain conditions to preserve UV convergence. The most restrictive conditions
arise from the kinetic and gradient contributions to the
$\dot H$ equation, whose adiabatic counterterms of orders $0$, $2$,
and $4$ scale as $k^4$, $k^2$, and $k^0$, respectively. Standard (namely, unfiltered) fourth-order adiabatic subtraction removes these contributions, leaving a UV remainder in the $\dot{H}$ integrand that decays as $k^{-2}$. Introducing the filters leaves additional UV contributions, since, at large momenta, $S_n-1=O(k^{-4/\sigma_n})$. Denoting by $\Delta I_{\dot H}$ the difference between the filtered and unfiltered subtraction integrands entering the $\dot H$ equation, we obtain the UV scaling
\begin{equation}
\Delta I_{\dot H}
= O\left(k^{4-(4/\sigma_0)}\right)
+ O\left(k^{2-(4/\sigma_2)}\right)
+ O\left(k^{-4/\sigma_4}\right) \, ,
\end{equation}
where faster-decaying contributions have been omitted.
Since the backreaction integrals are performed over $d\ln k$,
UV convergence requires $\sigma_0<1$, $\sigma_2<2$, and $\sigma_4>0$. Requiring, more
stringently, that the filters do not spoil the $k^{-2}$
(or faster) UV decay obtained with the standard adiabatic
subtraction gives
$\sigma_0\leq 2/3$, $\sigma_2\leq 1$, and $\sigma_4\leq 2$. The field variance entering the inflaton
equation imposes weaker constraints. 
For definiteness, we choose
$(\sigma_0,\sigma_2,\sigma_4)=(0.5,0.5,1)$
in the following analysis.

Before we apply this prescription to our test scenarios in the next Section, let us comment on one final aspect of the WKB prescription proposed here. While the bare quantities satisfy the conservation law~\eqref{eq:cons}, or equivalently~\eqref{eq:consk} for the fluctuation sector, the filters introduced above do not preserve this relation exactly. Requiring the subtracted quantities to satisfy~\eqref{eq:consk} implies that
\begin{equation}
    \Delta_{\rm ad}\equiv\frac{d}{dt}\rho_{\delta\phi,\,{\rm ad}}+3 H (\rho_{\delta\phi,\,{\rm ad}}+p_{\delta\phi,\,{\rm ad}})-\frac{\dot\varphi}{2}   \frac{d^3 V}{d \varphi^3} \int d \ln k \; \calP_{\delta \phi,\,{\rm ad}}=0.
    \label{eq:violation-cons}
\end{equation}
The time dependence of the filters generically induces a non-vanishing  $\Delta_{\rm ad}$ through terms proportional to $\dot{S}_i$. Restoring exact conservation would require embedding the finite filtering
prescription into a consistent set of counterterms for the Klein--Gordon
and Friedmann equations, which goes beyond the scope of the present work.  The compatibility of the resulting background evolution with the Friedmann constraint is quantified in Appendix~\ref{sec:consistency}. Our choice  $(\sigma_0,\sigma_2,\sigma_4)=(0.5,0.5,1)$ is such that the filters vary smoothly around the transition region and the conservation-violation estimator $\Delta_{\rm ad}$ defined in Eq.~\eqref{eq:violation-cons} remains negligible compared with the physical backreaction source. Our results are largely insensitive to moderate variations of these widths.

\section{Backreaction for illustrative benchmark models}
\label{sec:example}

We now apply the mean-field system of Section~\ref{sec:system}, together with the adiabatic subtraction of Section~\ref{sec:adiabatic}, to assess whether the homogeneous, spatially averaged backreaction can appreciably modify two representative models of different physical mechanisms enhancing small-scale fluctuations. 

\begin{figure}[t]
\centering
\includegraphics[width=\textwidth]{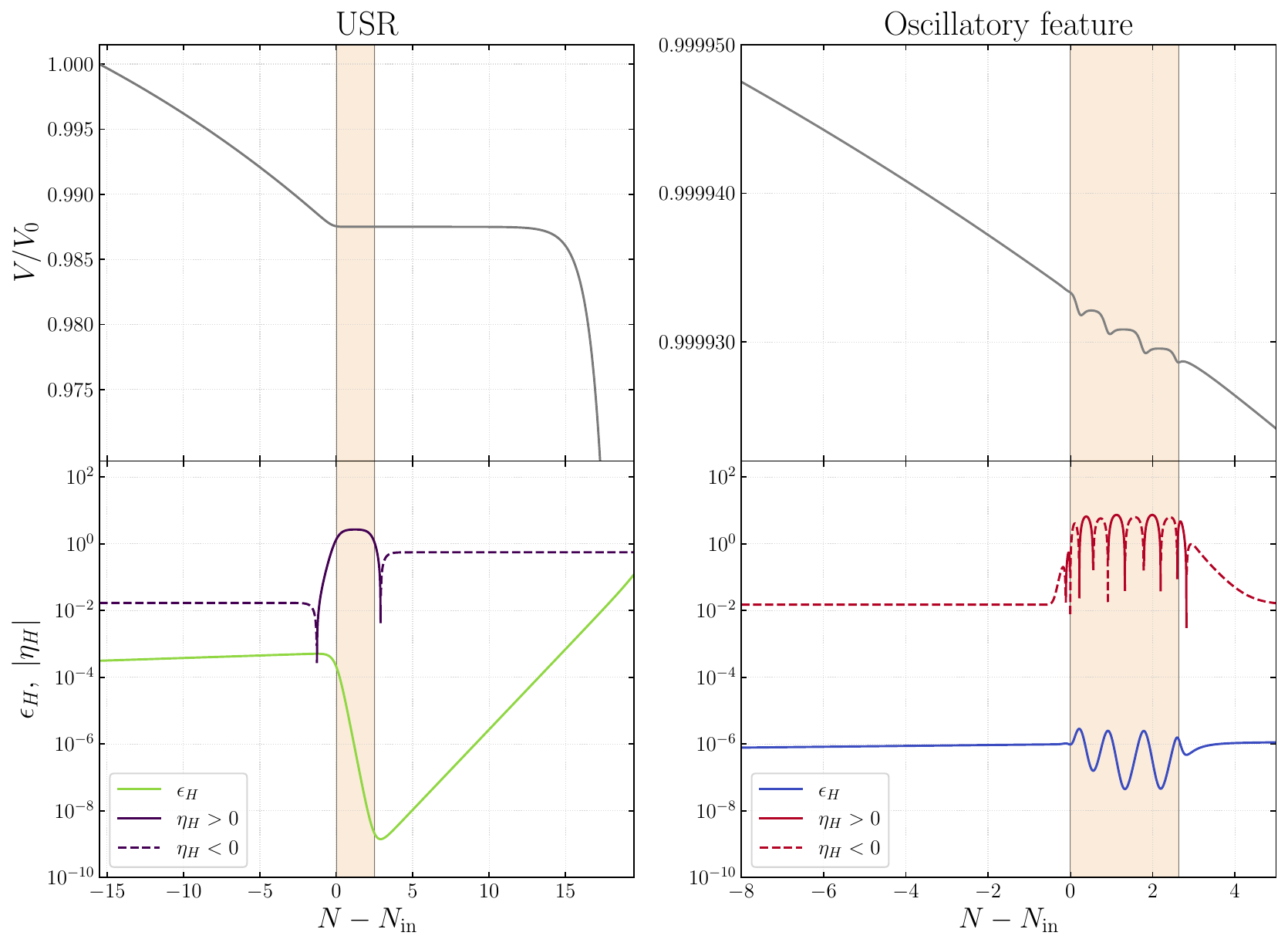}
\caption{The two benchmark models used in our analysis: USR (left) and OF (right). Upper panels show the potential $V[\varphi(N)]/V_0$ along each (non-backreacted) background trajectory. Both horizontal axes show $N-N_{\rm in}$, measured from the onset of the respective feature. The shaded bands and vertical lines in both rows indicate the duration of the USR stage (left) and the OF (right). Both panels display $V/V_0$, with a finer vertical scale in the OF panel to resolve its small variations. Lower panels show $\epsilon_H$ and $|\eta_H|$; the $\eta_H$ curve is solid where $\eta_H>0$ and dashed where $\eta_H<0$. USR has a single smooth episode in which $|\eta_H|$ reaches a few and $\epsilon_H$ falls exponentially. The OF benchmark model instead has a short burst of rapid oscillations reaching $|\eta_H|\sim6$, while $\epsilon_H$ remains of order $10^{-6}$.}
\label{fig:benchmark}
\end{figure}

The first example is a transient USR stage, during which the inflaton velocity and $\epsilon_H$ decrease exponentially in time, while the inflaton fluctuations, and hence the curvature perturbation, are strongly amplified over a finite range of scales.  The second example is the OF model. Here, the slow-roll parameter  $\epsilon_H$ remains small at all times, but higher-order parameters become transiently large, inducing a very sharp oscillatory feature in the background quantities. The modes crossing the horizon around the time of the feature, and eventually also those that can efficiently resonate with the background oscillations, are excited with respect to their Bunch--Davies state through a mechanism that is akin to copious particle production.
These two models are thus distinct physical scenarios in which to test the importance of the quadratic terms in the background equations~\eqref{eqs-bck}. The corresponding inflaton potentials are shown in Fig.~\ref{fig:benchmark}.

\subsection{Background evolution without backreaction}
\label{subsec:background}

In order to build physical intuition about the models considered, we first discuss the background evolution in the absence of backreaction. For simplicity, we will work with two benchmarks here and present them side by side in order to show the qualitative differences between the two scenarios in the application of our formalism introduced in the previous Section. A more in-depth analysis of each model separately is provided in the next Section, and the details of the implementation, as well as the numerical values of the parameters of the models are given in Appendix~\ref{app:benchmark_models}. 

For the USR benchmark, we reconstruct a potential from a smooth three-stage profile of $\eta_H$, following Ref.~\cite{Franciolini:2022pav}. Two attractor phases bracket a transient non-attractor stage with $\eta_H=2.709$, close to the typical USR value of $3$. During this stage, the inflaton velocity and $\epsilon_H$ decrease rapidly, amplifying the curvature perturbations over a finite range of scales. The final attractor phase makes the inflaton accelerate again. The prescribed background is used only to reconstruct the potential; the subsequent evolution, including backreaction, is determined dynamically based on the potential alone. The reconstruction procedure and all benchmark parameters are given in Appendix~\ref{subsec:benchmark_usr}. 
For definiteness, in Figs.~\ref{fig:benchmark}, \ref{fig:wkb-validity}, and \ref{fig:usr-backreaction}, we show the results with a USR duration such that the tree-level curvature power spectrum peaks at $\mathcal P_{\zeta,\mathrm{tree}}^{\max}=10^{-3}$.
Let us stress that we choose to start from a parameterized model of $\eta_H$ rather than specify the scalar field potential only for illustrative purposes. Our benchmark has a smooth transition from a slow-roll attractor to a near-USR phase, followed by a return to an attractor. Such an evolution could be realized by a potential with a sufficiently smooth plateau and a parabolic exit whose curvature differs from that of the plateau (so that the cubic derivative of the potential is negative, as in case II of Appendix~\eqref{app:benchmark_models}). 

Scenarios with sharper USR features can be easily reverse-engineered, although the reconstructed $V(\varphi)$ may not be easily obtained from a more explicitly parametrized/motivated scalar field potential~\cite{Barnert:2026ijs}. Nevertheless, the relevant physics discussed in this work is not going to be affected by our specific choice.

The OF benchmark model instead consists of a small-field quadratic hilltop potential with a transient oscillatory feature, which is localized around a finite field interval. As the inflaton crosses these wiggles, the higher derivatives of the potential vary rapidly while $\epsilon_H$ remains small. The resulting oscillations of the effective mass resonantly amplify a band of fluctuations, including subhorizon modes, producing a localized peak in the spectrum. In this Section, we adopt case C of Ref.~\cite{Inomata:2022yte} as our benchmark, but increase the value of the amplitude of the oscillatory correction in the scalar field potential. The explicit potential, the smooth envelope of the feature, and all parameter choices are given in Appendix~\ref{subsec:benchmark_oscillatory}.

Figure~\ref{fig:benchmark} illustrates the distinct background dynamics of the two models. The USR potential develops a broad, nearly flat region, across which $\epsilon_H$ falls by several orders of magnitude before re-entering the SR stage and growing back to a larger value. In the OF benchmark, the small modulations of the potential instead induce rapid excursions of $\eta_H$, reaching $|\eta_H|\sim6$, while $\epsilon_H$ remains of order $10^{-6}$.

\subsection{Backreaction contributions in the mean-field approximation}
\label{subsec:perturbations}

For the background evolution, we integrate the first and third equations of~\eqref{eqs-bck}, with the regularized power spectra~\eqref{P01-sub}. We rewrite them as  
\begin{eqnarray} 
&& \ddot{\varphi} + 3 H \dot{\varphi} \left[ 1 + \int d \ln k \; {\cal B}_\varphi \left( t ,\, k \right) \right] + \frac{d V}{d \varphi}  \simeq 0  \;, \nonumber\\
&& \dot{H} \simeq - \frac{\dot{\varphi}^2}{2 M_{\rm Pl}^2} \left[ 1 + \int d \ln k \; {\cal B}_H \left( t ,\, k \right) \right] \;,
\label{eq:back_simple}
\end{eqnarray} 
where we have defined the (dimensionless) backreaction terms 
\begin{eqnarray} 
{\cal B}_\varphi \equiv \frac{1}{3 H \dot{\varphi}} \frac{1}{2} \frac{d^3 V}{d \varphi^3} \calP_{\delta \phi,\,{\rm reg}} \;\;\;,\;\;\; 
{\cal B}_H \equiv \frac{1}{\dot{\varphi}^2} \, \left[ \calP_{\delta \dot\phi,\,{\rm reg}} + \frac{1}{3} \frac{k^2}{a^2}  \calP_{\delta \phi,\,{\rm reg}} \right] \;. 
\label{calB}
\end{eqnarray}

\begin{figure}[t]
\centering
\includegraphics[width= \textwidth]
{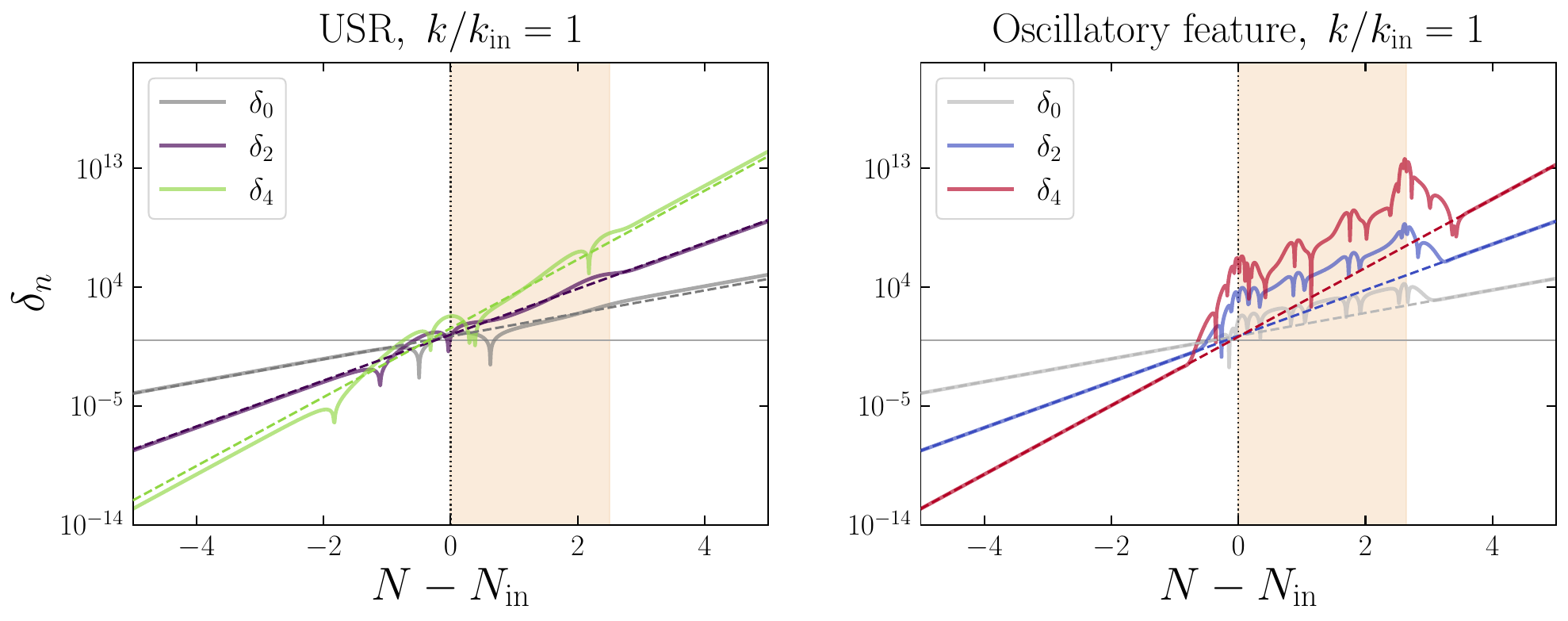}
\caption{Evolution of the WKB-validity estimators $\delta_{0,2,4}$ introduced in Eq. \eqref{eq:wkb-validity-estimator} as a function of $N - N_{\rm in}$, for the mode of momentum $k = k_{\rm in}$ that exits the horizon at the beginning of the USR (left panel) or the OF (right panel) region. The solid curves show the numerical evolution of the estimators, while the thin dashed curves show the fitting Eq.~\eqref{eq:wkb-estimator-fit},  with the parameters $(c_0,c_2,c_4)$ fitted to $(1.9,2.5,7.6)$ in the USR case and to $(2.045,2.022,1.769)$ in the OF case. As in Fig.~\ref{fig:benchmark}, the shaded bands indicate the duration of the USR stage and the oscillatory feature. In both panels the horizontal gray line marks $\delta_n=1$. The asymptotic form approximate very well the numerical evolution of the estimators outside the bands. The agreement is particularly striking for the OF case, for which, far from the feature, the potential is the pure quadratic hilltop and $\epsilon_H$ is negligible, and the $c_i$ fitting functions are accurate to five significant digits on both the pre- and post-feature plateaus.}
\label{fig:wkb-validity}
\end{figure}

The power spectra of the scalar field perturbations can be used to compute that of the curvature perturbation, which has a more physical interpretation.
In the decoupling limit and at the linear level adopted here for perturbations, the curvature perturbation on comoving slices is related to the inflaton fluctuation in spatially flat gauge by the simple linear relation\footnote{We denote by $\zeta_{\rm GI}$ the full nonlinear, gauge-invariant curvature perturbation, which is the physical observable. The variable $\zeta$ used here is its linear approximation, related linearly to the inflaton fluctuation $\delta\phi$ in spatially flat gauge, so that $\zeta_{\rm GI}=\zeta+\mathcal O(\delta\phi^2)$. The nonlinear contributions to this relation generate additional corrections to the full one-loop power spectrum of $\zeta_{\rm GI}$. In principle, these may be as large as the backreaction effects we compute~\cite{Inomata:2026csq}. Their inclusion goes beyond the scope of our paper.}
\begin{equation}
\zeta_k=-\frac{H}{\dot\varphi}\,\delta\phi_k \;.
\label{zeta-linear}
\end{equation}
Accordingly, we define
\begin{equation}
\calP_\zeta\equiv\frac{k^3}{2\pi^2}|\zeta_k|^2
=\frac{H^2}{\dot\varphi^2}\calP_{\delta\phi} \;,
\qquad
\calP_{\dot\zeta}\equiv\frac{k^3}{2\pi^2}|\dot\zeta_k|^2 \;,
\label{eq:Pzeta-def}
\end{equation}
with the same definitions applied to the adiabatic and regularized spectra. For each model, we denote the onset of the USR stage or the oscillatory feature by $N_{\rm in}$ and normalize momenta to the corresponding comoving Hubble scale, $k_{\rm in}\equiv a(N_{\rm in})H(N_{\rm in})$. We plot spectra as functions of $k/k_{\rm in}$. In the following plots, the vertical black dashed line marks the instantaneous Hubble-crossing scale $k=aH$.

\begin{figure}[t]
\centering
\begin{subfigure}[b]{\textwidth}
\centering
\includegraphics[width=\textwidth]{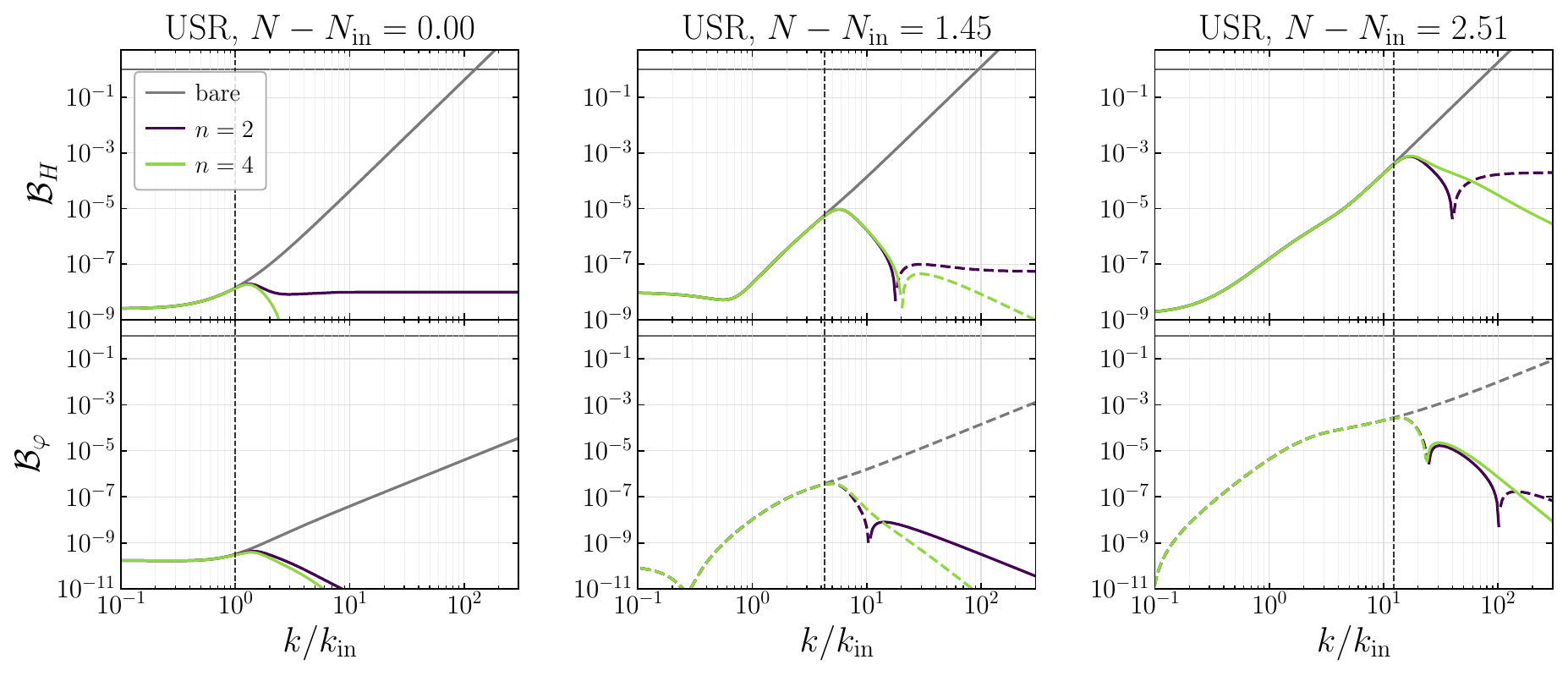}
\end{subfigure}
\vspace{3mm}
\begin{subfigure}[b]{\textwidth}
\centering
\includegraphics[width=\textwidth]{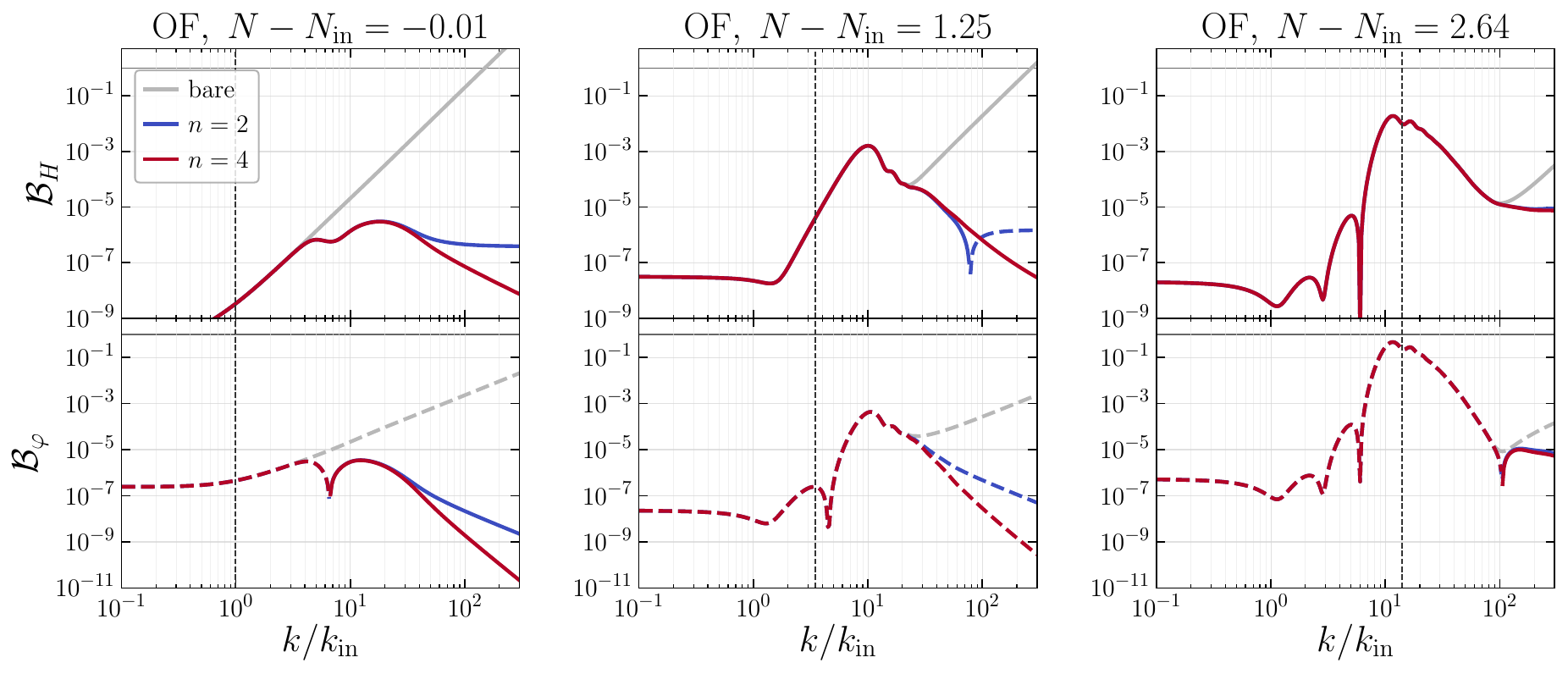}
\end{subfigure}
\caption{Backreaction integrands ${\cal B}_H$ (top row of each panel) and ${\cal B}_\varphi$ (bottom row) for the USR in the upper block and OF in the lower block, at three representative times, chosen at the onset/middle/end of the respective feature. In the legends, ``bare'' denotes the integrands without WKB subtraction, while ``second'' and ``fourth'' denote the integrands after second- and fourth-order WKB subtraction, respectively. As in the previous figures, a quantity is shown with solid lines (respectively, dashed lines) when it is positive (respectively, negative). The vertical black dashed line marks $k=aH$, while the thick
horizontal grey line marks unity for reference.}
\label{fig:usr-backreaction}
\end{figure}

Fig.~\ref{fig:wkb-validity} displays the validity of our approximation for different adiabatic orders, and confirms the goodness of the fitting functions~\eqref{eq:wkb-estimator-fit}, which are employed in the filters \eqref{eq:wkb-smooth-filters}. With these filters, we can compute the effects of the backreaction, which we show in Fig.~\ref{fig:usr-backreaction} for the two scenarios. There, we compare the bare (namely, without WKB subtraction) backreaction integrands with their second- and fourth-order subtractions at the beginning, midpoint, and end of the USR stage and of the OF interval. In all panels, the bare integrands exhibit the same asymptotic subhorizon behavior, $|{\cal B}_\varphi|\propto k^2$ and ${\cal B}_H\propto k^4$, irrespective of the background dynamics. These powers follow directly from the Bunch--Davies vacuum: for sufficiently large $k$, $v_k\simeq e^{-ik\tau}/\sqrt{2k}$, so that $\calP_{\delta\phi}\simeq k^2/(4\pi^2a^2)$ and $\calP_{\delta\dot\phi}\simeq k^4/(4\pi^2a^4)$. 
Fig.~\ref{fig:usr-backreaction} explicitly shows that $\mathcal{B}_\varphi$ and $\mathcal{B}_H$ require, respectively, a second- and a fourth-order adiabatic expansion in order to remove the UV divergences, as discussed in Section~\ref{sec:adiabatic}.

Once the UV region is correctly regularized, we see that the behavior of the backreaction is qualitatively different in the two scenarios considered. In particular, in the USR benchmark, adiabatic subtraction already removes power around horizon crossing. The subtracted integrands consequently develop a peak around the scales crossing the horizon, followed by a decaying UV tail. The subtraction thus reduces power exactly in the same range of scales that would otherwise contribute substantially to the inferred backreaction. 
On the other hand, in the OF benchmark, even before the subtraction we can notice a significant bump around the scales that are resonantly amplified, which is the signature of particle production and survives adiabatic subtraction. We note that a similar behavior occurs in models of axion inflation, as discussed for instance in Refs.~\cite{Garcia-Bellido:2023ser,Sharma:2024nfu,Caravano:2022epk,Figueroa:2023oxc,Figueroa:2024rkr}.

The relative magnitude and sign of the two sources also clarify their dynamical role. Toward the end of both features, the impact of $|{\cal B}_\varphi|$ dominates over that of $|{\cal B}_H|$. After the onset of the feature, the enhanced contribution to ${\cal B}_\varphi$ is always negative in the USR benchmark, as indicated by the dashed curves, while it oscillates in sign for the OF benchmark, following the sign of $d^3V/d\varphi^3$. A negative contribution to ${\cal B}_\varphi$ corresponds to a backreaction effective force that pushes the inflaton along its trajectory, while a positive contribution opposes its motion. The corresponding backreaction therefore tends to accelerate the inflaton relative to the unbackreacted solution in our USR benchmark, while it can either accelerate or slow down the inflaton in the OF model.\footnote{In App.~\ref{subsec:benchmark_usr} we discuss different USR scenarios in which $d^3V/d\varphi^3$ takes the opposite sign at the end of USR.}

This difference reflects the mechanism by which the curvature perturbation is enhanced, $\mathcal{P}_\zeta = (H^2/\dot\varphi^2)\,\mathcal{P}_{\delta\phi}$, cf.\ Eq.~\eqref{eq:Pzeta-def}. In USR, the growth of $\mathcal{P}_\zeta$ is primarily driven by the rapid decrease of the background inflaton velocity, rather than by a resonant amplification of the inflaton fluctuations: subhorizon modes remain close to the adiabatic vacuum, and the physical contribution to the integrands peaks around horizon crossing, where it directly adjoins the UV vacuum tail. In the OF model, instead, the oscillatory feature directly amplifies the inflaton fluctuations through resonant excitation, including while the relevant modes are still subhorizon, producing a peak that stands well above the vacuum contribution at the same scales. As we show in Section~\ref{sec:results}, this difference controls the sensitivity of the unsubtracted backreaction to the
UV cutoff.

\label{subsec:adiabatic-results}

\begin{figure}
\centering
\includegraphics[width=0.92\textwidth]{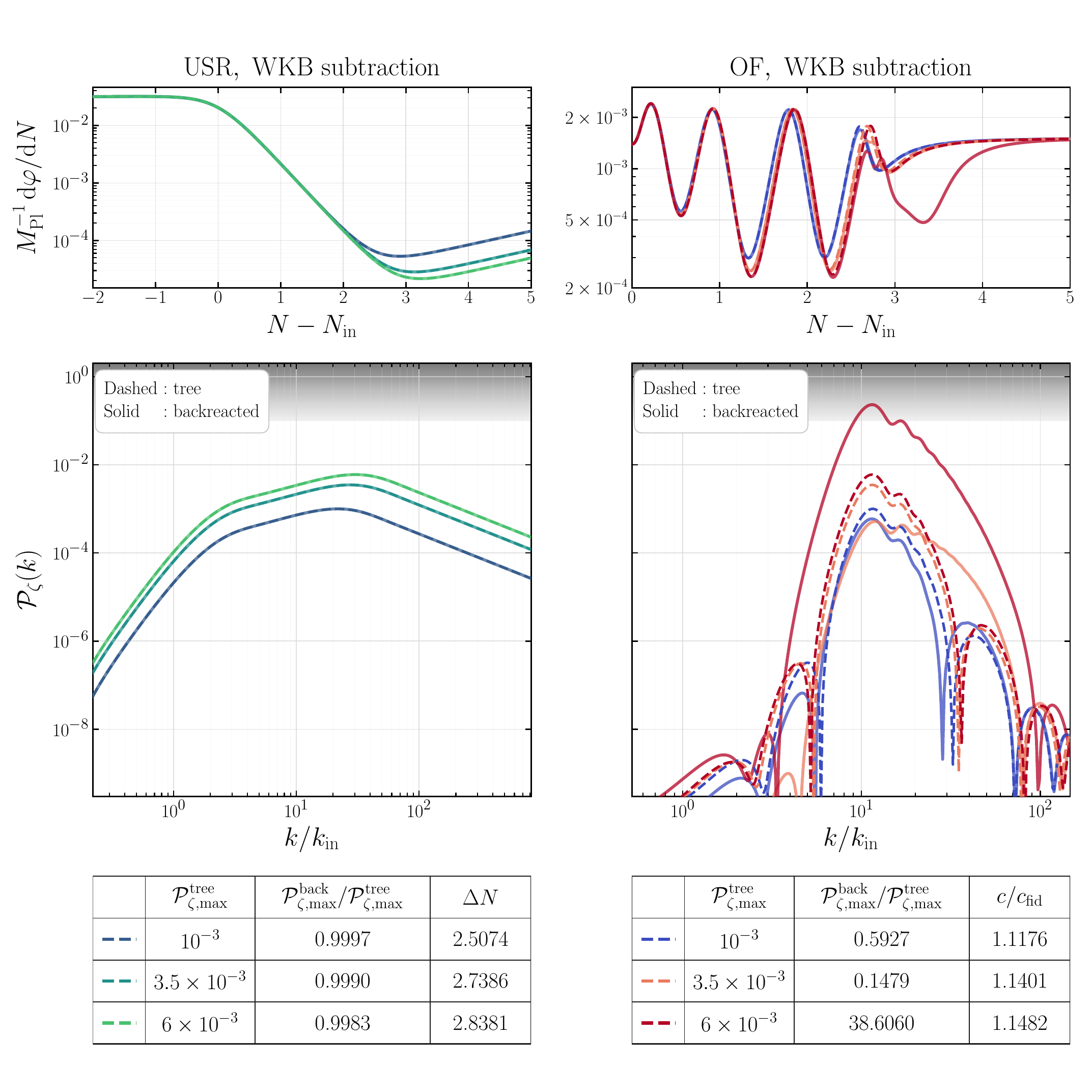}
\caption{Mean-field results with WKB subtraction. Top: field velocity for the USR (left) and OF model (right). Bottom-left: curvature spectra for the USR benchmark, with nominal tree-level peak amplitudes reported in the table. Bottom-right:  corresponding curvature spectra for the OF benchmark, for the values of $c/c_{\rm fid}$ reported in the table. Dashed and solid curves denote evolution without and with mean-field backreaction, respectively. The three corresponding cases in the two models share the same tree-level peak amplitude, while the ratio of the backreacted peak amplitude to the corresponding tree-level peak amplitude is reported in the table. The gray shading marks $\mathcal P_\zeta\gtrsim10^{-1}$ where we conservatively expect PBH overproduction is difficult to avoid even including non-Gaussianity \cite{Ferrante:2022mui,Young:2022phe,Gow:2022jfb,Iovino:2024tyg}.
}
\label{fig:adiabatic-results}
\end{figure}

Figure~\ref{fig:adiabatic-results} collects the regularized power spectra of the curvature perturbation for the two scenarios, computed using the backreacted evolution of the inflaton fluctuations. In both cases, we evolve the background and linear modes self-consistently with the adiabatically subtracted sources of Section~\ref{sec:adiabatic}. For each case, we show three values of the relevant parameter controlling the amplitude of the peak in the tree-level power spectrum, which realize $\mathcal P_{\zeta,\mathrm{tree}}^{\max} =\{1,\,3.5,\,6\}\times10^{-3}$ in each model. As this procedure removes the UV divergences, the backreaction integrals converge in the UV.

We start by discussing the USR model, and vary the parameter controlling the amplitude of the peak, i.e., $\Delta N$. From the plot, the backreacted spectra are indistinguishable from the corresponding tree-level counterparts. Indeed, the backreaction after subtracting the UV divergences remains very small and its effect on the background and perturbed equations is negligible, at the level of ${\cal O}(10^{-3})$ at most. The reduced impact of the backreaction terms is due to the mildly enhanced perturbations, as well as the significant reduction of power at $k \gtrsim aH$ produced by adiabatic subtraction. Larger backreaction is possible, in scenarios with higher enhancement of perturbations and larger $d^3 V/d\varphi^3$, as we discuss in App.~\ref{subsec:benchmark_usr}.

The situation is different in the OF model. As discussed in the previous Section, particle production triggered by the feature in the potential generates a pronounced peak in the backreaction, which survives adiabatic subtraction. This peak can produce a significant backreaction, which affects the background dynamics. In particular, the background begins to deviate from the solution without backreaction after the first oscillation. For $\mathcal P_{\zeta,\mathrm{tree}}^{\max} = 6\times10^{-3}$, backreaction slows the final oscillation of the field velocity around the attractor, delaying its relaxation to the attractor by approximately one e-fold. The power spectrum visually changes compared to the tree-level one. For the same tree-level amplitudes considered in the USR model, for which the power spectrum is nearly unchanged, here the effect of the backreaction on the evolution of the perturbations is such that the power spectrum is reduced by $\mathcal{O}(1)$ for $\mathcal P_{\zeta,\mathrm{tree}}^{\max} =\{1,\,3.5\}\times10^{-3}$ and is even increased by a factor ${\cal O} (40)$ for  $\mathcal P_{\zeta,\mathrm{tree}}^{\max} =6\times10^{-3}$. Although the tree-level peak is only $\mathcal P_{\zeta,\mathrm{tree}}^{\max}=6\times10^{-3}$, in this highest-amplitude case the backreaction terms $\lvert\mathcal{B}_\varphi\rvert$ and $\lvert\mathcal{B}_H\rvert$ transiently exceed unity, violating the assumptions of our mean-field approach and signalling a breakdown of perturbative validity. This is consistent with the sizeable backreaction integrands already visible in the lowest-amplitude case shown in Fig.~\ref{fig:usr-backreaction}. Moreover, the background trajectory approaches the regime in which field trapping is expected, as illustrated in the last figure. This proximity to trapping, together with the strong backreaction, provides a qualitative explanation for the large enhancement of the power spectrum compared with the two lower-amplitude cases.
Our findings suggest that backreaction is more relevant for models where the amplification of the perturbations is produced by an explosive mechanism, where it occurs when the modes are still within the horizon and significantly deviate from the adiabatic vacuum.

Finally, we stress that the correction to the background during the earlier phase of the inflationary evolution before the USR and the OF transient stage is negligible. As such---given that the only effect on the perturbations in our mean-field approach is through changes in the background---the power spectrum of curvature perturbations at large scales is not modified by backreaction.

\section{Comparison with the lattice}
\label{sec:results}

After having presented the impact of backreaction on the background dynamics and on the spectrum of curvature perturbations for a few benchmarks, we now compare our analysis with new non-perturbative lattice simulations of benchmarks studied in recent works. Our goal is to assess the extent to which our method can reproduce these results and then quantify their potential sensitivity to unsubtracted UV contributions in the lattice dynamics. Given the qualitative differences between the two models, in this Section we discuss them separately in the two Subsections~\label{subsec:comparison-USR} and~\label{subsec:comparison-OF} below. Throughout this Section, we will use  \texttt{InflationEasy}\footnote{\href{https://github.com/caravangelo/inflation-easy}{https://github.com/caravangelo/inflation-easy}} (see Ref.~\cite{Caravano:2025klk}) as the benchmark lattice code\footnote{For the time evolution we use its second-order leapfrog integrator. The latter, had to be refined compared to the initial version implemented in in the most recent release of the \texttt{InflationEasy}, which gave incorrect results that were inconsistent with the other integrator implemented in the code, i.e. the fourth-order Runge--Kutta integrator. The fix consists in making sure that the field values remain half a time step ahead of the field velocities when the step size changes, and the Hubble friction term is evaluated at the centre of the step. We have verified that, with this integrator, our results agree with those obtained with the fourth-order Runge--Kutta integrator. In addition to the correction discussed here, the code includes a fix for a separate bug that caused the numerical derivatives of the potential to be read incorrectly. This fix was already present in the version we used as the starting point for our modifications. For these reasons, we compare our mean-field results with new lattice simulations performed by us for this work using the corrected code.}. 

Lattice simulations do not include metric perturbations, and evolve the scalar field in real space by solving the classical Klein--Gordon equation without truncating its nonlinear interactions, and extract power spectra through a discrete Fourier transform of the evolved field configurations. The finite lattice spacing restricts the range of comoving momenta represented on the grid, providing a UV regulator. At the same time, the discretized Laplacian modifies the dispersion relation: the effective wavenumber governing mode evolution generally differs from the nominal Fourier wavenumber, particularly near the lattice cutoff, with the relation depending on the finite-difference discretization of the Laplacian. In particular, consider the standard second-order finite-difference discretization of the Laplacian on a cubic lattice of $N^3$ points with comoving side $L$ and spacing $\Delta x=L/N$. A Fourier mode with integer wavevector $\mathbf m$ is an eigenfunction of the discrete Laplacian, with eigenvalue $-k_{\mathrm{eff}}^2(\mathbf m)$, where
\begin{equation}
k_{\mathrm{eff}}^2(\mathbf m)=\frac{4}{\Delta x^2}\sum_{i=1}^{3}\sin^2\left(\frac{\pi m_i}{N}\right).
\label{eq:keff}
\end{equation}
The maximum effective wavenumber is therefore
\begin{equation}
k_{\mathrm{eff,max}}=\frac{2\sqrt{3}}{\Delta x}=\frac{2}{\pi}k_{\mathrm{lat,max}},
\end{equation}
which thus acts as an effective UV cutoff for the lattice dynamics. Moreover, the modes of a cubic lattice are not distributed in momentum like those of a continuum sphere, which is the integration domain naturally implemented in our mean field approach. Matching lattice results to continuum momentum integrals therefore requires accounting for both the dispersion relation and the mode counting; a sharp cutoff of the continuum integral, regardless of its value, does not reproduce the lattice sums at all times. Below, we explain how both effects are taken into account in our calculation.

Since standard cosmological lattice codes do not subtract UV contributions, modes within the resolved UV range may affect the final result if their contributions are appreciable. To provide a fair comparison, our mean-field computation is performed without any adiabatic subtraction, with UV divergences instead regulated by restricting the momentum modes in a way that closely reproduces what is done by the lattice, as we now explain.

We have tested that if we perform the spherical integrals in $\mathcal{B}_\varphi$, $\mathcal{B}_H$ and fix $k_{\rm max}=k_{\mathrm{eff,max}}$, our mean-field approach can reproduce the trend observed in the quantities evaluated by the lattice. Such a qualitative agreement is reassuring, but the results obtained with the two methods differ by $\mathcal{O}(10\%)$, which is not satisfactory. In particular, as mentioned above, the lattice integration in real space on a discrete grid is not equivalent to the spherical integration over the $k-$sphere in our mean-field approach. To isolate physical effects from differences in the numerical schemes, we thus need to adapt the integration over wavenumbers of the mean-field quantities $\mathcal{B}_\varphi$, $\mathcal{B}_H$ to sum over exactly the modes present on the lattice. To this end, we can simply re-weight the $k-$integrals in our mean-field approach as follows
\begin{equation}
\int {\rm d}\ln k\,\frac{k^3}{2\pi^2}\,F(k)
\;\longmapsto\;
\frac{1}{L^3}\sum_{\mathbf m\neq0}F\big(k_{\rm eff}(\mathbf m)\big)
\simeq
\Delta\sum_i w_i\,\frac{\bar k_i^3}{2\pi^2}\,F(\bar k_i)=\frac{1}{L^3}
\sum_i N_i\,\,F(\bar k_i) \;,
\label{eq:lattice-mode-sum}
\end{equation}
where
\begin{equation}
    w_i\equiv \frac{N_i}{L^3\,\Delta\,}\frac{2\pi^2}{\bar k_i^3},
\end{equation}
and $N_i$ is the number of lattice modes in the logarithmic bin $i$, in which $\bar k_i$ the mean of the modes with $k_{\rm eff}$. The logarithmic bins all have an equal (narrow) width $\Delta$. The weight $w_i$ is thus the ratio between the number of lattice modes in the bin and the number of continuum modes in the same momentum interval. 

We compare with lattice simulations on $N^3$ grids with $N=64$, $128$, $256$ and $512$, all in a box of the same comoving size, with the corresponding quantities computed in our mean-field approach. Increasing $N$ adds modes in the UV, and corresponds to a larger UV cutoff. In addition to $\langle\varphi_N\rangle$ and $\mathcal{P}_\zeta$, which we presented in the previous Section, here we also plot the field variance $\langle\delta\varphi^2\rangle$ and the sum of the kinetic and gradient energies defined as,
\begin{equation}
    \langle\delta\phi^2\rangle
    =\int {\rm d}\ln k\,\mathcal P_{\delta\phi},
    \qquad
    \rho_{\delta\phi}^{K+G}
    =\frac12\int {\rm d}\ln k
    \left[
        \mathcal P_{\delta\dot\phi}
        +\frac{k^2}{a^2}\mathcal P_{\delta\phi}
    \right],
    \label{eq:var-rhoKG}
\end{equation}
where the momentum integrals are replaced by the lattice
mode sum of Eq.~\eqref{eq:lattice-mode-sum}, with
$k^2\to k_{\rm eff}^2$ in the gradient contribution.
We present the fluctuation energy as a fraction of the
total energy density, $\rho_{\delta\phi}^{K+G}/(3H^2M_{\rm Pl}^2)$. These quantities are useful, as they are more directly related to the backreaction $k-$integrals and thus better encapsulate eventual UV dependencies.

Figure~\ref{fig:lattice-case-I} shows the comparison for the largest target amplitude in the linear theory, $\mathcal P_{\zeta,\mathrm{tree}}^{\max}=1$, together with the tree-level and the WKB-regularized results.  We have tested that we obtain similar results for decreasing  $\mathcal P_{\zeta,\mathrm{tree}}^{\max}$, but we chose to present only this value to keep the discussion simple, and because it produces the strongest backreaction, making the variations visible by eye.

\subsection{Comparison for the Ultra-Slow-Roll case}

\label{subsec:comparison-USR}

We start from the USR model. In the main text, we compare our results with Case I of Ref.~\cite{Caravano:2024moy}, which realizes Wands duality~\cite{Wands:1998yp} through a USR phase induced by a local minimum followed by a local maximum in the potential. Near the maximum, the potential is approximately symmetric and well described by an inverted parabola, with small higher-order derivatives.  This scenario is particularly useful to build intuition, since, as we verify below, non-linearities are small and thus can be captured by our mean field result. We therefore expect our method to be able to reproduce the results obtained with the lattice. Cases II and III of Ref.~\cite{Caravano:2024moy}, discussed in App.~\ref{subsec:benchmark_usr}, feature a larger cubic derivative of either sign at the local maximum and thus probe departures from this approximately quadratic form.

\begin{figure}[t]
\centering
\includegraphics[width=\textwidth]{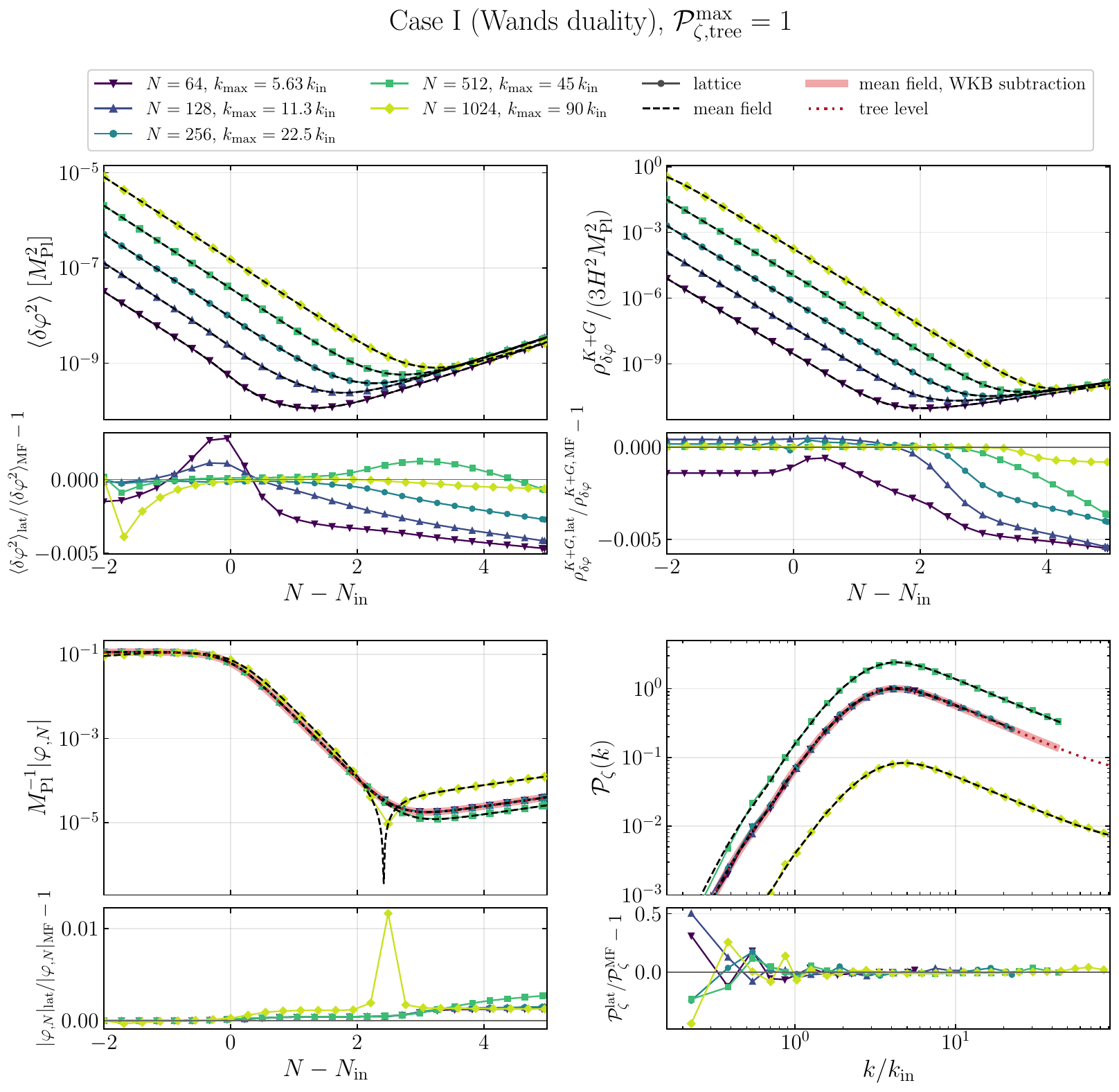}
\caption{
USR Case I, $\mathcal P_{\zeta,\mathrm{tree}}^{\max}=1$. Lattice simulations with $N^3$ points in a box of fixed comoving size $L \simeq 39.4 / k_{\rm in}$ (colored lines with markers; $k_{\max}$ is the largest effective momentum of each lattice) and bare mean field summed over the same lattice modes, Eq.~\eqref{eq:lattice-mode-sum} (black dashed). Top: variance of the field fluctuations and their kinetic plus gradient energy density in units of $3H^2M_{\rm Pl}^2$. Bottom: field velocity and curvature power spectrum at the end of the evolution, where the red band and the red dotted line show the mean field with WKB subtraction and the tree-level result, respectively. The lower panels show the relative difference between lattice and mean field.
}
\label{fig:lattice-case-I}
\end{figure}

In Fig.~\ref{fig:lattice-case-I}, we follow the setup of Ref.~\cite{Caravano:2024moy} and use a comoving box of size $L\simeq39.4/k_{\rm in}$, evolved from $N_{\rm in}-2$ to $N_{\rm in}+5$, so that the smallest momentum on the lattice is $k_{\rm IR}=2\pi/L\simeq0.16\,k_{\rm in}$ and the effective UV cutoff ranges from $k_{\rm eff,max}\simeq5.6\,k_{\rm in}$ for $N=64$ to $45\,k_{\rm in}$ for $N=512$. Once the lattice modes are accounted for as explained above, the mean field computation reproduces the lattice results  at all lattice sizes: all the quantities agree at better the percent level throughout the evolution,\footnote{The larger scatter of the lattice spectrum at the largest scales is due to sample variance, as only a few lattice modes contribute there.} showing that our mean field can very accurately reproduce the lattice for all $N$ considered. Another feature that can be appreciated in the figure is the departure from the tree-level result of the UV-regulated quantities, which grows with the UV cutoff and even modifies the results by a factor $\mathcal{O}(1)$ for  $N=512$. This, together with the matching with our UV-regulated mean-field approach and the perfect agreement of the linear theory with the UV-subtracted mean-field, strikingly proves the UV sensitivity of the lattice, which should not be misled as a modification to the tree-level results induced by nonlinear interactions. This conclusion is one of the main results of our paper. To our knowledge, the UV dependence of lattice simulations of inflation of USR has not been explicitly proved elsewhere.

In particular, from Fig.~\ref{fig:lattice-case-I}, we can also explain how the UV modes influence the dynamics. At the beginning of the evolution, when all modes are in their vacuum state, the bare variance and energy density of the fluctuations are dominated by the largest momenta: doubling $N$, which doubles $k_{\max}$, increases them by factors of four and sixteen, i.e.\ as $k_{\max}^2$ and $k_{\max}^4$ (top panels).\footnote{We note that the Hubble parameter is initialized by including the variance of linear perturbations, see Refs.~\cite{Caravano:2025klk,Caravano:2021pgc}, which is itself sensitive to the UV cutoff. We make sure to initialize lattice and mean-field in a consistent manner in our comparison.} These UV contributions redshift away as the modes are stretched, and after USR the variance and energy density are set by the amplified modes, which all but the smallest ($N = 64$) lattice resolve, so that they become nearly independent of $N$.

The effect of the UV modes on the background, however, persists. The fluctuations act on the background through two channels: the ``effective force'' from the field variance, $\propto V'''\langle\delta\varphi^2\rangle$, which in this benchmark accelerates the field and suppresses the power spectrum, and the energy density of the fluctuations, which changes the Hubble rate and thereby slows the field down and enhances the power spectrum. Both contributions are dominated by the unsubtracted UV vacuum modes, but they have different sensitivities to the UV cutoff: the field variance grows as $k_{\max}^2$, while the fluctuation energy density grows as $k_{\max}^4$. The latter therefore becomes increasingly important as $k_{\max}$ increases. On the two smaller lattices the effective force from the field variance dominates and the peak of the power spectrum is suppressed by less than one percent. On the larger ones the energy of the UV modes takes over: at $N=512$ the field velocity after USR drops well below its tree-level value and the peak of the power spectrum more than doubles, on the lattice as in the mean field. The backreaction found on the lattice in this benchmark is thus set by the unsubtracted UV vacuum energy, and it changes with the lattice spacing. The dependence of the results on the UV is even more pronounced for $N=1024$. Interestingly, in that case, we observe that the backreaction is so large that it changes the sign of the field velocity, significantly altering the physics of the model. Our mean-field approach perfectly reproduces this behavior, which is an unphysical artifact, and a consequence of including the UV divergences. In passing, this proves that the leading UV effect affecting the lattice, at least in this USR benchmark, is encoded in the 1-loop backreaction of the perturbations on the background.

We stress once more that, when these UV vacuum contributions are removed by adiabatic subtraction, the effective backreaction remains small, and the curvature power spectrum is indistinguishable from the tree-level one. Our results therefore provide a cautionary lesson for lattice simulations of inflation: without a proper treatment of UV vacuum contributions, the inferred backreaction can depend strongly on the lattice spacing, whose choice is thus far from innocuous.

\begin{figure}[t]
\centering
\includegraphics[width=\textwidth]{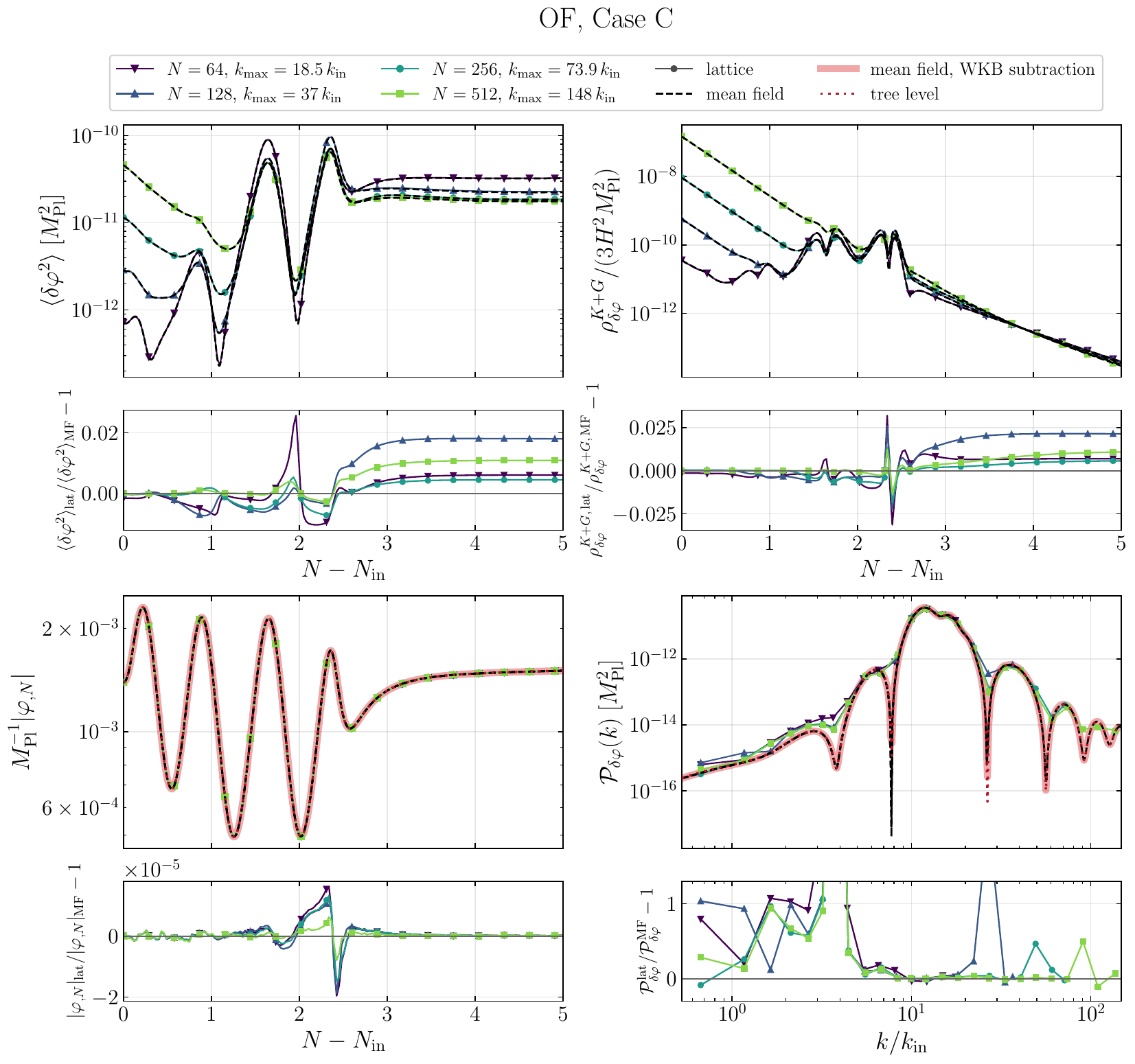}
\caption{OF Case C in the setup of Ref.~\cite{Caravano:2024tlp}. Lattice simulations  with $N^3$ grid points, for $N=64$, $128$, $256$ and $512$ (colored lines with markers), in a box of comoving size $L=12/k_{\rm in}$ are compared with the bare mean field evaluated using the same lattice momentum distribution, Eq.~\eqref{eq:lattice-mode-sum} (black dashed lines). Here $k_{\max}$ denotes the largest effective momentum of each lattice. The upper panels show the fluctuation variance and the kinetic plus gradient energy density, normalized to $3H^2M_{\rm Pl}^2$. The lower panels show the field velocity and the power spectrum of $\delta\phi$ at $N-N_{\rm in}=5$. The red band and red dotted line indicate the WKB-subtracted mean field and tree-level result, respectively. The residual panels show the relative differences between the lattice and bare mean field.}
\label{fig:lattice-OF}
\end{figure}

\subsection{Comparison for the Oscillatory Feature case}

\label{subsec:comparison-OF}

We now turn to the OF model, focusing on Case C of Ref.~\cite{Inomata:2022yte}, referred to as Case I in Ref.~\cite{Caravano:2024tlp}. Both studies found this scenario to be perturbative, with non-linear corrections producing only small changes to the power spectrum. The other scenarios considered in those works enter the non-perturbative regime. As we show in Appendix~\ref{app:benchmark_models}, our formalism also breaks down in those cases and is therefore not applicable to them.

In Fig.~\ref{fig:lattice-OF}, we compare our bare mean-field results with lattice simulations performed following the set up of Ref.~\cite{Caravano:2024tlp}, using $N^3$ spatial grid points with $N=64$, $128$, $256$ and $512$, in a comoving box of size $L=12/k_{\rm in}$ evolved from $N_{\rm in}$ to $N_{\rm in}+5$. The results are presented in the same format as Fig.~\ref{fig:lattice-case-I}, with the power spectrum of $\delta\phi$ shown in place of $\mathcal P_\zeta$. This choice is made both to allow for a more direct comparison with Refs.~\cite{Caravano:2024tlp,Inomata:2022yte}, and because the dominant effect--unlike in USR--in these types of model is to amplify directly $\delta\phi$.

Unlike in USR, the backreaction in the OF model is dominated by the modes resonantly amplified by the feature, rather than by the UV vacuum modes, within the range of lattice cutoffs considered here. This is consistent with the pronounced enhancement of $\mathcal P_{\delta\phi}$ over a finite momentum range visible in Fig.~\ref{fig:lattice-OF}. Although the bare fluctuation variance and energy density are initially dominated by vacuum modes and increase with lattice resolution, the contribution of these UV modes to the backreaction remains subdominant to that of the resonantly amplified modes after the feature. Consequently, the physical origin of the backreaction is relatively insensitive to the UV cutoff in this regime. At sufficiently large cutoffs, however, the unsubtracted UV vacuum contribution would eventually become important. This is consistent with Ref.~\cite{Inomata:2022yte}, which identified a range of loop momenta in which the one-loop correction is dominated by resonantly amplified modes and shows little sensitivity to the UV cutoff.

The only significant differences between the simulation and the tree-level result occur around the zeros of the power spectrum, where the latter becomes extremely small. These zeros are an artifact of the tree-level calculation and we expect it to disappear once the 1-loop corrections to $\mathcal P_{\delta\phi}$ induced by higher-order interactions are included. Indeed, around the oscillations at wavenumbers smaller than the first peak, the lattice spectrum closely resembles the 1-loop result of Ref.~\cite{Inomata:2022yte}. The fact that our mean-field treatment--which includes 1-loop effects only through backreaction--remains very close to the tree-level result further supports the interpretation that the lattice genuinely captures non-linear corrections at scales where the tree-level power is strongly suppressed.

\section{Discussion and conclusions}
\label{sec:conslusions}

The recent debate on the validity of perturbation theory in single-field inflationary models that enhance the power spectrum at small scales has highlighted several technical and conceptual issues. These include identifying the relevant interactions, regularizing UV divergences, and consistently accounting for the effect of fluctuations on the inflationary background. Our analysis focuses on the latter two questions. We do not attempt to compute the one-loop power spectrum of curvature perturbations, which is left for future work. Instead, our goal is to incorporate the backreaction of fluctuations into the background and mode evolution in a form suited to numerical computation.

To this end, we have adopted a mean-field approach that includes the backreaction of perturbations on the homogeneous background at leading perturbative order. This framework allows us to identify the sensitivity to the divergent UV contribution of the adiabatic vacuum. Removing this contribution through standard adiabatic subtraction, we find that backreaction is milder than previously inferred in some of the scenarios considered. The precise finite magnitude of the backreaction obtained after adiabatic subtraction remains scheme dependent until physical renormalization conditions are specified; our results should therefore be regarded as quantitative estimates within the prescription adopted here, while their parametric and qualitative features are considerably more robust.

Our results clarify when non-linear calculations without explicit subtraction of UV vacuum contributions can provide reliable estimates of backreaction. A sufficiently clear separation between the particle-production peak and the adiabatic vacuum tail allows the momentum integral to capture the amplified modes while keeping the unsubtracted vacuum contribution negligible. Without this separation, the inferred backreaction can depend significantly on the UV cutoff, even when the non-linear evolution is fully included, as in lattice calculations.

In the USR models, the curvature perturbation grows mainly because the inflaton velocity decreases, while the inflaton fluctuations evolve only gradually away from the adiabatic vacuum; this makes the separation less clear and requires a careful subtraction of the UV vacuum contribution. After WKB subtraction, we find that backreaction remains small for realistic scenarios in which the enhanced $\mathcal P_\zeta$ does not lead to PBH overproduction, unless the third derivative of the potential is tuned to be large.

Among the scenarios studied here, a clear separation is instead realized in the OF model, where the sharp oscillatory feature drives strong particle production in a distinct band of modes. This resonant amplification can produce sizeable physical backreaction even at lower peak amplitudes than in USR. The importance of backreaction therefore depends on the amplification mechanism as well as on the amplitude of the curvature spectrum. When the resulting non-linear effects become large, our mean-field approximation is no longer adequate, motivating lattice simulations with appropriate control of the UV contribution.

\paragraph{Relation to other treatments of backreaction.}
Our treatment is closely related to that of Ref.~\cite{Inomata:2024lud}. In the spatially flat gauge and in the decoupling limit, the backreaction on the inflaton background can be derived perturbatively from the in-in formalism and expressed as a correction to the Klein--Gordon equation. In this sense, our backreacted Klein--Gordon equation implements the tadpole condition
\begin{equation}
    \langle \delta\phi(t,\mathbf{x})\rangle = 0,
    \qquad
    \varphi(t)=\langle\phi(t,\mathbf{x})\rangle ,
\end{equation}
as we show explicitly in Appendix~\ref{sec:Bphi_allscale} (see also Refs.~\cite{Sloth:2006az,Sloth:2006nu}). Compared with a perturbative treatment in which the loop-induced shift of the background is computed around a fixed classical solution, our approach incorporates this shift directly into the subsequent evolution. The background field and the fluctuation modes are therefore evolved self-consistently, so that the modification of the background induced by the fluctuations feeds back into their later evolution. In addition, we consistently include the backreaction of the fluctuation energy density and pressure on the Hubble rate through the Friedmann equations. In this sense, our framework provides a self-consistent implementation of the tadpole-induced background shift, rather than a computation of a fixed-order correction to the curvature power spectrum. An alternative renormalization prescription keeps the chosen background solution fixed and cancels the tadpole by means of a local linear counterterm~\cite{Pimentel:2012tw,Fumagalli:2024jzz,Ballesteros:2024zdp,Inomata:2025pqa,Braglia:2025cee,Braglia:2025qrb,Ballesteros:2025nhz,Kristiano:2025ajj,Inomata:2025bqw}. In this latter approach, the evolution of the perturbations in the interaction picture is also left unaffected, and the effects on the backreaction are encoded in quadratic and higher-order counterterms that are unavoidably generated by non-linearly realized time diffeomorphisms.  

Our approach, where the Klein--Gordon equation is explicitly sourced by nonlinear fluctuation terms, provides a clearer framework to compare with lattice simulations, where the nonlinear scalar-field equation is evolved directly. In fact, while our framework retains only the homogeneous mean-field contribution of these nonlinearities, we find that this is enough to draw some interesting conclusions. We have compared our mean-field approach with the direct evolution of the KG and Friedmann equations on the lattice using the \texttt{InflationEasy} code. We find that when non-linearities are small enough so that our mean-field approach can be reliably used, we can perfectly reproduce the quantities computed by the lattice for both the USR and the OF model. Importantly, in order to compare with the lattice, where UV divergences are not removed, we retain them in our mean-field approach. The matching of the two approaches sheds light on the UV dependence of the lattice, which should be accurately checked in each model considered. This dependence is particularly relevant for the USR model and less so for the OF. Our mean-field approach proves to be very useful to trace this difference back to a separation of scales in the process of enhancing perturbations, which is observed in the OF but not in the USR model. Furthermore, it is interesting to stress that the mean-field approach proved to be sufficient to capture non-linear corrections to the tree level predictions in some of the cases considered in this work.


\acknowledgments

We thank Angelo Caravano, Giordano Cintia, Keisuke Inomata, Gabriele Perna, Lucas Pinol, Vincent Vennin for interesting discussions, and Guillermo Ballesteros, Valerie Domcke, Keisuke Inomata, Lucas Pinol, and Lorenzo Sorbo for comments on a draft of this paper.
M.B. has received funding from the European Union’s Horizon 2020
research and innovation programme under the Marie Skłodowska-Curie grant
agreement No 101205460.
G.F. acknowledges support from the Italian Ministero dell'Università e della Ricerca through the FIS 3 project FIS-2024-02546 “PRIMAVERA” (CUP C53C25001070001), the Departments of Excellence grant 2023–2027 “Quantum Frontier”, as well as from Istituto Nazionale di Fisica Nucleare (INFN) through the Theoretical Astroparticle Physics (TAsP) project.
N.I.  and M.P. also acknowledge support from INFN through the TAsP project.

\appendix


\section{Slow-roll relations and the linearized inflaton equation}
\label{app:linear-dynamics}
In this Appendix, we collect the background identities and derive the linearized mode equation with and without metric perturbations, making explicit their relation in the decoupling limit.

\subsection{Exact relations involving the slow-roll parameters}
\label{sec:slow}
We first derive relations involving the slow-roll parameters $\epsilon_H$ and $\eta_H$ defined in Eq.~\eqref{def-veh}. We use the number of e-folds $N$, defined through $dN=H\,dt$. All these relations are exact in the limit of negligible backreaction of the inflaton (and metric) perturbations on the background dynamics (namely, they do not assume that the slow-roll parameters are small). We start from the background equations with no backreaction 
\begin{eqnarray}
\ddot{\varphi} + 3 H \dot{\varphi} + \frac{d V}{d \varphi} = 0 \;\;,\;\; 
\dot{H} = - \frac{\dot{\varphi}^2}{2 M_{\rm Pl}^2} \;\;,\;\; 
H^2 = \frac{1}{3 M_{\rm Pl}^2} \left( \frac{\dot{\varphi}^2}{2} + V \right) \;, 
\label{eq-bck}
\end{eqnarray} 
where, as is well known, the second relation can be obtained from the first and third ones due to a nontrivial Bianchi identity. Using the definition~\eqref{def-veh} of $\epsilon_H$, we can express this second relation as 
\begin{equation}
\dot{\varphi}^2 = 2 \epsilon_H \, H^2 \, M_{\rm Pl}^2 \;, 
\label{phidot-eps}
\end{equation}
which, inserted in the third equation of~\eqref{eq-bck}, results in  
\begin{equation}
V  = \left( 3 - \epsilon_H \right) M_{\rm Pl}^2 H^2 \;. 
\label{eqV}
\end{equation} 

We can then combine the derivative of the definition of $\epsilon_H$ with the definition of $\eta_H$ to write 
\begin{equation}
\dot{\epsilon}_H = 2 H \epsilon_H \left( \epsilon_H - \eta_H \right) \;. 
\label{eps-dot} 
\end{equation} 

We also have 
\begin{equation} 
\eta_H =  - \frac{\ddot{H}}{2 H \dot{H}} = - \frac{\frac{d}{d t} \left( - M_{\rm Pl}^2 \dot{H} \right)}{H \left( - 2 M_{\rm Pl}^2 \dot{H} \right)} = - \frac{\frac{d}{d t} \frac{\dot{\varphi}^2}{2}}{H \dot{\varphi}^2} = - \frac{\ddot{\varphi}}{H \dot{\varphi}} \;, 
\end{equation} 
which we combine with the first of \eqref{eq-bck} to write 
\begin{equation}
\eta_H =  - \frac{\ddot{\varphi}}{H \dot{\varphi}} = \frac{3 H \dot{\varphi} + \frac{d V}{d \varphi}}{H \dot{\varphi}} = 3 + \frac{1}{H \dot{\varphi}} \, \frac{d V}{d \varphi}\;.
\label{Vp} 
\end{equation} 
Differentiating this relation leads to 
\begin{equation}
\frac{d^2 V}{d \varphi^2} = H^2 \left[ \left( 3 - \eta_H \right) \left( \epsilon_H + \eta_H \right) + \frac{d \eta_H}{d N} \right] \;. 
\label{Vpp}
\end{equation} 
Equation~\eqref{phidot-eps} expresses the inflaton velocity in terms of the slow-roll parameters, while Eqs.~\eqref{eqV},~\eqref{Vp}, and~\eqref{Vpp} give the corresponding expressions for the potential and its first two derivatives.

From the definition of $\epsilon_H$, we can also express the second derivative of the scale factor as 
\begin{equation}
\frac{a'}{a} = a \, H \;\; \Rightarrow \;\; \frac{a''}{a} = a^2 H^2 \left( 2 - \epsilon_H \right) \;. 
\label{ap-app}
\end{equation}

\subsection{Linearized equation for the inflaton perturbations}
\label{sec:eq:MS}
Including metric perturbations in spatially flat gauge, the variable $v\equiv a\,\delta\phi$ obeys the Mukhanov--Sasaki equation
~\cite{Mukhanov:1985rz,Sasaki:1986hm},
\begin{equation}
v'' + \left( k^2 - \frac{z''}{z} \right) v = 0 \;\;\;,\;\;\; z \equiv \frac{a \dot{\varphi}}{H} \;.
\end{equation}
Employing the background identities derived above, we obtain $\frac{z'}{z} = a H \left( 1 + \epsilon_H - \eta_H \right)$, and
\begin{equation}
\frac{z''}{z} = a^2 H^2 \left( 2 + 2 \epsilon_H - 3 \eta_H + 2 \epsilon_H^2 - 3 \epsilon_H \eta_H + \eta_H^2 - \frac{d \, \eta_H}{d N} \right)
\end{equation}
so that the Mukhanov--Sasaki equation takes the form
\begin{equation}
v'' + \left[ k^2 - a^2 H^2 \left( 2 + 2 \epsilon_H - 3 \eta_H  + 2 \epsilon_H^2 - 3 \epsilon_H \eta_H + \eta_H^2 - \frac{d \eta_H}{d N} \right) \right] v = 0 \;.
\label{eqv-dg}
\end{equation}
Neglecting metric perturbations altogether---i.e., varying the action~\eqref{S-phi} for the same $v\equiv a\,\delta\phi$ on an unperturbed FLRW background, and using~\eqref{Vpp} and~\eqref{ap-app} as before---gives instead
\begin{equation}
v'' + \left[ k^2 - a^2 H^2 \left( 2 - 4 \epsilon_H - 3 \eta_H  + \epsilon_H \eta_H +\eta_H^2 - \frac{d \eta_H}{d N} \right) \right] v = 0 \;.
\label{eqv-nodg}
\end{equation}
This is the mode equation on an unperturbed FLRW geometry. Using Eqs.~\eqref{phidot-eps} and~\eqref{Vp}--\eqref{ap-app}, Eqs.~\eqref{eqv-dg} and~\eqref{eqv-nodg} can be combined into the field-space expression~\eqref{eq:mu} quoted in the main text. In particular, the difference induced by the metric perturbations is (see also \eqref{eq:mu})
\begin{equation}
\left.\mu^2\right|_{\alpha=1}-\left.\mu^2\right|_{\alpha=0}
=-2a^2H^2\epsilon_H\left(3+\epsilon_H-2\eta_H\right)\;.
\label{eq:metric-mass-difference}
\end{equation}
Thus the metric terms carry an explicit factor of $\epsilon_H$, including terms proportional to $\epsilon_H\eta_H$, and are suppressed in the decoupling regime considered here. For $\epsilon_H\ll1$ with $\eta_H$ finite, both equations give
\begin{equation}
\mu^2=-a^2H^2\left[2-3\eta_H+\eta_H^2-\frac{d\eta_H}{dN}
+{\rm O}(\epsilon_H,\epsilon_H\eta_H)\right]\;.
\label{mu-no-eps}
\end{equation}
This slow-roll representation clarifies the decoupling limit. The numerical evolution instead uses the field-space expression~\eqref{eq:mu} with $\alpha=0$, which reduces to
\begin{equation}
\mu^2(\varphi,\dot\varphi,H)
=-a^2H^2\left(
2-\frac{\dot\varphi^2}{2M_{\rm Pl}^2H^2}
-\frac{1}{H^2}\frac{d^2V}{d\varphi^2} \right) \;.
\label{eq:mu-nometric}
\end{equation}
We have verified that neglecting the metric contributions does not significantly affect our numerical results. Equations~\eqref{eq:mu} and~\eqref{eq:mu-nometric} involve only the reconstructed potential and its derivatives, together with the dynamically evolved $\dot\varphi$ and $H$. 

\section{Consistency of our regularization procedure}
\label{sec:consistency}
As mentioned in the main text, we numerically evolve the backreacted
Klein--Gordon equation and the backreacted equation for $\dot H$.
We assess their compatibility with energy conservation by comparing
the evolved $\dot H$ with the value required by differentiating the
Friedmann constraint for $H^2$. All quantities entering this comparison
are evaluated along the same numerical trajectory.

We evolve numerically
\begin{equation}
    \dot{H}_{\rm num} \equiv - \frac{1}{2 M_{\rm Pl}^2} \left\{ \rho_{\varphi}+p_{\varphi} +\rho_{\delta\phi,\,{\rm reg}}+p_{\delta\phi,\,{\rm reg}} \right\}. 
\end{equation}
On the other hand, differentiating the second equation of~\eqref{eqs-bck} would require:
\begin{equation}
    \dot{H}_{\rm th} \equiv  \frac{ \dot{\rho}_{\varphi} +\dot{\rho}_{\delta\phi,\,{\rm reg}}}{6 M_{\rm Pl}^2 H}\Bigg\vert_{\rm num}. 
\end{equation}
We therefore define the residual:
\begin{equation}
   \Delta_H\equiv \Biggl\lvert1-\frac{\dot{H}_{\rm th}}{\dot{H}_{\rm num}}\Biggr\rvert=\Biggl\lvert\frac{\Delta_{\rm ad}}{6 \mpl^2 H \dot{H}}\Biggr\rvert_{\rm num},
\end{equation}
where the right-hand side is evaluated on the numerically evolved dynamics, and $\Delta_{\rm ad}$ is defined in Eq.~\eqref{eq:violation-cons}, given explicitly by
\begin{align}
\Delta_{\rm ad}
&=\frac{H}{2}\int d\ln k\,\Bigg[
\frac{d}{dN}\mathcal{P}_{\delta\dot\varphi,\rm ad}\big|_4
+6\mathcal{P}_{\delta\dot\varphi,\rm ad}\big|_4
+\frac{k^2}{a^2}
\frac{d}{dN}\mathcal{P}_{\delta\varphi,\rm ad}\big|_4
+V_{,\varphi\varphi}
\frac{d}{dN}\mathcal{P}_{\delta\varphi,\rm ad}\big|_2
\Bigg]\\
&\equiv \Delta_{{\rm ad},\,S_i}+\Delta_{{\rm ad},\,{S}_{i,\,N}},
\end{align}
where, for clarity, we have split $\Delta_{\rm ad}$ into terms where the derivatives act directly on the filters and those where the derivatives act on the mode functions. Equivalently, we can write:
\begin{equation}
    \Delta_H\equiv \Delta_{H,\,S_i}+\Delta_{H,\,{S}_{i,\,N}}.
\end{equation}
 We have verified that $\Delta_H$ is typically smaller than $\mathcal{O}(0.001)$ in the models we considered, implying consistency at better than the $0.1\%$ level throughout the whole evolution. A peak of $1.87\%$ is reached in the OF model, but only transiently. In both cases, we see that most of the contribution to $\Delta_H$ is given by $\Delta_{H,\,{S}_{i,\,N}}$, i.e. the term containing the time derivatives of the filters.

\begin{figure}[t]
\centering
\includegraphics[width= \textwidth]
{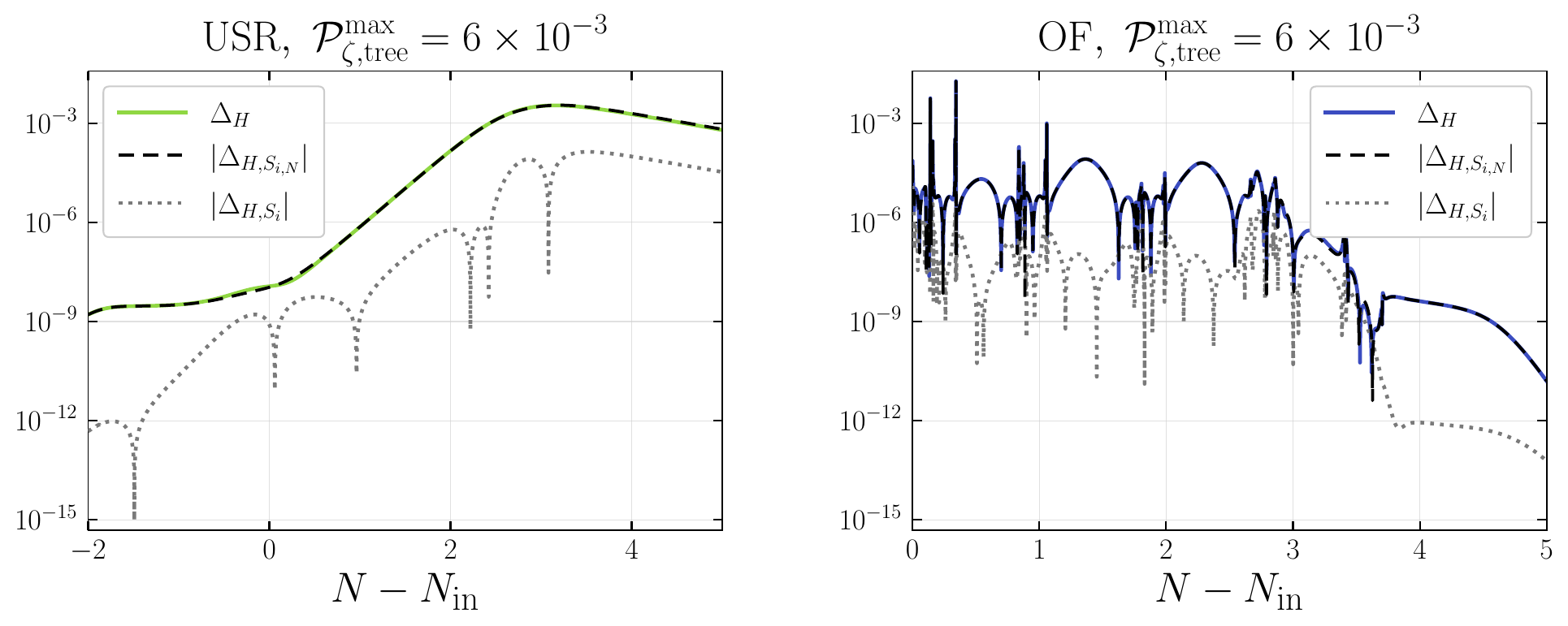}
\caption{The quantity $\Delta_H$ defined in Eq. (B.3) and its separate contributions for the USR (left) and OF (right) benchmarks, with $\mathcal P_{\zeta,\mathrm{tree}}^{\max}=6\times10^{-3}$. The violation of energy conservation remains small throughout the evolution, reaching at most $0.34\%$ in USR and $1.87\%$ in OF, with the latter attained during a narrow transient spike. The maxima are dominated by the time dependence of the filters.}
\label{fig:Delta_H}
\end{figure}

\section{The benchmark USR and OF models in detail and further results}\label{app:benchmark_models}

In this Appendix, we provide additional details about the models considered in the main text, their dynamics, and a more comprehensive exploration of the backreaction in benchmarks considered in the literature.

\subsection{Ultra-Slow-Roll}\label{subsec:benchmark_usr}

\paragraph{The benchmark model.} 
For simplicity, rather than choosing an analytic potential and tuning its parameters until it produces a suitable non-attractor phase, we reverse-engineer a potential that realizes a smooth, otherwise standard USR history~\cite{Franciolini:2022pav}. We begin by prescribing a three-stage profile for $\eta_H$,
\begin{equation}
\begin{aligned}
\eta_H(N) ={}& \frac{\eta_{\rm in}+\eta_{\rm out}}{2}
+\frac{\eta_{\rm USR}-\eta_{\rm in}}{2}
\tanh\!\left(\frac{N-N_{\rm in}}{\delta N}\right)
+\frac{\eta_{\rm out}-\eta_{\rm USR}}{2}
\tanh\!\left(\frac{N-N_{\rm end}}{\delta N}\right) \;.
\end{aligned}
\label{etaH-param}
\end{equation}
The first and last plateaus describe the attractor phases before and after USR, while the middle plateau controls the non-attractor evolution. Closely following the benchmark of Ref.~\cite{Franciolini:2022pav}, but omitting its additional intermediate $\eta_H=0$ stage, in our benchmark model considered in the main text we choose $\eta_{\rm in}=-1.6875\times10^{-2}$, $\eta_{\rm USR}=2.709$, $\eta_{\rm out}=-0.554$, $N_{\rm in}=15.5$, $N_{\rm end}\simeq18.01$, and $\delta N=0.5$. The intermediate value is close to the ideal USR limit $\eta_H=3$ and drives a rapid decrease of $\epsilon_H$; the final negative value brings the system back toward an attractor solution.

At $N=0$ we set $\varphi=0$ and choose $\dot{\varphi}>0$. The initial slow-roll parameter and the normalization of the potential are fixed by
$\epsilon_H(0)\simeq r/16=3.125\times10^{-4}$ and
$V_0/M_{\rm Pl}^4\simeq(3\pi^2/2)rA_s=1.55\times10^{-10}$, using $n_s=0.965$, $r=0.005$, and $A_s=2.1\times10^{-9}$. The value of $n_s$ fixes the initial plateau to $\eta_{\rm in}\simeq(n_s-1+r/4)/2$.

The relations collected in Appendix~\ref{sec:slow} allow us to reconstruct the potential directly from this profile. Defining $x\equiv\varphi/M_{\rm Pl}$ and $u\equiv\ln(V/V_0)$, and using the monotonic branch with $\dot{\varphi}>0$, we integrate
\begin{equation}
\frac{d\epsilon_H}{dN}=2\epsilon_H\left(\epsilon_H-\eta_H\right) \;,
\qquad
\frac{dx}{dN}=\sqrt{2\epsilon_H} \;,
\qquad
\frac{du}{dN}=-2\epsilon_H\frac{3-\eta_H}{3-\epsilon_H} \;.
\end{equation}
The resulting monotonic relation $x(N)$ can be inverted, giving $V(\varphi)=V_0\exp[u(N(x))]$. We tabulate the reconstructed potential and its first three derivatives and interpolate them as functions of $\varphi$. We numerically reconstruct the potential using a grid in the number of e-folds with $\Delta N=2.5\times10^{-3}$.
Figure~\ref{fig:benchmark} summarizes the resulting benchmark. The reconstructed potential remains monotonic and develops only a mild shoulder around the non-attractor region. As the field crosses this feature, $|\eta_H|$ becomes of order unity and $\epsilon_H$ falls by several orders of magnitude; the final attractor phase then drives its recovery.
The prescribed histories $\eta_H(N)$, $\epsilon_H(N)$, and $H(N)$ are used only for this one-time reconstruction. Once $V(\varphi)$ has been obtained, these template histories are discarded. From that point onward, the potential and its derivatives are the only model-dependent inputs: $H$, $\epsilon_H$, and $\eta_H$ are determined anew by the dynamical equations and can therefore respond self-consistently to the backreaction.

As emphasized in Ref.~\cite{Barnert:2026ijs}, tuning the background evolution to obtain the desired features can produce sharp structures in the higher derivatives of the reconstructed potential, even for smooth transitions in $\eta_H$. We deliberately adopt this construction to explore the phenomenology of such potentials. Their higher derivatives, and hence the backreaction, can be sensitive to the smoothing adopted in the numerical reconstruction. Once the potential and its smoothing scale are fixed, however, the predictions are well defined within the regime of validity of our approximation. We leave applications to specific microphysically motivated inflationary models for future work.

\paragraph{Results for other benchmarks.}
We extend the comparison of Section~\ref{sec:results} to Cases II and III of Ref.~\cite{Caravano:2024moy}, characterized by, respectively,  repulsive and attractive inflaton self-interactions,  corresponding to, respectively, a negative and a positive cubic derivative of the potential at the end of USR.~\footnote{While in this work $\dot{\phi} > 0$ is assumed, the opposite choice $\dot{\phi} < 0$ is made in Ref.~\cite{Caravano:2024moy}. The two choices are equivalent up to the $\phi \to - \phi$ change. This, in particular, changes the sign of the cubic derivative of the potential, so that cases II and III are characterized by, respectively, positive and negative cubic potential derivative in ~\cite{Caravano:2024moy}. For the models considered here, and with our convention $\dot\varphi>0$,
the sign of $V'''$ at the end of USR is that of
$\eta_{\rm USR}+\eta_{\rm out}-3$, where $\eta_{\rm USR}$ and
$\eta_{\rm out}$ are the plateau values of $\eta_H$ before and after the transition out of USR, see Eq.~\eqref{etaH-param}.} We reconstruct the potentials using $\eta_{\rm II}=3$ for Case II and $\eta_{\rm II}=4.5$ for Case III, while keeping $\eta_{\rm I}=0$, $\eta_{\rm III}=-0.5$, and $\delta N=0.5$, and choose the USR duration for each case as in Ref.~\cite{Caravano:2024moy}. The lattice simulations and the mean-field computation are performed as for Case I in Section~\ref{sec:results}.

These two cases differ from Case I in the exit from USR. In Case I, the potential near the local maximum is close to an inverted parabola, and $V_{,\varphi\varphi\varphi}$ is small. In Cases II and III, the exit is sharper and $V_{,\varphi\varphi\varphi}$ is large, negative in Case II and positive in Case III. The corresponding backreaction force on the inflaton is $-\frac12 V_{,\varphi\varphi\varphi}\langle\delta\phi^2\rangle$. It therefore accelerates the field in Case II and slows it down in Case III, suppressing or enhancing $\mathcal P_\zeta$, respectively. Our mean field includes this effect on the homogeneous background, but evolves each mode linearly on that background. The lattice, instead, evolves the full non-linear Klein--Gordon equation at each spatial grid point, retaining the non-linear interactions in the evolution of the fluctuations themselves. At quadratic order in the fluctuations, the perturbation equation of Section~\ref{sec:system} contains the term $\frac12 V_{,\varphi\varphi\varphi}(\delta\phi^2-\langle\delta\phi^2\rangle)$, controlled by the same cubic derivative that enters the mean-field backreaction. Cases II and III therefore allow us to test the mean field where non-linearities beyond the homogeneous backreaction can become relevant.

While Fig.~\ref{fig:lattice-case-I} shows results for $\mathcal P_{\zeta,\mathrm{tree}}^{\max}=1$, in this appendix we present Cases II and III for $\mathcal P_{\zeta,\mathrm{tree}}^{\max}=0.1$. This choice is motivated by the stronger non-linear effects in Cases II and III, which are not fully captured by our mean-field treatment. Indeed, at the larger amplitude, which we have also simulated but do not show, the lattice and mean-field results differ by large factors, especially in Case III. This is because, in Case III, many field trajectories get trapped in the local minimum of the potential, which makes the power spectrum less informative on the full dynamics. The smaller amplitude allows us to compare the two approaches in a regime where these additional non-linear effects are less pronounced.

\begin{figure}[t]
\centering
\includegraphics[width=\textwidth]{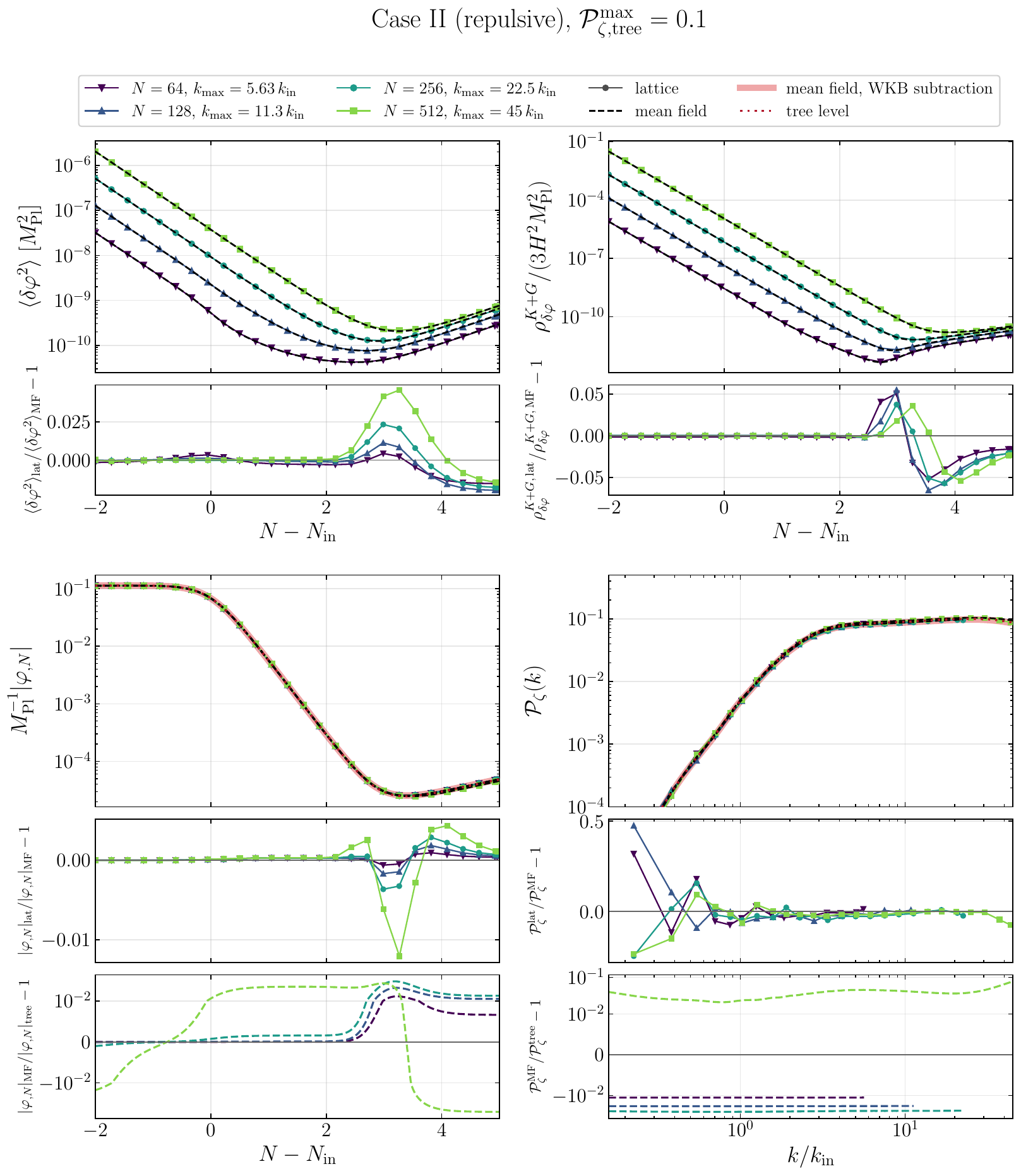}
\caption{USR Case II (repulsive), $\mathcal P_{\zeta,\mathrm{tree}}^{\max}=0.1$, in the same format as Fig.~\ref{fig:lattice-case-I}. The insets zoom on the late-time variance and energy, shown on a linear scale.}
\label{fig:lattice-case-II}
\end{figure}

Let us first consider Case II, shown in Figure~\ref{fig:lattice-case-II}. The upper panels of the figure show the bare (i.e., without WKB subtraction) variance and energy density of the fluctuations. We see that these quantities are dominated by the largest lattice momenta and grow as $k_{\rm max}^2$ and $k_{\rm max}^4$, respectively. This induces a dependence of these two quantities on the number of lattice points $N$. The same dependence was noted in Case I, shown in the main text, during the earlier part of the evolution, but it was then reduced by the mode amplification that takes place in the USR phase. The smaller value of $\mathcal P_{\zeta,\mathrm{tree}}^{\max}$ considered here results in a smaller amplification of these modes, so that the UV sensitivity of the bare variance and energy density persists until the end of the simulation. 

The lower panels of the figure show the effect of the backreaction on the inflaton velocity and on $\mathcal P_\zeta$. The results shown arise from the competition between the two backreaction channels. The energy of the UV modes increases Hubble friction, which slows the field down and enhances $\mathcal P_\zeta$. Conversely, as discussed above, in Case II the effective force $-\frac12 V_{,\varphi\varphi\varphi}\langle\delta\phi^2\rangle$ accelerates the field and suppresses $\mathcal P_\zeta$. The effective force grows as $k_{\rm max}^2$, while the energy grows as $k_{\rm max}^4$. As a result, the effective force dominates for $N\leq256$, where $\mathcal P_\zeta$ is suppressed by a few percent with respect to the tree level, while for $N=512$ the energy of the UV modes takes over and $\mathcal P_\zeta$ is enhanced by a similar amount. After WKB subtraction, the mean-field $\mathcal P_\zeta$ coincides with the tree-level one at the percent level (red band). 

Lattice and mean field agree to better than $1\%$ for all quantities and lattice sizes up to $N-N_{\rm in} \simeq 2.5$, shortly before the end of USR. Around the end of USR, where $|V_{,\varphi\varphi\varphi}|$ is largest, short-lived differences appear: the lattice variance exceeds the mean-field one by up to $\simeq 5\%$, the fluctuation energy differs by up to $\pm 6\%$, and the field velocity by at most $\simeq 1\%$. 
The transients in the variance and in the velocity grow with $N$, which indicates that the non-linear interactions are partly fed by the UV modes. These transients decay within about one e-fold. At the end of the simulation, the lattice variance and energy are $\simeq2\%$ below the mean-field values for all $N$, and the field velocity agrees to $\simeq 0.1\%$. As a consequence, the lattice and mean-field power spectra agree within a few percent for $k\gtrsim k_{\rm in}$. The larger scatter of the lattice spectrum at the largest scales is statistical, as only a few lattice modes contribute there. 

In summary, in Case II, the non-linearities beyond the homogeneous backreaction remain small, and the mean field captures reasonably well the lattice dynamics, including its UV dependence.

\begin{figure}[t]
\centering
\includegraphics[width=\textwidth]{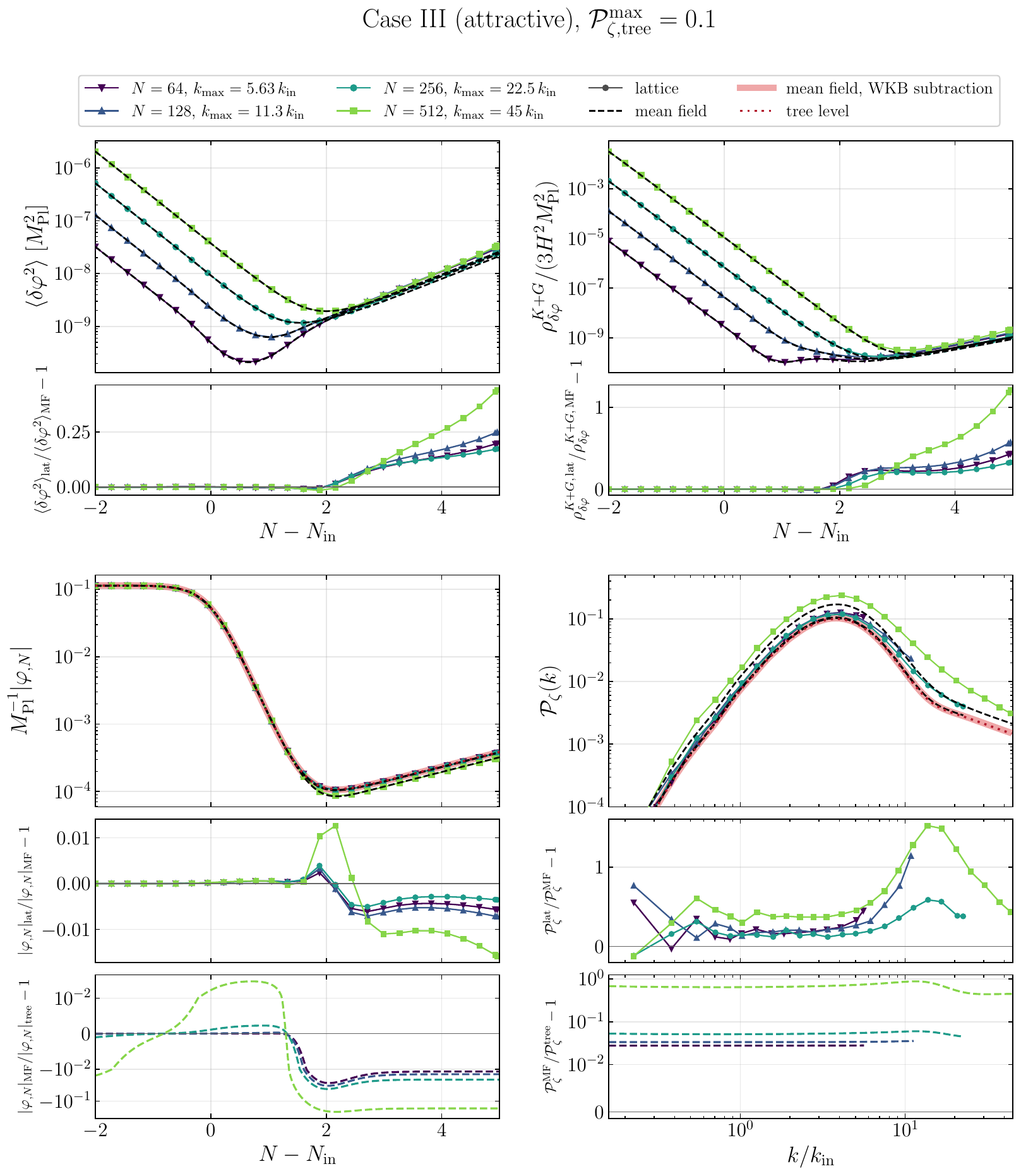}
\caption{USR Case III (attractive), $\mathcal P_{\zeta,\mathrm{tree}}^{\max}=0.1$, in the same format as Fig.~\ref{fig:lattice-case-II}.}
\label{fig:lattice-case-III}
\end{figure}

We now consider Case III, shown in Figure~\ref{fig:lattice-case-III}. As in Case II, the upper panels show that, during USR, the bare variance and energy density of the fluctuations are dominated by the largest lattice momenta and grow as $k_{\rm max}^2$ and $k_{\rm max}^4$, respectively. Contrary to Case II, however, this UV sensitivity does not persist until the end of the simulation. After USR, the amplified modes dominate, and the late-time variance and energy density become nearly independent of the number of lattice points $N$, as in Case I. Although $\mathcal P_{\zeta,\mathrm{tree}}^{\max}$ is the same as in Case II, the field velocity after USR is about one order of magnitude larger here (see the lower-left panel), so that the same $\mathcal P_\zeta$ corresponds to a larger field variance, $\langle\delta\phi^2\rangle \propto \epsilon_H \mathcal P_\zeta$. 

The lower panels of the figure show the effect of the backreaction on the inflaton velocity and on $\mathcal P_\zeta$. As we discussed, in this case both backreaction channels contribute to slowing the field down and, therefore, to enhance $\mathcal P_\zeta$, which is indeed greater than the tree level one for all lattice sizes. The magnitude of the increase is however very sensitive to the lattice size. In the mean field computation, the enhancement at the peak is a few percent for $N\leq256$ and $70\%$ for $N=512$. After WKB subtraction, the mean-field $\mathcal P_\zeta$ again coincides with the tree-level one at the percent level (red band).

Lattice and mean field agree for all quantities and lattice sizes up to $N-N_{\rm in}\simeq 1.5$, which corresponds to the end of USR, sharper here than in Case II. After the end of USR, the field velocity still agrees at the percent level, which shows that the backreaction on the homogeneous background is correctly captured. The fluctuations, however, grow more on the lattice. At the end of the simulation, the lattice variance exceeds the mean-field one by $\simeq 17$--$25\%$ for $N\leq256$ and by $\simeq45\%$ for $N=512$, and the fluctuation energy is larger by $\simeq35$--$60\%$ for $N\leq256$ and by up to about a factor of two for $N=512$. Correspondingly, the lattice power spectrum lies $10$--$25\%$ above the mean-field one at $k\lesssim5\,k_{\rm in}$ for all lattice sizes, and up to a factor of a few above it at smaller scales. Since this excess is present for all lattice sizes, including the smallest one, where the UV contribution is negligible, it does not originate from the UV vacuum modes. We therefore attribute it to the term $\frac12 V_{,\varphi\varphi\varphi}(\delta\phi^2-\langle\delta\phi^2\rangle)$ in the evolution of the fluctuations, which is not captured by our mean field and which becomes relevant for the attractive self-interaction of Case III. The growth of the excess with the number of lattice points $N$, especially at $k\gtrsim5\,k_{\rm in}$, shows however that this non-linear term is itself fed by the UV modes, so that the two effects are not fully separable on the lattice.

In summary, in Case III, the mean field still captures well the backreaction on the homogeneous background, but the non-linearities in the evolution of the fluctuations beyond the homogeneous backreaction become significant, and enhance the fluctuations on the lattice by tens of percent.

The lattice results exhibit a clear dependence on the number of grid points $N$, and hence on the UV cutoff at fixed box size, in all three cases. This dependence is also reproduced by our bare mean-field calculation. However, we cannot attribute the departures from the linear solution entirely to UV modes, as additional contributions from non-linear dynamics beyond the mean-field approximation appear to be present. These contributions are negligible in Case I, small and transient in Case II, and more significant in Case III, where they enhance the fluctuations by tens of percent. On the contrary, after adiabatic subtraction, the mean-field correction to $\mathcal P_\zeta$ remains at the percent level in all three cases, as shown by the red bands.

\subsection{Oscillatory Feature model}
\label{subsec:benchmark_oscillatory}

\paragraph{The benchmark model.} As a second benchmark we use the resonant-feature potential of Ref.~\cite{Inomata:2022yte}, a quadratic hilltop with a short train of wiggles superimposed on it,
\begin{equation}
V(\varphi)=V_0\left[
1-\frac{1-n_s}{2}\frac{\varphi^2}{2M_{\rm Pl}^2}
+2c\,\epsilon_0\,D(\varphi)
\left(-1+\cos\frac{\varphi-\varphi_0}{\sqrt{2\epsilon_0}\,\Lambda M_{\rm Pl}}\right)\right] \;,
\label{eq:V-osc}
\end{equation}
where the wiggles are switched on and off by the smooth top hat
\begin{equation}
D(\varphi)=
\frac{1}{2}\left[1+\tanh\frac{\varphi-\varphi_0}{\sqrt{2\epsilon_0}\,\Lambda M_{\rm Pl}}\right]
\frac{1}{2}\left[1+\tanh\frac{\varphi_s-\varphi}{\sqrt{2\epsilon_0}\,\Lambda M_{\rm Pl}}\right] \;,
\qquad
\epsilon_0=\frac{\varphi_0^2}{2M_{\rm Pl}^2}\left(\frac{1-n_s}{2}\right)^2 \;.
\label{eq:D-osc}
\end{equation}
The feature therefore occupies the field interval $\varphi_0<\varphi<\varphi_s$, and both the period of the wiggles and the width of the switch-on/off are set by the single scale $\sqrt{2\epsilon_0}\,\Lambda M_{\rm Pl}$. The mechanism is parametric resonance rather than a non-attractor phase: a mode whose comoving frequency matches the rate at which the background crosses successive wiggles is amplified, so the curvature spectrum develops a narrow peak at $k$ a factor of order $\sqrt{c}/\Lambda$ above the horizon scale at $\varphi_0$, rather than at the horizon scale itself as in USR.

We use the parameters of Ref.~\cite{Inomata:2022yte},
\begin{equation}
n_s=0.97 \;,\qquad
\epsilon_0=10^{-6} \;,\qquad
\frac{\varphi_0}{M_{\rm Pl}}=9.428\times10^{-2} \;,\qquad
\frac{V_0}{M_{\rm Pl}^4}=3.338\times10^{-13} \;,
\label{eq:params-osc}
\end{equation}
together with the three cases of Table~\ref{tab:osc-cases}, which differ in the wiggle amplitude $c$, the wiggle scale $\Lambda$, and the extent $\varphi_s$ of the feature.

\begin{table}[t]
\centering
\begin{tabular}{cccccc}
\toprule
case & $c$ & $\Lambda$ & $\varphi_s/M_{\rm Pl}$ &
$(\varphi_s-\varphi_0)/(\sqrt{2\epsilon_0}\Lambda M_{\rm Pl})$ & wiggles \\
\midrule
A & $0.1392$ & $0.04$ & $0.0956942$ & $25.0$ & $4$ \\
B & $0.2113$  & $0.10$ & $0.0971084$ & $20.0$ & $3$ \\
C & $0.19$  & $0.10$ & $0.0972923$ & $21.3$ & $3$ \\
\bottomrule
\end{tabular}
\caption{The three cases of the OF model, Eqs.~\eqref{eq:V-osc}--\eqref{eq:D-osc}. The last two columns give the length of the feature in units of the wiggle scale and the resulting number of full oscillations. The parameters are chosen as in Ref.~\cite{Inomata:2022yte}, apart from the value of $c$ in cases A and B (see the text for details).}
\label{tab:osc-cases}
\end{table}

\begin{figure}
\centering
\begin{subfigure}[b]{\textwidth}
\centering
\includegraphics[width=\textwidth]{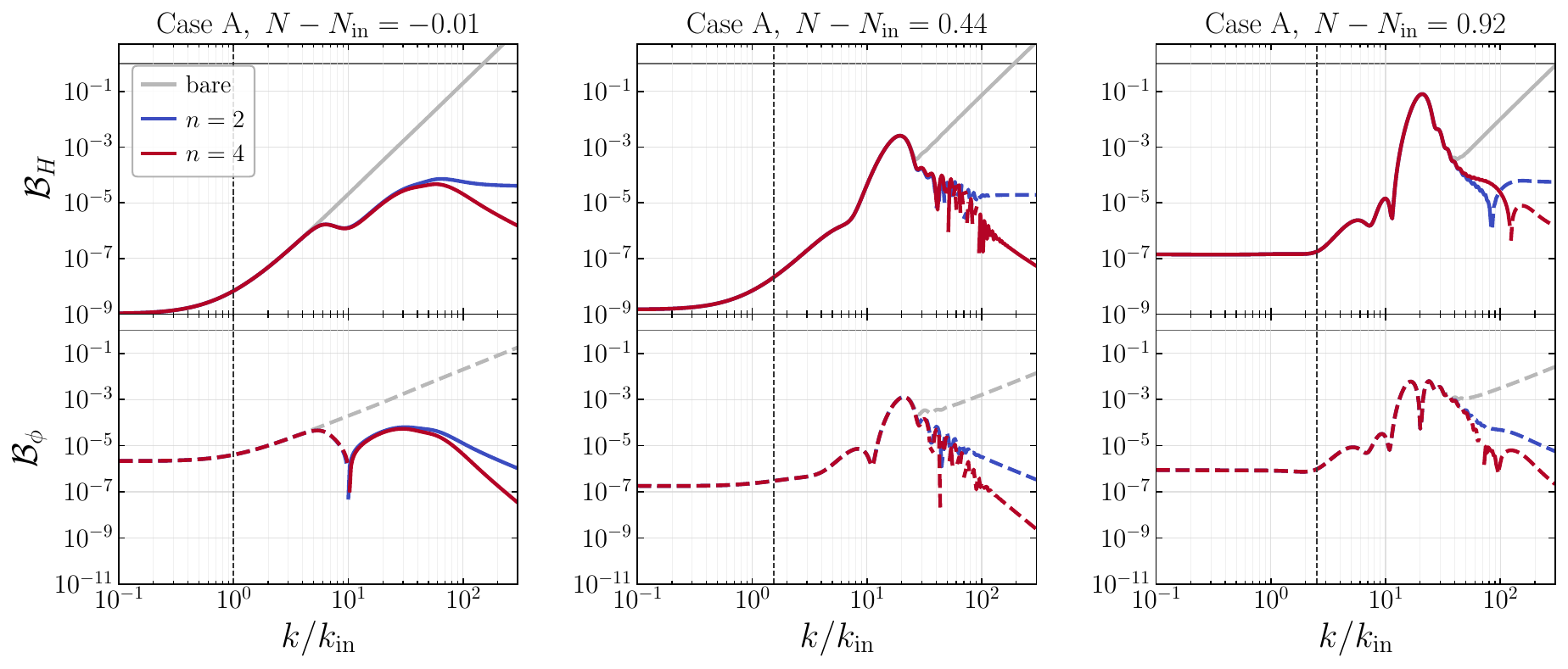}
\end{subfigure}
\vspace{1mm}
\begin{subfigure}[b]{\textwidth}
\centering
\includegraphics[width=\textwidth]{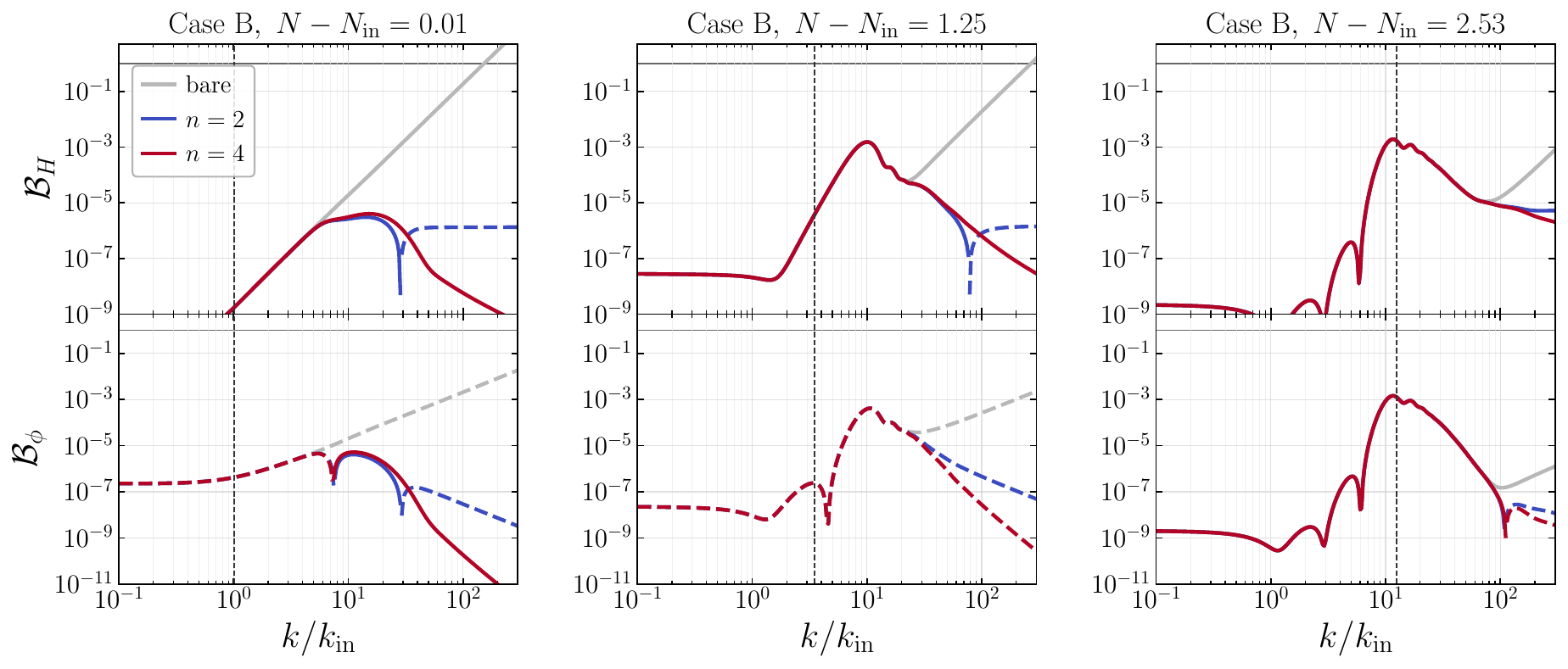}
\end{subfigure}
\caption{Backreaction integrands $\mathcal B_H$ (top row of each panel) and $\mathcal B_\varphi$ (bottom row) versus $k/k_{\rm in}$ for cases A and B of Table~\ref{tab:osc-cases}. The left, central and right column display the integrands at, respectively, the onset, the middle and the end of each feature. The integrands have been obtained with numerical evolutions with mean field backreaction with, respectively, no WKB subtraction, second order, and fourth order subtractions for the gray, blue and red curves (with solid segments for positive values and dashed for negative). The three curves coincide across the full range of momenta where perturbations have been enhanced, showing that the adiabatic subtraction does not affect the peak. The vertical dashed line marks $k=aH$ and the horizontal line marks unity. }
\label{fig:osc-BHBphi-k}
\end{figure}

\begin{figure}
\centering
\includegraphics[width=\textwidth]{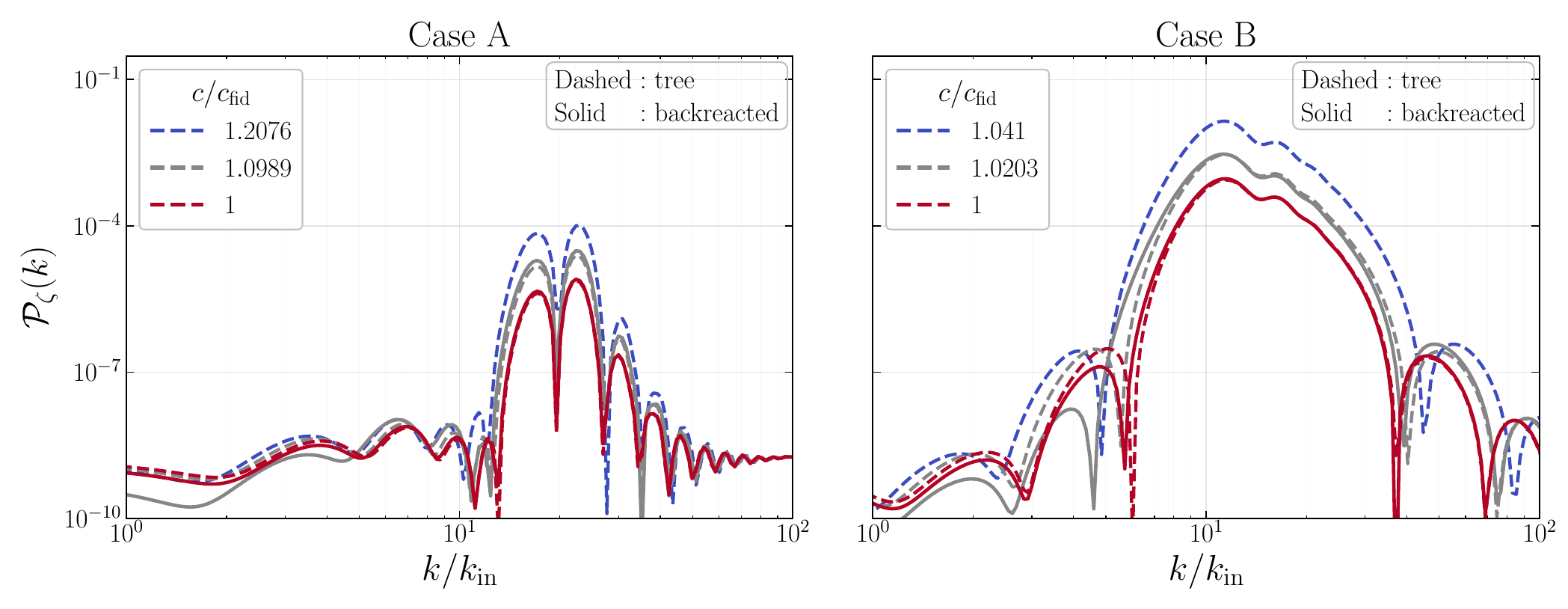}
\caption{Curvature power spectra for the oscillatory-feature Cases A (left) and B (right). Different colors denote the values of $c/c_{\mathrm{fid}}$ (where $c_{\mathrm{fid}}$ is the value given in Table~\ref{tab:osc-cases}) indicated in the legends. Dashed curves show the tree-level spectra, while solid curves include homogeneous backreaction with WKB-filtered adiabatic subtraction. Solid curves are shown only for integrations that reach the final evaluation time.}
\label{fig:Cases_A_B_scan}
\end{figure}

The background is initialized in slow roll at $\varphi_{\rm in}/M_{\rm Pl}=0.070$, well before the feature, and integrated on the branch with $\dot\varphi>0$; the field then crosses the wiggles and rolls for a further four e-folds past $\varphi_s$. As shown in Fig.~\ref{fig:benchmark}, the resulting $\epsilon_H$ stays close to $\epsilon_0=10^{-6}$ throughout, while $\eta_H$ executes one large, rapidly oscillating excursion per wiggle, reaching large values of $|\eta_H|\sim6$; the overall variation of the potential itself, $10^6(V/V_0-1)\sim-70$, is five orders of magnitude smaller than in the USR benchmark.

The benchmark used in the main text is case C of Table \ref{tab:osc-cases}. This benchmark is the only one of the three listed in the table for which the effective mass in the mode equation, $|\mu^2|\simeq3c\,(aH)^2/\Lambda^2$, stays small enough ($\lesssim60\,(aH)^2$, against $\sim400\,(aH)^2$ in case A). As a consequence, backreaction is small. 

\paragraph{Results for other benchmarks.} 
Here, we discuss Cases A and B of Table \ref{tab:osc-cases}, which exhibit stronger backreaction than Case C. The parameters of all three cases are taken from Ref.~\cite{Inomata:2022yte}, except for $c$, which controls the amplitude of the feature. For Cases A and B, we reduce this parameter relative to the values adopted in Ref.~\cite{Inomata:2022yte}, from $0.203$ to $0.1392$ in Case A and from $0.22$ to $0.2113$ in Case B. The reason for this reduction is that, for the values chosen in~\cite{Inomata:2022yte}, the field amplification in these two cases results in a too strong backreaction for our numerical algorithm to converge. This problem could be solved by decreasing the amplitude of the $c$ parameter. 

Figure~\ref{fig:osc-BHBphi-k} shows the backreaction integrands for Cases A and B at three different times during the oscillatory feature. We note that the excited modes are concentrated in a narrower range of scales in Case A, reflecting the higher oscillation frequency of its potential, which is more than twice that of Case B. In both cases, the resonant amplification of the perturbations produces pronounced peaks in the integrands. As for the OF benchmarks discussed in the main text, these peaks occur at momentum scales well separated from the UV regime where the unphysical vacuum contributions become dominant. Adiabatic subtraction removes these UV contributions while leaving the physically amplified modes unaffected. Thanks to this separation of scales, even bare lattice simulations can remain insensitive to UV artifacts, provided that the hard UV cutoff is appropriately chosen.

In Figure~\ref{fig:Cases_A_B_scan}, we compare the tree-level and backreacted curvature power spectra for three different values of the feature amplitude $c$. The smallest value of $c$ shown in each panel corresponds to that given in Table~\ref{tab:osc-cases} and used in Figure~\ref{fig:osc-BHBphi-k}. As $c$ increases, the resonant amplification becomes stronger, and backreaction effects become more significant. For the largest values of $c$ shown, our numerical evolution with backreaction fails to converge, and therefore no backreacted spectrum is displayed. Interestingly, Case A requires a substantially larger reduction of $c$ relative to Ref.~\cite{Inomata:2022yte} than Case B to obtain a convergent evolution, consistent with its stronger resonant production.

\section{All-scale backreaction beyond the decoupling limit}
\label{sec:Bphi_allscale}
In this Appendix, we extend the calculation of the backreaction on the background evolution by including metric perturbations and abandoning the decoupling limit.

\subsection{Klein\texorpdfstring{--}{--}Gordon equation} We start from the Klein--Gordon equation. Here, we use an alternative method, namely the requirement that the one-point function of the scalar field vanishes, extending Refs.~\cite{Sloth:2006az,Sloth:2006nu} without taking any slow-roll or super-horizon limit, including time derivatives of $\delta\phi$ as well as gradient terms.  Our goal is ultimately to compare it with the simple expression~\eqref{eq:KG_backreaction} in the decoupling limit derived in the main text, which, unlike our treatment here, does not take into account metric perturbations. 

Our starting point is the action~\eqref{S-phi}, together with the splitting~\eqref{eq:definition_phi_background}. Working in the spatially flat gauge
\begin{equation}
h_{ij}=a^2\delta_{ij},
\qquad
N=1+\alpha_{1}+\alpha_{2}+\dots,
\qquad
N^i=\partial^i\chi_{1}+\partial^i\chi_2+\dots,
\end{equation}
we have the following solutions for the (non-dynamical) lapse and shift at  first order:
\begin{align} \alpha_1 &= \frac{\dot\varphi}{2HM_{\rm Pl}^2}\, \delta\phi, \label{eq:alpha1_constraint} \\ \partial^2\chi_1 &= -\frac{\dot\varphi}{2HM_{\rm Pl}^2}\,
\delta\dot\phi
+
\frac{1}{2M_{\rm Pl}^2}
\left(
\frac{\ddot\varphi}{H}
-
\frac{\dot\varphi\dot H}{H^2}
\right)\,\delta\phi ,\label{eq:chi1_constraint} \end{align}
and no second-order lapse or shift is needed for our purposes.

After inserting the lapse constraint,  the linear action is:
\begin{equation}
S_1
=
\int dt\,d^3x\,a^3
\Biggl[
\dot\varphi\,\delta\dot\phi
-
V'(\varphi)\delta\phi
\Biggr].
\label{eq:S1}
\end{equation}

Integrating Eq.~\eqref{eq:S1} with respect to time, and dropping a total time derivative interaction $\propto{\rm d}/{\rm d}t \left[a^3\dot\varphi\,\delta\phi
\right]$ that does not contribute to correlation functions of $\delta\phi$~\cite{Braglia:2024zsl}, we recast the linear action as
\begin{equation}
S_1
=
-
\int d^4x\,a^3\,
[\ddot\varphi+3H\dot\varphi+V'(\varphi)]\,
\delta\phi\equiv
-
\int d^4x\,a^3\,
{\cal E}_\varphi\,
\delta\phi ,
\label{eq:S1_tadpole}
\end{equation}
which would be zero in the absence of perturbations, where the background field obeys the classical equations of motion.
The goal of this Appendix, however, is to derive the equation for the scalar field taking into account the 1-loop back-reaction term on the evolution, which is most transparently defined through the tadpole
condition. This condition identifies $\varphi$ with the expectation value of the full field:
\begin{equation}
\varphi(t)=\langle\phi(t,\mathbf{x})\rangle\qquad \Leftrightarrow\qquad\langle\delta\phi(t,\mathbf{x})\rangle=0.
\label{eq:tadpole}
\end{equation}
We compute the tadpole using the operator formulation of the in-in formalism~\cite{Maldacena:2002vr,Weinberg:2005vy}. We work in the interaction picture, where operators are evolved using the free Hamiltonian, which we identify here as the quadratic Hamiltonian for $\delta\phi$, and states are evolved using the interacting Hamiltonian, which, up to cubic order, is simply
\begin{equation}
H_I=H_I^{(1)}+H_I^{(3)}=-L_I^{(1)}-L_I^{(3)},    
\end{equation}
and where we can use the linear relation between the conjugate momentum to $\delta\phi$ and its velocity, i.e. $\pi_{\delta\phi}=a^3\delta{\dot \phi}$. 

The in-in one-point function at 1-loop is thus
\begin{align}
\langle\delta\phi(t,\,\mathbf{x})\rangle_{\rm 1,loop}
&=i
\int_{-\infty}^{t}dt'\,
\left\langle
\left[
H_I (t'),\delta\phi_I(x)
\right]
\right\rangle,
\label{eq:tadpole_inin}\\
&=
-\int_{-\infty}^{t}dt'\,a^3(t')
\,
G_R(t,t')
\left[
{\cal E}_{\varphi}(t')
+
B_\phi(t')
\right],
\label{eq:tadpole_complete}
\end{align}
where
\begin{equation}
B_\phi
=-\frac{1}{a^3}
\left\langle
\frac{\delta S_3}{\delta(\delta\phi)}
\right\rangle,
\label{eq:Bphi_definition}
\end{equation}
and
\begin{equation}
G_R(t,t')
\equiv
i\,\Theta(t-t')
\left[
u_{k=0}(t)u_{k=0}^*(t')
-
u_{k=0}^*(t)u_{k=0}(t')
\right]
\end{equation}
is the retarded Green function constructed out of the interaction-picture mode functions $u_k(t)$ evolved with the free Hamiltonian. The limit $k\to0$ has to be taken at the end of the calculation and it is a consequence of the $\delta^{(3)}(\mathbf{x})$ enforced by the one-point function~\cite{Seery:2007we,Braglia:2025cee,Braglia:2025qrb}.
Imposing the tadpole condition~\eqref{eq:tadpole} therefore gives
\begin{equation}
\ddot\varphi+3H\dot\varphi+V'(\varphi)+B_\phi=0.
\label{eq:KG_with_Bphi}
\end{equation}

Below, we derive the explicit form of $B_\phi$ from its definition in Eq.~\eqref{eq:Bphi_definition} and compute explicitly the extra pieces that we neglected in the main text.
We start from the cubic action in the perturbations, which, before substituting the lapse constraint, can be written as~\cite{Arroja:2008ga}:
\begin{align}
S_3
=
\int dt\,d^3x\,a^3
\Bigg\{&
-\frac12\alpha_1\,\delta\dot\varphi^2
-\frac12\alpha_1\,(\partial\delta\varphi)^2
-\delta\dot\varphi\,
\partial_i\chi_1\,\partial_i\delta\varphi
\nonumber\\
&
+\dot\varphi\,\alpha_1\,
\partial_i\chi_1\,\partial_i\delta\varphi
+\dot\varphi\,\alpha_1^2\,\delta\dot\varphi
+2HM_{\rm Pl}^2\alpha_1^2\,\partial^2\chi_1
\nonumber\\
&
+\frac{M_{\rm Pl}^2}{2}\alpha_1
\left[
(\partial^2\chi_1)^2
-
(\partial_i\partial_j\chi_1)^2
\right]
\nonumber\\
&
+\left(
3M_{\rm Pl}^2H^2
-\frac12\dot\varphi^2
\right)\alpha_1^3
-\frac12V''\,\alpha_1\,\delta\varphi^2
-\frac16V'''\,\delta\varphi^3
\Bigg\}.
\label{eq:S3}
\end{align}
By substituting the constraints~\eqref{eq:alpha1_constraint} and~\eqref{eq:chi1_constraint} into
Eq.~\eqref{eq:S3}, we get
\begin{align}
S_3
=
\int dt\,d^3x\,a^3
\Bigg\{
&
-\frac{\dot\varphi}{4HM_{\rm Pl}^2}
\delta\phi\,\delta\dot\phi^2
-\frac{\dot\varphi}{4HM_{\rm Pl}^2}
\delta\phi\,(\partial\delta\phi)^2
\nonumber\\
&-\delta\dot\phi\,
\partial_i\partial^{-2}
\Bigg[
-\frac{\dot\varphi}{2HM_{\rm Pl}^2}\delta\dot\phi
+
\frac{1}{2M_{\rm Pl}^2}
\left(
\frac{\ddot\varphi}{H}
-
\frac{\dot\varphi\dot H}{H^2}
\right)
\delta\phi
\Bigg]
\partial_i\delta\phi
\nonumber\\
&+
\Bigg[
\frac{3\dot\varphi^3}{8HM_{\rm Pl}^4}-\frac{\dot\varphi^5}{16H^3M_{\rm Pl}^6}
-\frac{\dot\varphi}{4HM_{\rm Pl}^2}V''
-\frac16V'''
\nonumber\\
&\hspace{12mm}
+
\frac{\dot\varphi^2}{8HM_{\rm Pl}^4}
\left(
\frac{\ddot\varphi}{H}
-
\frac{\dot\varphi\dot H}{H^2}
\right)
\Bigg]\delta\phi^3
\nonumber\\
&+
\frac{\dot\varphi^3}{8H^2M_{\rm Pl}^4}
\delta\phi^2\delta\dot\phi
\nonumber\\
&+
\frac{\dot\varphi}{4H}\delta\phi
\Bigg\{
\Bigg[
-\frac{\dot\varphi}{2HM_{\rm Pl}^2}\delta\dot\phi
+
\frac{1}{2M_{\rm Pl}^2}
\left(
\frac{\ddot\varphi}{H}
-
\frac{\dot\varphi\dot H}{H^2}
\right)\delta\phi
\Bigg]^2
\nonumber\\
&\hspace{4mm}
-
\Bigg[
\partial_i\partial_j\partial^{-2}
\Bigg(
-\frac{\dot\varphi}{2HM_{\rm Pl}^2}\delta\dot\phi
+
\frac{1}{2M_{\rm Pl}^2}
\left(
\frac{\ddot\varphi}{H}
-
\frac{\dot\varphi\dot H}{H^2}
\right)\delta\phi
\Bigg)
\Bigg]^2
\Bigg\}
\Bigg\},
\label{eq:S3_delta_phi_only}
\end{align}
which is entirely written in terms of $\delta\phi$ and recovers the result of~\cite{Maldacena:2002vr}.

We now want to vary this cubic action with respect to $\delta\phi$ and take its spatial average. The calculation is tedious, but straightforward, although there are two subtleties:
\begin{enumerate}
    \item We have to perform a few integrations by parts. Spatial integration by parts is innocuous: they produce boundary terms that vanish at spatial infinity. On the other hand, temporal integration by parts is in principle not innocuous, as total time derivative interactions can contribute to in-in correlators; see, e.g.,~\cite{Burrage:2011hd,Arroja:2011yj,Rigopoulos:2011eq,Braglia:2024zsl,Kawaguchi:2024lsw}. When varying the term $S^{(3)}\supset\,\delta\phi^2\delta\dot\phi\,(\dot\varphi^3/8H^2M_{\rm Pl}^4)$, a total time derivative is produced. However, the resulting interaction does not contain any (powers of) $\delta\dot\phi$ inside the total time derivative and therefore does not contribute to the tadpole~\cite{Braglia:2024zsl}.
    \item The cubic action includes non-local interactions arising from the linear Arnowitt--Deser--Misner (ADM) constraints. For those interactions we will use the identity (see e.g.~\cite{Seery:2007we}):
    \begin{equation}
\left\langle
(\partial_i\partial^{-2}f)
(\partial_i g)
\right\rangle
=
-\langle fg\rangle,
\label{eq:inverse_identity}
\end{equation}
and
\begin{equation}
\left\langle
(\partial_i\partial_j\partial^{-2}f)
(\partial_i\partial_j\partial^{-2}g)
\right\rangle
=
\langle fg\rangle.
\end{equation}
\end{enumerate}

Combining all the contributions we get the expression for $B_\phi$:
\begin{align}
B_\phi
={}&
\frac{\dot\varphi}{4HM_{\rm Pl}^2}
\left[
\left\langle\delta\dot\phi^2\right\rangle
+
\left\langle(\partial\delta\phi)^2\right\rangle
\right]
-\frac{1}{a^3}\frac{d}{dt}
\left[
a^3\frac{\dot\varphi}{2HM_{\rm Pl}^2}
\left\langle\delta\phi\,\delta\dot\phi\right\rangle_{\rm s}
\right]+
\Biggl\{
\frac{1}{a^3}\frac{d}{dt}
\left[
a^3\frac{\dot\varphi^3}{8H^2M_{\rm Pl}^4}
\right]
\nonumber\\
&
-3\left[
\frac{3\dot\varphi^3}{8HM_{\rm Pl}^4}-\frac{\dot\varphi^5}{16H^3M_{\rm Pl}^6}
-\frac{\dot\varphi}{4HM_{\rm Pl}^2}V''
-\frac16V'''
+\frac{\dot\varphi^2}{8HM_{\rm Pl}^4}
\left(
\frac{\ddot\varphi}{H}
-\frac{\dot\varphi\dot H}{H^2}
\right)
\right]
\Biggr\}
\left\langle\delta\phi^2\right\rangle
\nonumber\\
&+
\frac{1}{a^3}\frac{d}{dt}
\left\{
a^3\left[
-\frac{\dot\varphi}{2HM_{\rm Pl}^2}
\left\langle\delta\phi\,\delta\dot\phi\right\rangle_{\rm s}
+\frac{1}{2M_{\rm Pl}^2}
\left(
\frac{\ddot\varphi}{H}
-\frac{\dot\varphi\dot H}{H^2}
\right)
\left\langle\delta\phi^2\right\rangle
\right]
\right\}
\nonumber\\
&-
\frac{1}{a^3}\frac{d}{dt}
\left[
a^3\frac{\dot\varphi}{4HM_{\rm Pl}^2}
\left\langle\delta\phi\,\delta\dot\phi\right\rangle_{\rm s}
\right]
-\frac{1}{4M_{\rm Pl}^2}
\left(
\frac{\ddot\varphi}{H}
-\frac{\dot\varphi\dot H}{H^2}
\right)
\left\langle\delta\phi\,\delta\dot\phi\right\rangle_{\rm s}
\nonumber\\
&-
\frac{\dot\varphi}{6HM_{\rm Pl}^2}
\left(
\frac{\ddot\varphi}{H}
-\frac{\dot\varphi\dot H}{H^2}
\right)
\left[
-\frac{\dot\varphi}{2HM_{\rm Pl}^2}
\left\langle\delta\phi\,\delta\dot\phi\right\rangle_{\rm s}
+\frac{1}{2M_{\rm Pl}^2}
\left(
\frac{\ddot\varphi}{H}
-\frac{\dot\varphi\dot H}{H^2}
\right)
\left\langle\delta\phi^2\right\rangle
\right]
\nonumber\\
&-
\frac{1}{a^3}\frac{d}{dt}
\left\{
a^3\frac{\dot\varphi^2}{6H^2M_{\rm Pl}^2}
\left[
-\frac{\dot\varphi}{2HM_{\rm Pl}^2}
\left\langle\delta\phi\,\delta\dot\phi\right\rangle_{\rm s}
+\frac{1}{2M_{\rm Pl}^2}
\left(
\frac{\ddot\varphi}{H}
-\frac{\dot\varphi\dot H}{H^2}
\right)
\left\langle\delta\phi^2\right\rangle
\right]
\right\},
\label{eq:Bphi_final}
\end{align}

where we have defined
\begin{align}
\left\langle
\delta\phi\,\delta\dot\phi
\right\rangle_{\rm s}\equiv\frac12
\left\langle
\delta\phi\,\delta\dot\phi
+
\delta\dot\phi\,\delta\phi
\right\rangle.
\end{align}

The expression above is rather cumbersome. It is valid at all scales, and no slow-roll approximation is made. Note that, upon expanding the time derivatives, some terms proportional to $\delta\ddot\phi$ appear. Those terms can be simplified using the free equations of motion of $\delta\phi$ in the interaction picture, so that the whole backreaction onto the Klein--Gordon equation only depends on the correlators
\begin{equation}
\left\langle\delta\phi^2\right\rangle,
\qquad
\left\langle
\delta\phi\,\delta\dot\phi
\right\rangle_{\rm s},
\qquad
\left\langle\delta\dot\phi^2\right\rangle,
\qquad
\left\langle(\partial\delta\phi)^2\right\rangle .
\end{equation}

A few checks are in order. To perform these checks, let us introduce the Hubble-flow slow-roll parameters defined as~\cite{Schwarz:2001vv} 
\begin{equation}
\epsilon_1=\epsilon_H\equiv\epsilon,\qquad\epsilon_{n+1}\equiv\frac1H\frac{\dd\ln\epsilon_n}{\dd t}\qquad{\rm for}\,\,n\geq1.
\end{equation} 
We have the following useful identities
\begin{align}
\dot\varphi^2
&=
2\epsilon H^2M_{\rm Pl}^2,
\qquad 
\frac{\ddot\varphi}{H\dot\varphi}
=
-\epsilon+\frac{\epsilon_2}{2},
\qquad 
\frac{\dddot\varphi}{H^2\dot\varphi}
=
2\epsilon^2
-\frac52\epsilon\epsilon_2
+\frac14\epsilon_2^2
+\frac12\epsilon_2\epsilon_3,
\\
V
&=
M_{\rm Pl}^2H^2(3-\epsilon),
\qquad
V'
=
-H\dot\varphi
\left(
3-\epsilon+\frac{\epsilon_2}{2}
\right),
\\
\frac{V''}{H^2}
&=
6\epsilon
-\frac32\epsilon_2
-2\epsilon^2
+\frac52\epsilon\epsilon_2
-\frac14\epsilon_2^2
-\frac12\epsilon_2\epsilon_3,
\label{eq:V2}
\\
\dot\varphi \frac{V'''}{H^3}
&=
\left(
4\epsilon^3
-12\epsilon^2
-9\epsilon^2\epsilon_2
+9\epsilon\epsilon_2
+3\epsilon\epsilon_2^2
+\frac72\epsilon\epsilon_2\epsilon_3
-\frac12\epsilon_2^2\epsilon_3
-\frac32\epsilon_2\epsilon_3
-\frac12\epsilon_2\epsilon_3^2
-\frac12\epsilon_2\epsilon_3\epsilon_4
\right ).
\label{eq:V3_decoupling}
\end{align}

Furthermore, let us write explicitly the equation of motion for the linear perturbations in terms of these slow-roll parameters:
\begin{equation}
\delta\ddot\phi
+
3H\delta\dot\phi
-
\frac{\nabla^2}{a^2}\delta\phi
+
m_{\rm eff}^2\delta\phi
=
0,
\label{eq:deltaphi_linear}
\end{equation}
with
\begin{equation}
m_{\rm eff}^2
=
V''
-
\frac{1}{a^3M_{\rm Pl}^2}
\frac{d}{dt}
\left(
\frac{a^3\dot\varphi^2}{H}
\right)=
H^2\frac{\epsilon_2}{4}
\left(
2\epsilon-\epsilon_2-2\epsilon_3-6
\right)
.
\label{eq:meff_hubble_flow}
\end{equation}
We will use this equation to replace terms containing $\delta\ddot\phi$ in the expression for $B_\phi$.

\paragraph{Neglecting metric perturbations.} We expect to recover correctly the one in the main text by neglecting metric perturbations. The easiest way to see this is that the terms in $B_\phi$ arising from the metric perturbations---the coupling to gravity of the scalar field---carry inverse powers of the reduced Planck mass. Neglecting such terms, we recover:
\begin{equation}
\label{eq:nometric}
B_\phi^{\rm no\,metric}\equiv\lim_{\mpl\to\infty}B_\phi
=\frac12V'''(\varphi)
\left\langle\delta\phi^2\right\rangle,
\end{equation}
as expected.

\paragraph{Decoupling limit.} In the decoupling limit, we assume that $\epsilon\ll1$, but impose no conditions on $\epsilon_{n>1}$ which could eventually become transiently large. First, we rewrite $B_\phi$ using the Hubble-flow slow-roll parameters and using the linear equations of motion for $\delta\phi$ as 
\begin{align}
B_\phi
={}&
\frac12V'''
\left\langle\delta\phi^2\right\rangle
\nonumber\\
&
+
\sigma_\varphi
\frac{\sqrt{2\epsilon}}{M_{\rm Pl}}
\Bigg[
\frac{\epsilon-6}{6}
\left\langle\delta\dot\phi^2\right\rangle
+
\frac{9-\epsilon}{6}
\left\langle(\partial\delta\phi)^2\right\rangle
+
\frac{H\epsilon_2(4\epsilon-3)}{24}
\left\langle
\delta\phi\,\delta\dot\phi
\right\rangle_{\rm s}
\nonumber\\
&
+
\frac{H^2}{8}
\left(
24\epsilon
+
18\epsilon\epsilon_2
-
8\epsilon^2
-
\epsilon\epsilon_2^2
-
18\epsilon_2
-
3\epsilon_2^2
-
6\epsilon_2\epsilon_3
\right)
\left\langle\delta\phi^2\right\rangle
\Bigg],
\label{eq:Bphi_hubble_flow_exact}
\end{align}
where, for the moment, we have left $V'''$ explicit. 

Taking the $\epsilon\ll1$ limit, we obtain:

\begin{align}
B_\phi^{\rm dec}
&=
\frac12V'''
\left\langle\delta\phi^2\right\rangle
\nonumber\\
&
\,\,\,\,\,\,\,+
\sigma_\varphi
\frac{\sqrt{2\epsilon}}{M_{\rm Pl}}
\Bigg[
-
\left\langle\delta\dot\phi^2\right\rangle
+
\frac32
\left\langle(\partial\delta\phi)^2\right\rangle
-
\frac{H\epsilon_2}{8}
\left\langle
\delta\phi\,\delta\dot\phi
\right\rangle_{\rm s}
\nonumber\\
&
\,\,\,\,\,\,\,
-
\frac{3H^2}{8}
\epsilon_2
\left(
6+\epsilon_2+2\epsilon_3
\right)
\left\langle\delta\phi^2\right\rangle
\Bigg]
+
\mathcal O(\epsilon^{3/2})\nonumber\\
&\simeq -
\sigma_\varphi
\frac{H^2}{4M_{\rm Pl}\sqrt{2\epsilon}}\,
\epsilon_2\epsilon_3
\left(
3+\epsilon_2+\epsilon_3+\epsilon_4
\right)
\left\langle\delta\phi^2\right\rangle\nonumber\\
&
\,\,\,\,\,\,\,+
\sigma_\varphi
\frac{\sqrt{2\epsilon}}{M_{\rm Pl}}
\Bigg[
-
\left\langle\delta\dot\phi^2\right\rangle
+
\frac32
\left\langle(\partial\delta\phi)^2\right\rangle
-
\frac{H\epsilon_2}{8}
\left\langle
\delta\phi\,\delta\dot\phi
\right\rangle_{\rm s}
\nonumber\\
&
\,\,\,\,\,\,\,
-
\frac{3H^2}{8}
\epsilon_2
\left(
6+\epsilon_2+2\epsilon_3
\right)
\left\langle\delta\phi^2\right\rangle
\Bigg]
+
\mathcal O(\epsilon^{3/2}),
\label{eq:Bphi_decoupling}
\end{align}
where $\sigma_\varphi\equiv{\rm sign}\, \dot\varphi=+1$ in our model and, in the second equality, we have explicitly kept only the leading-order terms for $V'''$ in the decoupling limit. From this form we can conclude that:
\begin{itemize}
    \item In the strict decoupling limit where $\epsilon\ll1$, but the higher-order slow-roll parameters are not necessarily small, and can reach $\mathcal{O}(1)$ as in the scenarios considered in the main text, $V'''$ is indeed the dominant interaction, as all the terms in the fifth and sixth lines in Eq.~\eqref{eq:Bphi_decoupling} are suppressed by one power of $\epsilon$ with respect to the fourth line, so we can safely neglect them in the main text. In this limit, we therefore recover the results obtained in the absence of metric perturbations in the main text, which we obtained a few lines above by formally sending $\mpl\to\infty$---see Eq.~\eqref{eq:nometric}. 
    
    On the other hand, during standard slow roll, where $\epsilon,\,\vert\epsilon_2\rvert,\,\vert\epsilon_3\rvert,\vert\epsilon_4\rvert\ll1$, $V'''$ is no longer necessarily the dominant interaction. Indeed, we have already clarified the distinction between decoupling limit and slow-roll approximation above Eq.~\eqref{def-veh}.
    \item In the super-horizon limit, neglecting time and spatial derivatives of $\delta\phi$, we recover the super-horizon and slow-roll limit of Ref.~\cite{Sloth:2006nu} (our $\epsilon_2$ is their $\eta$). 
    \item During exact USR, we have $V'(\varphi)=0$ exactly so that $\ddot\varphi+3H\dot\varphi=0$. This equation can be used to derive
    \begin{align}        \epsilon_2&\,\,\underset{{\rm USR}}{=}\,\,-6+2\epsilon,\,\\\epsilon_3&\,\,\underset{{\rm USR}}{=}\,\,2\epsilon,\,\\\epsilon_4&\,\,\underset{{\rm USR}}{=}\,\,\epsilon_2\,\,\underset{{\rm USR}}{=}\,\,-6+2\epsilon.\,
    \end{align}
    In this limit one can verify that $V'''=0$ identically\footnote{To verify that, one has to keep all the slow-roll parameters, without taking any slow-roll or decoupling approximation.}. The dominant contribution during exact USR is therefore:
    \begin{equation}
B_\phi^{\rm USR,\,dec}
=
\sigma_\varphi
\frac{\sqrt{2\epsilon}}{M_{\rm Pl}}
\left[
-\left\langle\delta\dot\phi^2\right\rangle
+
\frac{3}{2}
\left\langle(\partial\delta\phi)^2\right\rangle
+
\frac{3H}{4}
\left\langle
\delta\phi\,\delta\dot\phi
\right\rangle_{\rm s}
\right]
+
\mathcal{O}\!\left(\epsilon^{3/2}\right).
\end{equation}
    where we have neglected the term proportional to $H^2\langle\delta\phi^2\rangle$ as it is suppressed by an additional power of $\epsilon$ compared to the terms displayed. We note, however, that we never have an exact stage of USR in the examples considered in our text.
\end{itemize}

\subsection{All-scale backreaction on the Friedmann equations}
\label{sec:BH_allscale}
We now move to the backreaction on the Friedmann equation. While the backreaction on the Klein--Gordon equation was easily defined by the tadpole condition, an analogous condition is not available for the Friedmann equations. We therefore compute it using the following method.

We consistently expand both sides of Einstein's equations,
\begin{equation}
    G_{\mu\nu}[g+\delta g]
    =
    \frac{1}{M_{\rm Pl}^2}
    T_{\mu\nu}[g+\delta g,\varphi+\delta\phi]
\end{equation}
up to second order in perturbations. The  $0^{\rm th}$ order gives the Friedmann equations. After averaging, the linear order perturbations, upon using the linear constraints, are proportional to $\langle\delta\phi\rangle$, which is zero because of the tadpole condition we imposed in the previous section. At second order, we can write schematically an effective stress tensor sourcing the background as
\begin{equation}
    \tau_{\mu\nu}
    =
    \left\langle T_{\mu\nu}^{(2)}\right\rangle
    -
    M_{\rm Pl}^2
    \left\langle G_{\mu\nu}^{(2)}\right\rangle.
    \label{eq:tmunuquad}
\end{equation}

Before moving on, we would like to note that a complementary canonical formulation of gravitational backreaction has recently been developed in Ref.~\cite{Cintia:2026qds}, where the quantum-corrected background is identified directly with the metric one-point function and its evolution is obtained from expectation values of the ADM constraint and evolution equations, similarly to what we did for the Klein--Gordon equation above. The analysis of~\cite{Cintia:2026qds} also emphasizes the dependence of the explicit representation of the backreaction on the background--fluctuation split and on the choice of slicing.

The quantities in Eq.~\eqref{eq:tmunuquad} were computed in Ref.~\cite{Abramo:1997hu} in gauge-invariant variables, and read
\begin{align}
\tau_{00}
={}&
M_{\rm Pl}^2
\left[
12H\left\langle\Phi\dot\Phi\right\rangle
-3\left\langle\dot\Phi^2\right\rangle
+9\left\langle(\partial\Phi)^2\right\rangle
\right]
\nonumber\\
&
+\frac12\left\langle\delta\dot\varphi_{\rm GI}^2\right\rangle
+\frac12\left\langle(\partial\delta\varphi_{\rm GI})^2\right\rangle
+\frac12V''
\left\langle\delta\varphi_{\rm GI}^2\right\rangle
\nonumber\\
&
+2V'
\left\langle\Phi\,\delta\varphi_{\rm GI}\right\rangle.
\label{eq:tau00_ABM}
\end{align}
In order to make connection with the ADM formalism in the spatially flat gauge, we note that:
\begin{align}
\Phi
&=
\alpha+\frac{d}{dt}(a^2\chi),
\\
\Psi
&=
-a^2H\chi,
\\
\delta\varphi_{\rm GI}
&=
\delta\phi+\dot\varphi a^2\chi.
\end{align}
One can easily verify that $\Phi=\Psi$ using the equations of motion for $\delta\phi$, which allows us to replace any instance of $\Phi$ with $\delta\phi$ using
\begin{equation}
\Phi
=
a^2\frac{\dot\varphi}{2M_{\rm Pl}^2}
\partial^{-2}
\left(
\delta\dot\phi -\frac{H\epsilon_2}{2}\delta\phi
\right).
\end{equation}

Defining
\begin{equation}
\rho_{\delta\phi}\equiv\tau_{00},\qquad{\rm and}\qquad
p_{\delta\phi}\equiv\frac{1}{3a^2}\delta^{ij}\tau_{ij},
\end{equation}
we find:
\begin{align}
\rho_{\delta\phi}
={}&
\frac12
\left\langle\delta\dot\phi^2\right\rangle
+
\frac{1}{2a^2}
\left\langle
\partial_i\delta\phi\,
\partial_i\delta\phi
\right\rangle
\nonumber\\
&
-
\frac{H^2}{8}
\left(
4\epsilon^2
-6\epsilon\epsilon_2
-12\epsilon
+\epsilon_2^2
+2\epsilon_2\epsilon_3
+6\epsilon_2
\right)
\left\langle\delta\phi^2\right\rangle
\nonumber\\
&
+
\frac{H^3\epsilon}{2}
\left(
2\epsilon^2
-3\epsilon\epsilon_2
-4\epsilon
+\epsilon_2\epsilon_3
+2\epsilon_2
+6
\right)
a^2
\left\langle
\delta\phi\,
\partial^{-2}
\left(
\delta\dot\phi
-\frac{H\epsilon_2}{2}\delta\phi
\right)
\right\rangle_{\rm s}
\nonumber\\
&
+
\frac{H^2\epsilon}{2}
\left(
\epsilon-\epsilon_2-7
\right)
a^2
\left\langle
\partial^{-2}
\left(
\delta\dot\phi
-\frac{H\epsilon_2}{2}\delta\phi
\right)
\left(
\delta\dot\phi
-\frac{H\epsilon_2}{2}\delta\phi
\right)
\right\rangle_{\rm s}
\nonumber\\
&
-
\frac{H^4\epsilon}{4}
\left(
2\epsilon^3
-3\epsilon^2\epsilon_2
-2\epsilon^2
+\epsilon\epsilon_2\epsilon_3
+\epsilon\epsilon_2
+10\epsilon
+30
\right)
a^4
\left\langle
\left[
\partial^{-2}
\left(
\delta\dot\phi
-\frac{H\epsilon_2}{2}\delta\phi
\right)
\right]^2
\right\rangle ,
\label{eq:rho_BR_reduced}
\end{align}
and
\begin{align}
p_{\delta\phi}
={}&
\frac12
\left\langle\delta\dot\phi^2\right\rangle
-
\frac{1}{6a^2}
\left\langle
\partial_i\delta\phi\,
\partial_i\delta\phi
\right\rangle
-
\frac{4}{3}H\epsilon
\left\langle
\delta\phi\,\delta\dot\phi
\right\rangle_{\rm s}
\nonumber\\
&
+
\frac{H^2}{24}
\left(
36\epsilon^2
-26\epsilon\epsilon_2
-60\epsilon
+3\epsilon_2^2
+6\epsilon_2\epsilon_3
+18\epsilon_2
\right)
\left\langle\delta\phi^2\right\rangle
\nonumber\\
&
-
\frac{H^3\epsilon}{2}
\left(
6\epsilon^2
-7\epsilon\epsilon_2
-4\epsilon
+\epsilon_2^2
+\epsilon_2\epsilon_3
-6
\right)
a^2
\left\langle
\delta\phi\,
\partial^{-2}
\left(
\delta\dot\phi
-\frac{H\epsilon_2}{2}\delta\phi
\right)
\right\rangle_{\rm s}
\nonumber\\
&
+
\frac{H^2\epsilon}{6}
\left(
7\epsilon
-3\epsilon_2
+22
\right)
a^2
\left\langle
\partial^{-2}
\left(
\delta\dot\phi
-\frac{H\epsilon_2}{2}\delta\phi
\right)
\left(
\delta\dot\phi
-\frac{H\epsilon_2}{2}\delta\phi
\right)
\right\rangle_{\rm s}
\nonumber\\
&
+
\frac{H^4\epsilon}{4}
\left(
6\epsilon^3
-7\epsilon^2\epsilon_2
+2\epsilon^2
+\epsilon\epsilon_2^2
+\epsilon\epsilon_2\epsilon_3
-3\epsilon\epsilon_2
-10\epsilon
+10
\right)
a^4
\left\langle
\left[
\partial^{-2}
\left(
\delta\dot\phi
-\frac{H\epsilon_2}{2}\delta\phi
\right)
\right]^2
\right\rangle,
\label{eq:Pi_BR_reduced}
\end{align}
where we have simplified the expression by expressing derivatives of the scalar field potential $V$ in terms of the slow-roll parameters. 
These expressions are rather long, and, as opposed to the backreaction terms on the Klein--Gordon equation, where we could simplify the inverse Laplacian via the spatial average, here some non-local terms are retained.

These expressions are nevertheless greatly simplified in the decoupling limit, where we get
\begin{align}
\rho_{\delta\phi}^{\rm dec}
={}&
\frac12
\left\langle\delta\dot\phi^2\right\rangle
+
\frac{1}{2a^2}
\left\langle
\partial_i\delta\phi\,
\partial_i\delta\phi
\right\rangle
-
\frac{H^2}{8}
\epsilon_2
\left(
6+\epsilon_2+2\epsilon_3
\right)
\left\langle\delta\phi^2\right\rangle
\nonumber\\
={}&
\frac12
\left\langle\delta\dot\phi^2\right\rangle
+
\frac{1}{2a^2}
\left\langle
\partial_i\delta\phi\,
\partial_i\delta\phi
\right\rangle
+
\frac12 V''
\left\langle\delta\phi^2\right\rangle
+
\mathcal{O}(\epsilon),
\end{align}
and
\begin{align}
p_{\delta\phi}^{\rm dec}
={}&
\frac12
\left\langle\delta\dot\phi^2\right\rangle
-
\frac{1}{6a^2}
\left\langle
\partial_i\delta\phi\,
\partial_i\delta\phi
\right\rangle
+
\frac{H^2}{8}
\epsilon_2
\left(
6+\epsilon_2+2\epsilon_3
\right)
\left\langle\delta\phi^2\right\rangle
\nonumber\\
={}&
\frac12
\left\langle\delta\dot\phi^2\right\rangle
-
\frac{1}{6a^2}
\left\langle
\partial_i\delta\phi\,
\partial_i\delta\phi
\right\rangle
-
\frac12 V''
\left\langle\delta\phi^2\right\rangle
+
\mathcal{O}(\epsilon),
\end{align}
where in the second equalities for each quantity we have used the expression~\eqref{eq:V2} at leading order in $\epsilon$. These expressions can be seen to be equivalent to the ones used in the main text.

\section{Equations in code variables}
\label{sec:code}

We collect here the dimensionless form of the equations used to evolve the background and the regularized inflaton fluctuations.

We choose the initial time $t_0$ such that $a(t_0)=1$ and $N(t_0)=0$, so that
\begin{equation}
a={\rm e}^N={\rm e}^{\int_{t_0}^{t}dt'\,H(t')} \;.
\label{a-code}
\end{equation}
We introduce the dimensionless variables
\begin{equation}
\begin{aligned}
x&\equiv\frac{\varphi}{M_{\rm Pl}} \;, &
y&\equiv\frac{1}{M_{\rm Pl}}\frac{d\varphi}{dN} \;, &
\widetilde V&\equiv\frac{V}{V_0} \;,
\widetilde V_0&\equiv\frac{V_0}{M_{\rm Pl}^4} \;, &
h&\equiv\frac{M_{\rm Pl}}{\sqrt{V_0}}H \;, &
\widetilde k&\equiv\frac{M_{\rm Pl}}{\sqrt{V_0}}k \;.
\end{aligned}
\label{code-variables}
\end{equation}
Here $V_0$ is the value of the potential at the initial time, and a comma followed by $x$ denotes differentiation with respect to $x$. Since $\widetilde k$ differs from $k$ only by a constant normalization, $d\ln\widetilde k=d\ln k$.

In these variables, the first and third equations in~\eqref{eqs-bck}, with the regularized spectra understood, become
\begin{equation}
\begin{aligned}
\frac{dy}{dN}+3y-\frac{y^3}{2}+\frac{\widetilde V_{,x}}{h^2}
&=\int d\ln\widetilde k\left(\frac{y^3}{2}{\cal B}_H-3y{\cal B}_\varphi\right) \;,\\
\frac{dh}{dN}+\frac{y^2}{2}h
&=-\int d\ln\widetilde k\,\frac{y^2h}{2}{\cal B}_H \;.
\end{aligned}
\label{eqsbck-code}
\end{equation}
The perturbations obey $v''+\omega^2v=0$, with $\omega^2\equiv(V_0/M_{\rm Pl}^2)\,\widetilde\omega^2$ and $\widetilde\omega^2\equiv\widetilde k^2+\widetilde\mu^2$. Writing
\begin{equation}
\widetilde{\mathcal H}\equiv\frac{M_{\rm Pl}}{\sqrt{V_0}}\mathcal H={\rm e}^Nh \;,
\qquad
\lambda\equiv\frac{\sqrt{V_0}}{M_{\rm Pl}} \;,
\end{equation}
Eq.~\eqref{eq:mu} takes the dimensionless form
\begin{equation}
\widetilde\mu^2\equiv\frac{\mu^2}{\lambda^2}
=-\widetilde{\mathcal H}^{\,2}\left[2-\frac{y^2}{2}-\frac{\widetilde V_{,xx}}{h^2}
-2\alpha\left(
\frac{3y^2}{2}-\frac{y^4}{4}
+y\frac{\widetilde V_{,x}}{h^2}
\right)\right] \;.
\label{M-code}
\end{equation}
The choices $\alpha=0$ and $\alpha=1$ correspond, respectively, to Eqs.~\eqref{eqv-nodg} and~\eqref{eqv-dg}. Here $y$ and $h$ are the live variables evolved by Eq.~\eqref{eqsbck-code}. In writing Eq.~\eqref{M-code}, and in Eq.~\eqref{eqv-code} below, we use the background identities at zeroth order in the fluctuation backreaction, in particular $\epsilon_H=y^2/2$. Their mean-field corrections are already quadratic in the fluctuations and would therefore modify the linear mode equation only at cubic order, and the resulting backreaction only at quartic order. They are consistently neglected in the present quadratic treatment.

We remove the rapidly oscillating vacuum phase by defining
\begin{equation}
\widetilde v\equiv\sqrt{2k}\,v\,{\rm e}^{ik\tau}
=\sqrt{2k}\,v\exp\left[i\widetilde k\int_0^N\frac{dn}{{\rm e}^nh(n)}\right] \;,
\qquad
\widetilde w\equiv\frac{d\widetilde v}{dN} \;.
\label{vw-code}
\end{equation}
Equation~\eqref{eq:MS} becomes
\begin{equation}
\frac{d\widetilde w}{dN}
+\left(1-\frac{y^2}{2}-2i\frac{\widetilde k}{\widetilde{\mathcal H}}\right)\widetilde w
+\frac{\widetilde\mu^2}{\widetilde{\mathcal H}^{\,2}}\widetilde v=0 \;.
\label{eqv-code}
\end{equation}
For later convenience, define
\begin{equation}
\mathcal A_k\equiv\widetilde w-
\left(1+i\frac{\widetilde k}{\widetilde{\mathcal H}}\right)\widetilde v \;.
\label{Ak-code}
\end{equation}
The bare spectra~\eqref{eq:definition_power_spectra} are
\begin{equation}
\calP_{\delta\phi}=
\frac{V_0}{4\pi^2M_{\rm Pl}^2}\frac{\widetilde k^2}{{\rm e}^{2N}}|\widetilde v|^2 \;,
\qquad
\calP_{\delta\dot\phi}=
\frac{V_0^2}{4\pi^2M_{\rm Pl}^4}\frac{\widetilde k^2h^2}{{\rm e}^{2N}}|\mathcal A_k|^2 \;.
\label{Pbare-code}
\end{equation}

We now translate the adiabatic prescription of Section~\ref{sec:adiabatic} into the same variables. Introducing the dimensionless conformal time
$\widetilde\tau\equiv\lambda\tau$, we define
\begin{equation}
\mathcal D\equiv\frac{d}{d\widetilde\tau}
=\widetilde{\mathcal H}\frac{d}{dN} \;.
\label{D-code}
\end{equation}
This is the same $\widetilde\mu^2$ of Eq.~\eqref{M-code} that enters the mode equation, so the mode equation and its subtraction below are guaranteed to use the same effective mass. At fixed comoving $\widetilde k$,
\begin{equation}
\partial_\tau\mu^2=\lambda^3\mathcal D\widetilde\mu^2 \;,
\qquad
\partial_\tau^2\mu^2=\lambda^4\mathcal D^2\widetilde\mu^2 \;,
\qquad
\mathcal D^2\widetilde\mu^2
=\widetilde{\mathcal H}^{\,2}
\left[\frac{d^2\widetilde\mu^2}{dN^2}
+\left(1+\frac{1}{h}\frac{dh}{dN}\right)
\frac{d\widetilde\mu^2}{dN}\right] \;.
\label{M-derivatives-code}
\end{equation}
These derivatives are taken along the background generated dynamically from $V(\varphi)$. An analytic evaluation of $\mathcal D^2\widetilde\mu^2$ requires the potential through $\widetilde V_{,xxxx}$, as accounted for in the interpolation described in Section~\ref{subsec:background}.

The adiabatic counterterms of Section~\ref{sec:adiabatic} are subtracted in the same variables. Collecting the adiabatic orders of Eq.~\eqref{explicit-FQ} and inserting the filters~\eqref{eq:wkb-smooth-filters}, which depend on momentum only through $\widetilde k/\widetilde{\mathcal H}=k/(aH)$, the two dimensionless subtracted combinations are
\begin{equation}
\begin{aligned}
\Delta_0&\equiv|\widetilde v|^2
-S_0\left[1
-S_2\frac{\widetilde\mu^2}{2\widetilde k^2}
+S_4\frac{3\widetilde\mu^{\,4}+\mathcal D^2\widetilde\mu^2}{8\widetilde k^4}\right] \;,\\[2mm]
\Delta_1&\equiv|\mathcal A_k|^2
-S_0\left[\frac{\widetilde k^2}{\widetilde{\mathcal H}^{\,2}}
+S_2\left(1+\frac{\widetilde\mu^2}{2\widetilde{\mathcal H}^{\,2}}\right)
-S_4\frac{\mathcal D^2\widetilde\mu^2+\widetilde\mu^{\,4}
-4\widetilde{\mathcal H}\mathcal D\widetilde\mu^2
+4\widetilde\mu^2\widetilde{\mathcal H}^{\,2}}
{8\widetilde k^2\widetilde{\mathcal H}^{\,2}}\right] \;,
\end{aligned}
\label{Delta-code}
\end{equation}
where the terms multiplied by $S_4$ are absent in the second-order subtraction. The counterterms in~\eqref{Delta-code} are real truncated series, not absolute squares, and therefore carry no modulus signs. As a normalization check, for $\widetilde\mu^2=0$ and $S_n=1$ they cancel the positive-frequency Minkowski mode $\widetilde v=1$, $\widetilde w=0$, giving $\Delta_0=\Delta_1=0$. The fourth-order regularized spectra are
\begin{equation}
\begin{aligned}
\left.\calP_{\delta\phi,\,{\rm reg}}\right|_4
&=\frac{V_0}{4\pi^2M_{\rm Pl}^2}
\frac{\widetilde k^2}{{\rm e}^{2N}}\Delta_0 \;,
\qquad
\left.\calP_{\delta\dot\phi,\,{\rm reg}}\right|_4
=\frac{V_0^2}{4\pi^2M_{\rm Pl}^4}
\frac{\widetilde k^2h^2}{{\rm e}^{2N}}\Delta_1 \;.
\end{aligned}
\label{Preg-code}
\end{equation}
Fourth order is essential here: as shown in Section~\ref{sec:adiabatic}, a second-order subtraction leaves the $\calP_{\delta\dot\phi}$ and gradient contributions logarithmically divergent in the UV.

Using the normalization of~\eqref{Preg-code} in the definitions~\eqref{calB}, with second-order subtraction for ${\cal B}_\varphi$ and fourth-order subtraction for ${\cal B}_H$, the backreaction integrands used in~\eqref{eqsbck-code} are
\begin{equation}
\begin{aligned}
{\cal B}_\varphi
&=\frac{\widetilde V_0}{24\pi^2}
\frac{\widetilde V_{,xxx}}{h^2y}
\frac{\widetilde k^2}{{\rm e}^{2N}}
\left.\Delta_0\right|_2 \;,
\qquad
{\cal B}_H
=\frac{\widetilde V_0}{4\pi^2}
\frac{\widetilde k^2}{y^2{\rm e}^{2N}}
\left(
\left.\Delta_1\right|_4
+\frac{\widetilde k^2}{3\widetilde{\mathcal H}^{\,2}}
\left.\Delta_0\right|_4
\right) \;.
\end{aligned}
\label{calB-code}
\end{equation}

Here $\left.\Delta_i\right|_n$ denotes subtraction through adiabatic order $n$, so that $\left.\Delta_i\right|_4=\Delta_i$ for $i=0,1$, with $\Delta_i$ given in~\eqref{Delta-code}, whereas
\begin{equation}
\left.\Delta_0\right|_2
=|\widetilde v|^2
-S_0\left[1-S_2\frac{\widetilde\mu^2}{2\widetilde k^2}\right] \;.
\end{equation}

The numerical evolution therefore consists of Eqs.~\eqref{eqsbck-code} and~\eqref{eqv-code}, with the mode integrals evaluated over the chosen grid in $\ln\widetilde k$ and with the filtered local counterterms~\eqref{Delta-code} subtracted mode by mode. No auxiliary adiabatic mode or adiabatic phase is evolved.

Finally, the background is initialized in slow roll with negligible backreaction and with $\widetilde V_{\rm in}=1$. We shift the field so that $x_{\rm in}=0$ and choose the branch with $y_{\rm in}>0$. Solving the slow-roll field equation together with the Friedmann constraint gives
\begin{equation}
\begin{aligned}
x_{\rm in}&=0, 
\qquad 
y_{\rm in}=\frac{3}{-\widetilde V_{,x,{\rm in}}}
\left(\sqrt{1+\frac{2\widetilde V_{,x,{\rm in}}^2}{3}}-1\right),
\qquad 
h_{\rm in}=\frac{1}{\sqrt6}
\sqrt{1+\sqrt{1+\frac{2\widetilde V_{,x,{\rm in}}^2}{3}}} \;.
\end{aligned}
\label{background-ic-code}
\end{equation}
Each mode is initialized sufficiently far inside the adiabatic regime,
$\widetilde k/\widetilde{\mathcal H}_{\rm in}\gg1$. In this limit the positive-frequency adiabatic state reduces to
\begin{equation}
\widetilde v_{\rm in}=1 \;,
\qquad
\widetilde w_{\rm in}=0 \;,
\label{mode-ic-code}
\end{equation}
up to corrections suppressed by powers of
$\widetilde{\mathcal H}_{\rm in}/\widetilde k$.

\bibliographystyle{apsrev}
\bibliography{paper-biblio}

\end{document}